\documentclass{aa}  

\usepackage{graphicx}
\usepackage{float}
\usepackage{txfonts}
\usepackage{lipsum}
\usepackage{subcaption}         
                                
\usepackage{lscape}             
                                
\usepackage{placeins}           
                                
\usepackage{hyperref}

\usepackage{natbib}
\bibpunct{(}{)}{;}{a}{}{,}

\newcommand{\sigdel}{$\sigma_{\rm \delta}$}
\newcommand{\fckc}{$f_{\rm ckc}$}
\newcommand{\qfgriz}{qf$_{\text{grism}}$}
\newcommand{\fracfield}{$23_{-1.8}^{+1.7}$\%}
\newcommand{\frachyp}{$59_{-10}^{+9}$\%}
\newcommand{\hsthyp}{\textit{HST}-Hyperion}

\usepackage{xcolor}

\newcommand{\orcid}[1]{\href{https://orcid.org/#1}{\includegraphics[width=10pt]{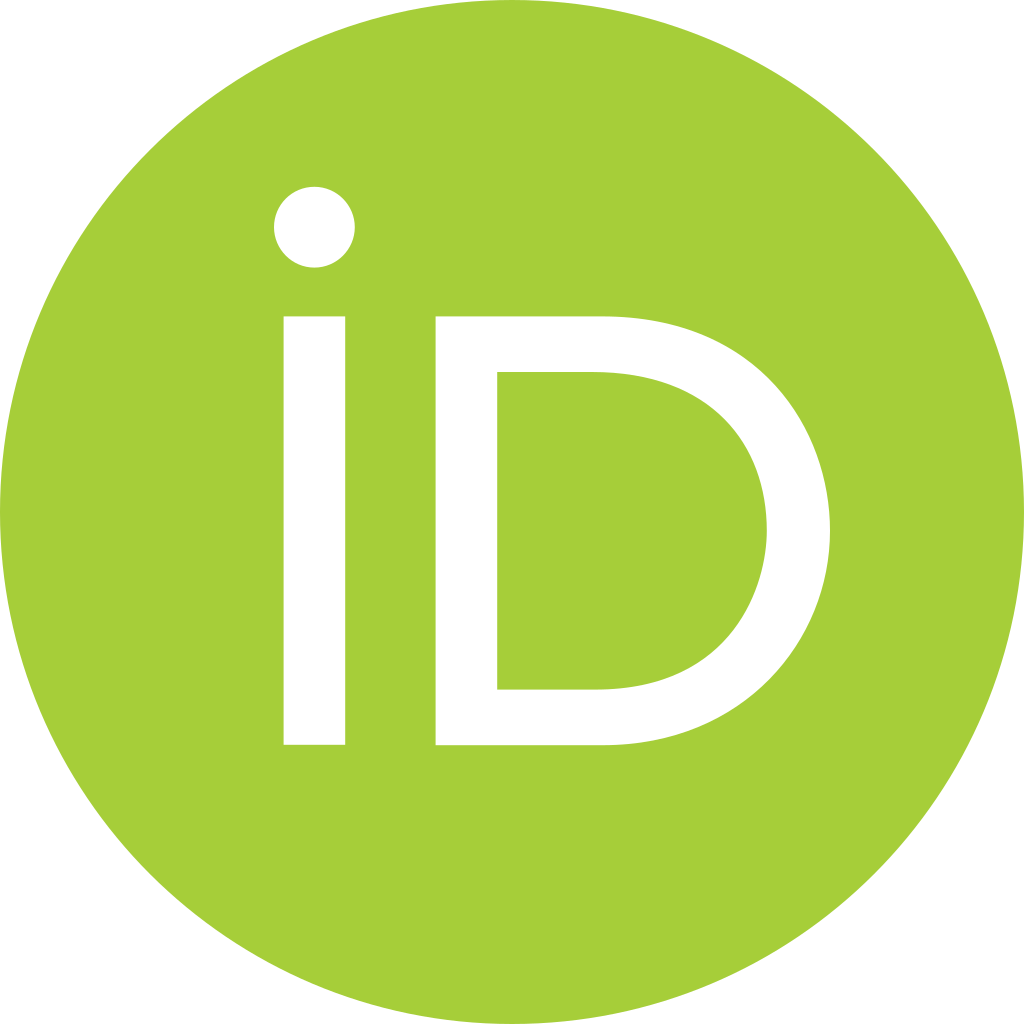}}}

\begin{document}

   \title{The \hsthyp \space Survey: Companion fraction and overdensity in a $z\sim2.5$ proto-supercluster}

   \author{F. Giddings\orcid{0009-0003-2158-1246} \inst{1}\thanks{email: finndg@hawaii.edu}
        \and B. C. Lemaux\orcid{0000-0002-1428-7036}\inst{2,3} 
        \and B. Forrest\orcid{0000-0001-6003-0541}\inst{3} 
        \and L. Shen\orcid{0000-0001-9495-7759}\inst{4,5}
        \and D. Sikorski\orcid{0000-0001-5796-2807} \inst{1}
        \and R. Gal\orcid{0000-0001-8255-6560}\inst{1} 
        \and O. Cucciati\orcid{0000-0002-9336-7551} \inst{6} 
        \and E. Golden-Marx\orcid{0000-0001-5160-6713}\inst{7} 
        \and W. Hu\orcid{0000-0003-3424-3230}\inst{4,5} 
        \and K. Ronayne\orcid{0000-0001-5749-5452}\inst{4} 
        \and E. Shah\orcid{0000-0001-7811-9042}\inst{3} 
        \and R. O. Amor\'{i}n\orcid{0000-0001-5758-1000}\inst{16}
        \and S. Bardelli\orcid{0000-0002-8900-0298}\inst{6} 
        \and D. C. Baxter\orcid{0000-0002-8209-2783} \inst{8} 
        \and L. P. Cassarà\orcid{0000-0001-5760-089X} \inst{9} 
        \and G. De Lucia\orcid{0000-0002-6220-9104}\inst{10,11} 
        \and F. Fontanot\orcid{0000-0003-4744-0188} \inst{10,11} 
        \and G. Gururajan \inst{11,12} 
        \and N. Hathi\orcid{0000-0001-6145-5090}\inst{13} 
        \and M. Hirschmann\orcid{0000-0002-3301-3321} \inst{14} 
        \and D. Hung\orcid{0000-0001-7523-140X}\inst{2} 
        \and L. Lubin\orcid{0000-0003-2119-8151} \inst{3} 
        \and D. B. Sanders\orcid{0000-0002-1233-9998} \inst{1} 
        \and D. Vergani\orcid{0000-0003-0898-2216} \inst{6}
        \and L. Xie\orcid{0000-0003-3864-068X} \inst{15} 
        \and E. Zucca\orcid{0000-0002-5845-8132}\inst{6}
        }

   \institute{Institute for Astronomy, University of Hawai‘i, 2680 Woodlawn Drive, Honolulu, HI 96822, USA
             \and Gemini Observatory, NSF NOIRLab, 670 N Aohoku Pl, Hilo, HI 96720 
             \and Department of Physics and Astronomy, University of California, Davis, One Shields Ave, Davis, CA 95616
             \and Department of Physics and Astronomy, Texas A\&M University, College Station, TX, 77843-4242 USA
             \and George P.\ and Cynthia Woods Mitchell Institute for Fundamental Physics and Astronomy, Texas A\&M University, College Station, TX, 77843-4242 USA
             \and INAF– Osservatorio di Astrofisica e Scienza dello Spazio di Bologna, Via Piero Gobetti 93/3, 40129 Bologna, Italy
             \and INAF - Osservatorio di Padova, Vicolo Osservatorio 5, 35122 Padova, Italy
             \and Department of Astronomy \& Astrophysics, University of California, San Diego, 9500 Gilman Dr, La Jolla, CA 92093, USA
             \and INAF – IASF Milano via A. Corti 12, 20133 Milano, Italy
             \and Italian National Institute for Astrophysics - Trieste Observatory via GB Tiepolo 11 - 34143 Trieste, Italy
             \and IFPU-Institute for Fundamental Physics of the Universe, Via Beirut 2, 34014 Trieste, Italy
             \and Scuola Internazionale Superiore Studi Avanzati (SISSA), Physics Area, Via Bonomea 265, 34136 Trieste, Italy
             \and Space Telescope Science Institute, 3700 San Martin Drive, Baltimore, MD 21218, USA
             \and Observatoire de Sauverny, Chemin Pegasi 51, 1290 Versoix, Switzerland
             \and Tianjin Astrophysics Center, Tianjin Normal University, Binshuixidao 393, 300384, Tianjin, China
             \and Instituto de Astrof\'{i}sica de Andaluc\'{i}a (CSIC), Apartado 3004, 18080 Granada, Spain
             }

   \date{Received March X, 2025}

  \abstract{We present a study of the galaxy merger and interaction activity within the Hyperion Proto-supercluster at $z \sim 2.5$ in an effort to assess the occurrence of galaxy mergers and interactions in contrast to the coeval field and their impact on the buildup of stellar mass in high-density environments at higher redshifts. For this work, we utilized data from the Charting Cluster Construction with VUDS and ORELSE Survey (C3VO) along with extensive spectroscopic and photometric datasets available for the COSMOS field --- including the \hsthyp \space Survey. To evaluate potential merger and interaction activity, we measured the fraction of galaxies with close kinematic companions (\fckc) both within Hyperion and the coeval field by means of a Monte Carlo (MC) methodology developed in this work that probabilistically employs our entire combined spectroscopic and photometric dataset. We validated our \fckc \space MC methodology on a simulated lightcone built from the GAlaxy Evolution and Assembly (GAEA) semi-analytic model, and we determined correction factors that account for the underlying spectroscopic sampling rate of our dataset. We find that galaxies in Hyperion have close kinematic companions $\gtrsim2.5\times$ more than galaxies in the field and measure a corrected \fckc \space $=$ \frachyp \space for Hyperion and a corrected \fckc \space $=$ \fracfield \space for the surrounding field; a $\gtrsim3\sigma$ difference. The enhancement in \fckc \space likely correlates to an enhancement in the merger and interaction activity within the high-density environment of Hyperion and matches the trend seen in other structures. The rate of merger and interactions within the field implied from our field \fckc \space measurement is well aligned with values measured from other observations in similar redshift ranges. The enhanced \fckc \space  measured within Hyperion suggests that merger and interaction activity play an important role in the mass growth of galaxies in denser environments at higher redshifts. }


   \titlerunning{Companion Fraction in Hyperion}
   \authorrunning{F. Giddings et al.}
   \maketitle
   \nolinenumbers

\section{Introduction}\label{sec:intro}

Galaxies embedded in large-scale structures are ideal settings for studying a wide variety of astrophysical phenomena. These large-scale structures, including galaxy clusters, help to map out the distribution of matter within our Universe and are interconnected at low redshifts through an underlying filamentary structure \citep{geller_mapping_1989,einasto_120-mpc_1997,colberg_clustering_2000,colless_2df_2001,evrard_galaxy_2002,dolag_simulating_2006}. Observations reveal that clusters form at the nodes of this underlying filamentary structure by accreting galaxies and groups of galaxies from the surrounding field \citep{frenk_galaxy_1996,eke_evolution_1998}.

Galaxy clusters occupy only a small volume fraction in the local Universe, but they have been studied extensively as their high galaxy densities offer a unique perspective into galaxy evolution in contrast to the evolution of ``coeval field'' galaxies of roughly the same age outside the cluster environment \citep{goto_morphology-density_2003,hansen_galaxy_2009,von_der_linden_star_2010}. Previous studies have clearly shown that cluster galaxies have experienced more rapid maturation than their non-cluster counterparts \citep{oemler_systematic_1974,butcher_evolution_1978,dressler_evolution_1984}, and this galactic maturation is influenced by the variety of mechanisms that affect a cluster galaxy \citep{treu_wide-field_2003,moran_reflections_2007}. These mechanisms can significantly alter the physical properties of these galaxies, and include, but are not limited to, ram pressure stripping \citep{gunn_infall_1972,hester_ram_2006,boselli_effect_2009}, galaxy strangulation \citep{peng_strangulation_2015}, and galaxy harassment \citep{moore_galaxy_1996,moore_morphological_1998}. These external drivers, combined with internal evolutionary processes, cause local cluster galaxies to exhibit early-type morphology more frequently \citep{dressler_galaxy_1980}, have older stellar populations \citep{smith_noao_2006,cooper_deep2_2008}, redder colors, higher overall stellar masses \citep{hogg_dependence_2004,kauffmann_environmental_2004}, and suppressed star formation rates (SFRs) \citep{lewis_2df_2002,gomez_first_2003,christlein_disentangling_2005,cooper_deep2_2008}.

Although studies have also probed intermediate-redshift galaxy clusters ($0.5 \lesssim z \lesssim 2$), their properties are more diverse and less well understood. In particular, the epoch of onset for many of the local density relations (e.g., the mass-density relation, SFR-density relation, morphology-density relation, etc.) are often disputed. Toward the lower end of this redshift range ($0.5 < z <1.5 $), most studies see the persistence of the local density relations \citep{patel_star-formation-rate-density_2011,lin_pan-starrs1_2014,ziparo_reversal_2014,lemaux_chronos_2017,lemaux_persistence_2019,tomczak_conditional_2019,old_gogreen_2020,mcnab_gogreen_2021}, but others see evidence of density relation reversals \citep{postman_morphology-density_2005,elbaz_reversal_2007,popesso_effect_2011} or that a galaxy's SFR and characteristics are largely independent of environment  \citep{grutzbauch_relationship_2011,muzzin_gemini_2012,koyama_evolution_2013,darvish_effects_2016}. However, toward the higher end of this redshift range ($z \gtrsim 1.5$) many studies show a consistent reversal of the local relations for cluster galaxies \citep{tran_reversal_2010,santos_reversal_2015,wang_discovery_2016,noirot_hst_2018}, but some of the hallmarks of lower-redshift clusters --- such as massive red late-stage galaxies --- are still present \citep{strazzullo_galaxy_2013}. These discrepancies are related to the variance in mechanisms that effect cluster galaxies and contribute to environmental quenching at intermediate redshifts \citep{muzzin_phase_2014,balogh_evidence_2016,van_der_burg_gogreen_2020,baxter_gogreen_2022,baxter_when_2023}, as well as the intrinsic variance of such structures \citep{chiang_ancient_2013} and the diversity of cluster and cluster galaxy selection criteria \citep{overzier_realm_2016}.  However, despite this diversity, we do see the emergence of a dependence of the stellar mass function (SMF) on local environment by at least $z \sim 1$ \citep{tomczak_glimpsing_2017}. Altogether, these observations suggest our current understanding of galaxy evolution within dense, high-redshift environments such as clusters is far from complete.

To shed light on the development and diversity of lower-redshift structure --- in particular, the evolution and eventual maturation of member galaxies --- we must investigate ``protoclusters'' (i.e., early-stage clusters at $z \gtrsim 2$). There is no single definition for what qualifies as or quantifies a protocluster in the literature \citep{muldrew_what_2015,contini_semi-analytic_2016}, but it is commonly accepted that protoclusters are the structures that will, at some stage, collapse into a galaxy cluster (i.e., a significantly massive virialized object at $z \geq 0$, see \citealt{overzier_realm_2016} for a substantial discussion). The identification and investigation of protoclusters is critical to understanding the formation and evolution of present-day galaxy clusters (e.g., \citealt{toshikawa_discovery_2012,toshikawa_first_2014,long_emergence_2020,calvi_probing_2021}) as they are predicted and shown to heavily contribute to the star formation rate density at early times \citep{chiang_galaxy_2017,lim_flamingo_2024,staab_protoclusters_2024}. Recent works have also found that some protoclusters begin to show evidence of environmental effects that eventually result in the formation of the largely quiescent systems we observe in the local Universe (e.g., \citealt{boselli_quenching_2016,foltz_evolution_2018,mcconachie_spectroscopic_2022}), as well as display enhanced rates of active galactic nucleus (AGN) activity \citep{shah_enhanced_2024} and increased stellar masses of member galaxies (\citealt{shimakawa_mahalo_2018,shimakawa_mahalo_2018-1,forrest_environmental_2024,sikorski25}).

\begin{figure*}
    \centering
    \includegraphics[width=\hsize]{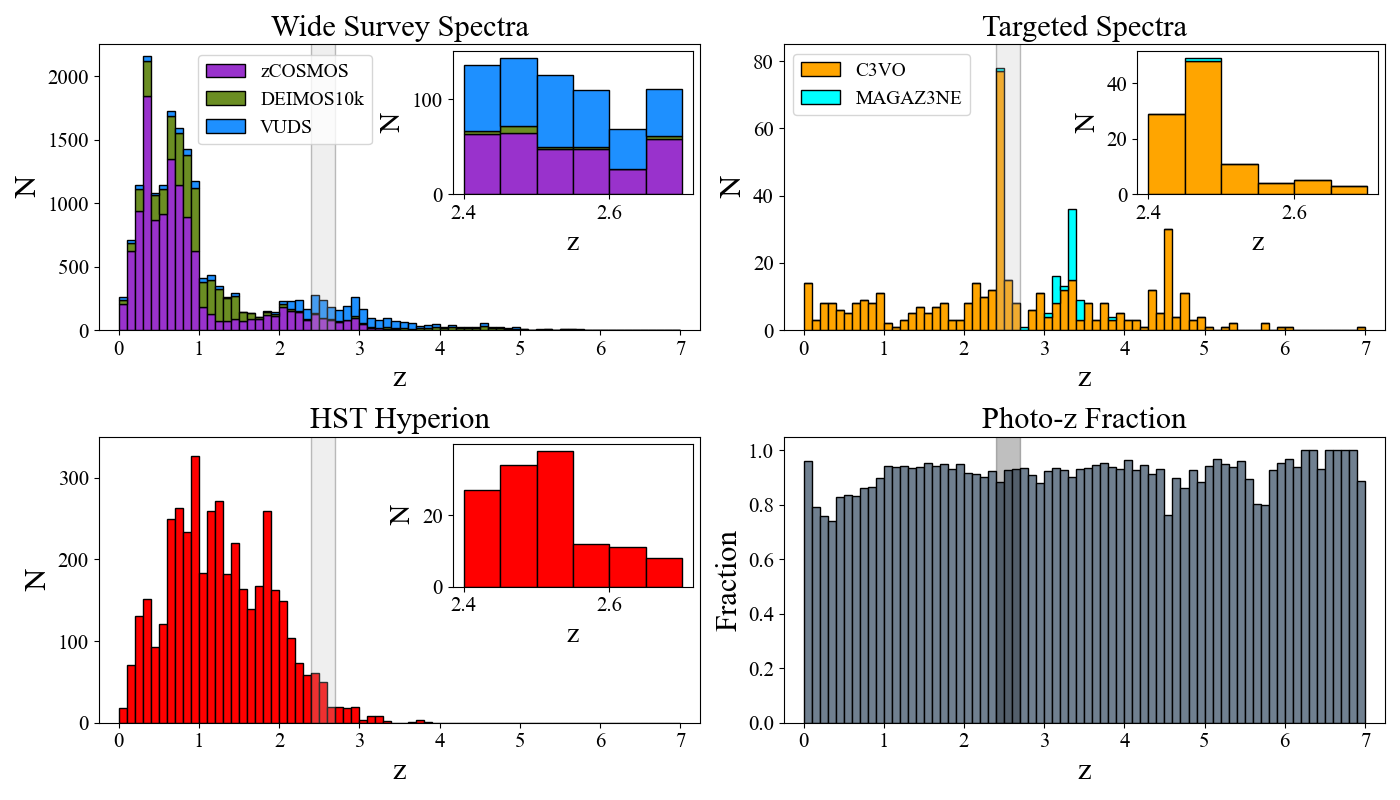}
    \caption{Source breakdown as a function of redshift for objects in our final catalog based on the cuts detailed in Section \ref{sec:sampsel}. Sources from different methods are shown in different panels: wide field spectroscopic surveys (top left), targeted spectra (top right), and HST Hyperion (bottom left). Rather than including a histogram of photometric redshift counts, we include the photometric redshift fraction (i.e the fraction of galaxies with only a photometric redshift measured) for reference on the bottom right. The location of Hyperion is included in each panel as a shaded gray region ($2.4 < z < 2.7$), and insets for this shaded region are included for the spectroscopic and grism redshift panels. Histograms with multiple survey sources are stacked.}
    \label{fig:zhists}
\end{figure*}

While members of large-scale structure, galaxies find themselves in close proximity to each other where they can interact and merge, so they evolve not as isolated systems, but as parts of the complex network of the cosmic web \citep{bond_how_1996}. These galaxy interactions and mergers oftentimes play a key role in a galaxy's evolution, driving many processes such as galactic mass growth, supermassive black hole (SMBH) accretion, AGN activity, morphological changes, and star formation triggering and quenching \citep{toomre_galactic_1972,hernandez-toledo_structural_2005,alonso_active_2007,woods_minor_2007,mesa_interacting_2014,satyapal_galaxy_2014,patton_galaxy_2016,garduno_galaxy_2021,ellison_galaxy_2022}. Protoclusters are naturally an ideal laboratory for investigating mergers and interactions at higher redshift and their dependence on environment, as they represent sites of large galaxy concentrations at these epochs, and their descendant galaxy clusters along with galaxy groups have been shown to impact the properties of merging galaxies and the rate of interactions --- both enhancing and prohibiting them depending on factors such as velocity dispersion \citep{alonso_galaxy_2004,ellison_galaxy_2010,das_galaxy_2021}. In particular, we see ``preprocessing'' effects at lower redshifts ($z \lesssim 1$) where many galaxies are quenched as they infall into structures, which often takes place in locations of similar mass, density, and dynamics as high-z protoclusters and includes merger and interaction activity \citep{hashimoto_influence_1998,kauffmann_environmental_2004,fujita_pre-processing_2004,lemaux_assembly_2012,werner_satellite_2022}. Thus, increased merger and interaction activity could be related to the increased stellar masses and enhanced AGN activity that we are beginning to observe and measure within the environments of galaxy protoclusters, and may offer an important insight into the development of large-scale structure and the mass buildup of their constituent galaxies at higher redshift. 

One of the best-mapped protoclusters at high redshift is the Hyperion proto-supercluster at $z\sim2.5$ \citep{diener_protocluster_2015,chiang_surveying_2015,casey_massive_2015,lee_shadow_2016,wang_discovery_2016,cucciati_progeny_2018}. First characterized by \citet{cucciati_progeny_2018}, Hyperion is an immense structure with an estimated total mass of M$_{\rm tot}\sim 4.8\times10^{15}$ M$_{\odot}$ extending over a volume of $60\times60\times150$ $h^{-1}_{70}$ comoving Mpc, making it comparable in size and mass to local superclusters. Hyperion also falls within the Cosmological Evolution Survey (COSMOS) field \citep{scoville_cosmic_2007}, so extensive data has been collected for the structure both spectroscopically \citep{lilly_zcosmos_2007,brammer_3d-hst_2012,le_fevre_vimos_2015,hasinger_deimos_2018,lemaux_vimos_2022,forrest_environmental_2024,forrest_hst-hyperion_2025} and photometrically \citep{laigle_cosmos2015_2016,weaver_cosmos2020_2022}. Even though it is seen at a lookback time of $\sim10$ Gyr, this wealth of data is comparable to that of some local structures making Hyperion a prime candidate to investigate the potential relation of merger and interaction activity to higher-z environment --- particularly accelerated mass growth through mergers. 

In this work, we present the results of a search for potential merging and interacting galaxies within the Hyperion proto-supercluster based on a combination of galaxies with spectroscopic and photometric redshifts in comparison to potential merger and interaction activity within the coeval field. To quantify the merger and interaction activity in both samples, we calculate a fraction of galaxies with close kinematic companions (\fckc) and attempt to disentangle what effect environment has on the incidence and strength of this potential activity. Our work is organized as follows: the data and selection methods employed in this study are described in Section \ref{sec:data} and Section \ref{sec:meth}, the results in Section \ref{sec:results}, a discussion of the implications in Section \ref{sec:diss}, and conclusions in Section \ref{sec:conc}. Throughout this paper, we utilize the AB magnitude system \citep{oke_secondary_1983} and employ a cosmology with $H_0 = 70$ km s$^{-1}$ Mpc$^{-1}$ and $\Omega_{\rm m,0} = 0.27$. All distances are in proper h$_{70}^{-1}$ units.

\begin{table}
    \caption{\label{tab:zsources} Breakdown of survey sources for all objects in the final catalog.}
    \centering
    \begin{tabular}{c|c|c|c|c}

         Survey & $z_{\rm type}$ & $N_{\rm gals, total}$ & $N_{\rm gals, \space 2<z<3}$ & Ref. \\
         \hline
         \hline
         VUDS & spec-z & 2139 & 975 & 1 \\
         zCOSMOS & spec-z & 11633 & 1110 & 2 \\
         DEIMOS10k & spec-z& 4373 & 93& 3 \\
         MAGAZ3NE & spec-z & 40 & 2& 4 \\
         C3VO & spec-z & 439 & 161 & 5 \\
         \hsthyp & grism-z & 4261 & 571 & 6 \\
         COSMOS2020 & photo-z& 197471 & 30592 & 7 \\
         \hline
         Total &  & 220356 & 33504 &
    \end{tabular}
    \tablefoot{See Section \ref{sec:sampsel} for details on our sample selection. For objects in multiple surveys (i.e., with multiple measured redshifts), we list here only the ``best'' available redshift (i.e., highest-quality spectroscopic or grism redshift if available, see Section \ref{sec:zMC}).}
    \tablebib{(1) \citet{le_fevre_vimos_2015}; (2) \citet{lilly_zcosmos_2007}; (3) \citet{hasinger_deimos_2018}; (4) \citet{forrest_massive_2020}; (5) \citet{lemaux_vimos_2022}; (6) \citet{forrest_hst-hyperion_2025}; (7) \citet{weaver_cosmos2020_2022}}
\end{table}

\section{Data}\label{sec:data}

The Charting Cluster Construction with VUDS and ORELSE Survey (C3VO; \citealt{shen_implications_2021,lemaux_vimos_2022}) is an ongoing survey that seeks to map out the growth of structure at $0.5 < z < 5$. This survey grew from the VIMOS Ultra Deep Survey (VUDS; \citealt{le_fevre_vimos_2013,le_fevre_vimos_2015,tasca_vimos_2017}) in combination with the Observations of Redshift Evolution in Large-Scale Environments Survey (ORELSE; \citealt{lubin_observations_2009}), and recently C3VO has worked to obtain additional visible and near-infrared wavelength spectroscopy of three extensively studied extragalactic fields in an effort to better characterize structure assembly and galaxy evolution within said structures. These fields are as follows: the Cosmic Evolution Survey (COSMOS; \citealt{scoville_cosmic_2007}) field, the Extended Chandra Deep Field South (ECDFS; \citealt{lehmer_extended_2005}), and the first field of the Canada-France-Hawai’i Telescope Legacy Survey (CFHTLS-D1). Due to the location of the Hyperion proto-supercluster, we focus on the COSMOS field portion of C3VO in this work. 

\subsection{Photometric data}
\label{sec:phot data}

The photometric data used for galaxy properties and redshifts in this study is drawn from COSMOS2020 Classic Catalog v2.0 \citep{weaver_cosmos2020_2022}. COSMOS2020 is the latest data release in a program that has long targeted the COSMOS field \citep{scoville_cosmic_2007,koekemoer_cosmos_2007,capak_first_2007,ilbert_cosmos_2009,ilbert_mass_2013,muzzin_public_2013} with extensive photometric observations. COSMOS2020 contains over 40 bands of multiwavelength observations ranging from the UV to IR. Far-UV and near-UV data is drawn from GALEX \citep{zamojski_deep_2007}, and U-Band data from Canada-France-Hawaii Telescope (CFHT) MegaCam \citep{boulade_megacam_2003} observations for the CFHT large area U-band deep survey (CLAUDS; \citealt{sawicki_cfht_2019}). Optical data comes from a combination of \emph{g},\emph{r},\emph{i},\emph{z},\emph{y} bands from the Subaru Hyper Suprime-Cam (HSC, \citealt{miyazaki_hyper_2018}) and the HSC Subaru Strategic Program (HSC-SSP; \citealt{aihara_second_2019}), along with Subaru Suprime-Cam data used in COSMOS2015 \citep{taniguchi_cosmic_2007,taniguchi_subaru_2015}. The YJHK$_s$ broad-band and NB118 narrow-band data from the fourth data release (DR4) of the UltraVISTA
survey \citep{mccracken_ultravista_2012,moneti_vizier_2023} are used for the near-IR, and the mid-IR data comprise of \emph{Spitzer} Infrared Array Camera (IRAC; \citealt{fazio_infrared_2004}) channel  1,2,3,4 images from the Cosmic Dawn Survey \citep{euclid_collaboration_euclid_2022}. Aside from IRAC/channel 3 and IRAC/channel 4, all bands reach a depth of $\sim$ 26 mag at 3$\sigma$ computed on PSF-homogenized images measured in empty 3" diameter
apertures (see Table 1 and Figure 3 of \citealt{weaver_cosmos2020_2022} for substantial information).

Sources for the COSMOS2020 Classic catalog were extracted using \texttt{SExtractor} \citep{bertin_sextractor_1996}, and the procedure to homogenize the PSF in the optical/near-infrared images is presented in \citet{laigle_cosmos2015_2016} and \citet{weaver_cosmos2020_2022}. The bright-star masks from HSC-SSP PDR2 \citep{coupon_bright-star_2018} are used to mask stars within the COSMOS field. For COSMOS2020, astrometric solutions were computed using the Gaia DR21 astrometric reference \citep{gaia_collaboration_gaia_2016}. Further details for the Classic catalog photometry can be found in \citet{weaver_cosmos2020_2022}.

\begin{figure}
    \centering
    \includegraphics[width=1\linewidth]{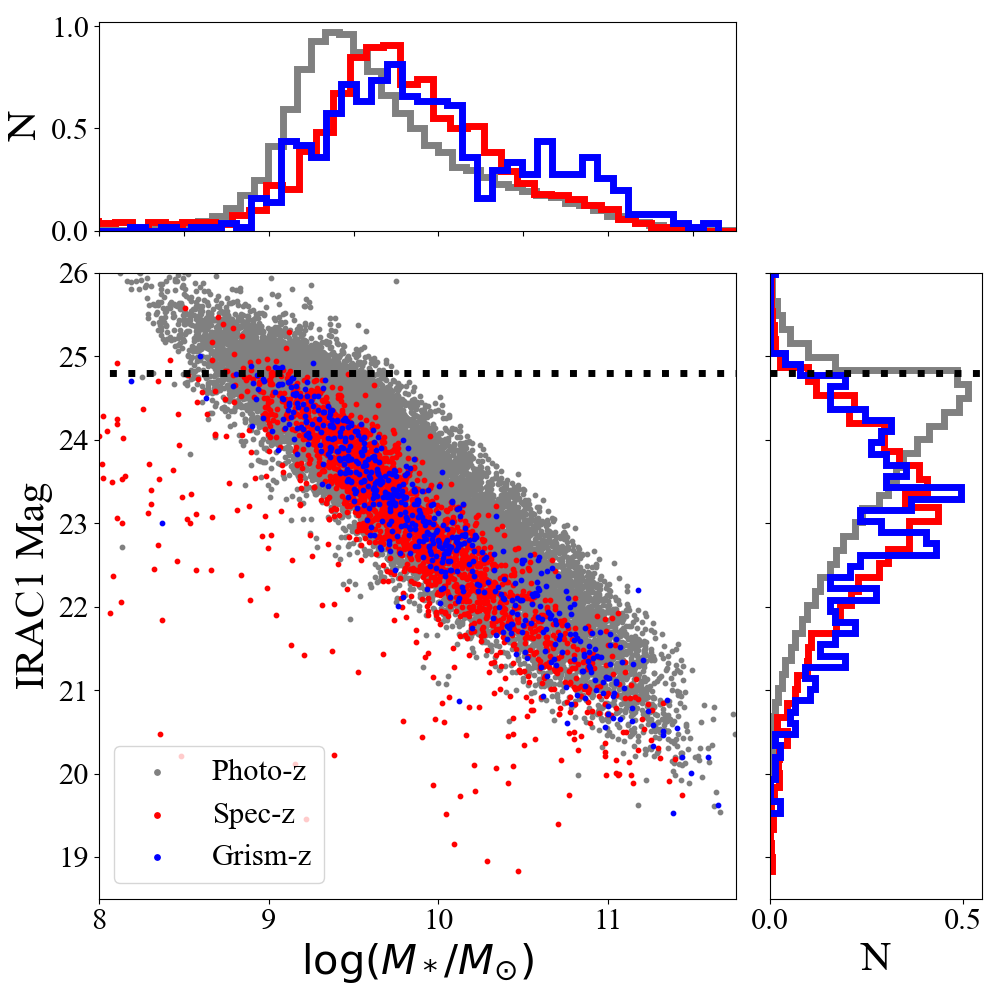}
    \caption{Galaxies in our final sample from $2 < z < 3$ plotted in stellar mass and IRAC ch1 magnitude space (lower left) along with individual normalized histograms of stellar mass (top) and IRAC ch1 magnitude (lower right). All values are drawn from COSMOS2020 aside from the redshifts where we employ the best spectroscopic or grism redshift when available. For reference, our imposed IRAC ch1 magnitude cut is plotted as the dotted black line (note: some galaxies in our final sample lie above this line as our IRAC cut is an or cut on both ch1 and ch2). As expected, our spectroscopic and grism redshift sources (spec-z and grism-z, respectively) are primarily at brighter magnitudes and higher stellar masses in comparison to the distribution of photometric redshift only sources. Our final sample has a rough stellar mass limit of $\log(M_*/M_{\odot}) \sim 9.3$ from $2 < z < 3$ based on our imposed IRAC cuts. }
    \label{fig:iracmass}
\end{figure}

\subsection{Spectroscopic data}
\label{sec:specdata}

The spectroscopic data employed in this study are detailed in \citet{lemaux_vimos_2022}, as well as \citet{forrest_environmental_2024} (particularly Appendix A). This includes public archival redshifts from a variety of surveys (see Section \ref{sec:arch spec}), C3VO observations (Section \ref{sec:c3vo spec}), and the new \hsthyp \space Survey (Section \ref{sec:hst spec}; \citet{forrest_hst-hyperion_2025}). Apart from the \hsthyp \space Survey --- which was obtained and cataloged independently --- these combined sources provide us with $\sim$ 40,000 galaxies with spectroscopic redshifts (spec-zs). From these sources, we employed in this work only those spec-z sources which have been matched to objects in the COSMOS2020 Classic Catalog v2.0 ($\sim94.4$\% of sources) following the method described in \citet{forrest_environmental_2024}. All spec-z sources from this catalog have assigned spec-z quality flags based on the VUDS quality flag scheme (with some complementary flags from the DEEP2 Galaxy Redshift Survey, see \citealt{newman_deep2_2013}). In this scheme,  $q_{\rm flag}$ = X2, X9, X3, or X4 correspond to acceptable quality flags (ranging from 70\% to 99.3\% confidence) for our study and $q_{\rm flag} =$ X0 or X1 correspond to poor quality flags (see \citealt{le_fevre_vimos_2015} as well as Appendix A of \citealt{lemaux_vimos_2022} for additional information). Of spectroscopic sources from \citet{forrest_environmental_2024} matched to COSMOS2020 sources, 66.7\% have moderate to high-quality spectroscopic redshift measurements (i.e., acceptable spec-z quality flags). 

While there are many additional spectroscopic redshift measurements available for the COSMOS field (see \citealt{khostovan_cosmos_2025}), we chose to only utilize those from the \citet{forrest_environmental_2024} compilation. We tested our main result (i.e., our reported \fckc \space values) against the inclusion of additional spectroscopic redshifts from the COSMOS Ly$\alpha$ Mapping and Tomography Observations (CLAMATO) Survey \citep{lee_first_2018,horowitz_second_2022} and MOSFIRE Deep Evolution Field (MOSDEF) Survey \citep{kriek_mosfire_2015}  ($\sim260$ additional redshifts from $2<z<3$ that match our sample selection criteria, see Section \ref{sec:sampsel}, representing a $\sim$9\% increase over our fiducial spec-$z$ sample) as these are the two largest surveys at $2<z<3$ from \citet{khostovan_cosmos_2025} that are not included in the \citet{forrest_environmental_2024} compilation. The inclusion of these two surveys constitutes $\sim42$\% of the galaxies in the redshift range of interest with secure redshifts that are omitted from \citet{forrest_environmental_2024}. We then performed an identical analysis using this composite catalog to that presented in this work (Section \ref{sec:meth} and subsections) and  find no significant change in our derived \fckc \space values for Hyperion and field. Due to the invariance of our primary result --- along with the fact that the \citet{forrest_environmental_2024} compilation contains more secure redshifts between $2<z<3$ that match our sample selection than the \citet{khostovan_cosmos_2025} DR1.0 compilation (2,912 versus 2,157 redshifts, respectively) --- we employed the \citet{forrest_environmental_2024} compilation only to ensure a more uniform selection function, as the inclusion of all sources from \citet{khostovan_cosmos_2025} would result in additional sources from at least 36 other surveys, which are subject to a wide range of selection biases. Additionally, we account for impact of the spectroscopic completeness of our underlying sample on our results regardless of the included surveys in Section \ref{sec:cone valid}.

\subsubsection{Archival spectroscopy}
\label{sec:arch spec}

Archival spectroscopic redshifts for the \citet{forrest_environmental_2024} catalog, and thus this work, are drawn from VUDS \citep{le_fevre_vimos_2015}, zCOSMOS \citep{lilly_zcosmos_2007}, DEIMOS10k \citep{hasinger_deimos_2018}, and the Massive Ancient Galaxies at $z > 3$ Near-Infrared Survey (MAGAZ3NE; \citealt{forrest_massive_2020}). The details for those studies are as follows:

\begin{itemize}
    \item VUDS utilized the VIMOS spectrograph \citep{lefevre_commissioning_2003} on the 8.2 m Very-Large Telescope (VLT) to target 1 deg$^2$ in 3 separate fields: COSMOS, ECDFS, and VVDS-02h (with 0.5 deg$^2$ solely in COSMOS). Spectroscopic targets for VUDS were chosen primarily on a inclusive combination of photometric redshifts and Lyman-break galaxy (LBG) color-color properties resulting in a sample of $\sim10^4$ targets covering 2 $< z_{\text{phot}} <$ 6.
    \item zCOSMOS targeted the COSMOS field with 600 hours of observations also using the VIMOS spectrograph, and consists of two sub-parts: zCOSMOS-bright, a magnitude-limited I-band I$_{\rm AB} <$ 22.5 survey covering the entire 1.7 deg$^2$ COSMOS ACS field in the redshift range 0.1 $< z <$ 1.2; and zCOSMOS-deep, a color-selected survey bxased on both the BzK criteria of \citet{daddi_new_2004} and the ultraviolet UGR ``BX'' and ``BM'' selection of \citet{steidel_survey_2004}, covering the central 1 deg$^2$ and the redshift range 1.3 $< z <$ 3.0.
    \item DEIMOS10k employed the DEIMOS spectrograph \citep{faber_deimos_2003} on Keck II to target the COSMOS field. DEIMOS10k targets were selected from a variety of input catalogs based on multiwavelength observations spanning from X-ray to IR, and resulted in a sample with magnitude distribution peaking at I$_{\rm AB} \sim$ 23 and K$_{\rm AB} \sim$ 21, and a redshift range of 0 $< z <$ 6. These sources are primarily included in this work to remove low redshift interlopers and constrain lower-redshift sources ($\sim4,300$ at $z<2$, see Figure \ref{fig:zhists}) as they sit in similar color phase space as many high redshift sources. Though subject to more complex selection from multiple input catalogs, DEIMOS10k imprints itself similarly on both the structure(s) and field populations from $2<z<3$ in this work as none of the sub-surveys of DEIMOS10k specifically target protocluster or structure galaxies.
    \item The MAGAZ3NE Survey employed Keck/MOSFIRE \citep{mclean_design_2010,mclean_mosfire_2012} to spectroscopically follow-up ultra-massive galaxies ($\log(M_*/M_\odot)> 11$ at $z >3$) and their surrounding environments. MAGAZ3NE targets in the COSMSOS field were selected for follow up based on the observed galaxy spectral energy distribution (SED), photometric redshift probability distribution (zPDF), stellar mass, and SFR from the UltraVISTA DR1 and DR3 catalogs \citep{muzzin_public_2013}.
\end{itemize}

\subsubsection{C3VO observations}
\label{sec:c3vo spec}

The C3VO spec-zs employed in this work are the result of observations with both Keck/DEIMOS and Keck/MOSFIRE that provide comprehensive mapping of six significant overdensities detected in VUDS, including Hyperion \citep{cucciati_progeny_2018}, as well as others reported in \citet{lemaux_vimos_2014,cucciati_discovery_2014,lemaux_vimos_2018,shen_implications_2021,forrest_elentari_2023,shah_identification_2024,staab_protoclusters_2024}. C3VO optical observations primarily target star-forming galaxies of all types down to \emph{i}$_{\rm AB} <$ 25.3 (or $<$ L$^*_{\rm FUV}$ at $z \sim 2.5$) and Lyman-$\alpha$ emitting galaxies to fainter magnitudes. The C3VO observations in the near-infrared (NIR) target sources to $m_{\rm H} < 24.5$, and, for these NIR observations, continuum redshifts are recoverable for the brighter sources (i.e., $m_{\rm H} <23.5$). These magnitude limits were chosen so continuum observations at modest signal-to-noise rations could be achieved, such that Ly$\alpha$ or other emission features are not required to measure a galaxy's redshift (though Ly$\alpha$-derived redshifts are incorporated when appropriate), which would bias resulting samples. The overall spectroscopic sampling of the combined C3VO observations and archival spectroscopy is roughly representative of the underlying photometric dataset to $\gtrsim10^{9.5}$ $\mathcal{M}_{\odot}$ in stellar mass, though with a bias toward bluer galaxies at fixed stellar mass and redshift \citep{lemaux_vimos_2022}.

\subsubsection{\hsthyp \space spectroscopy}
\label{sec:hst spec}

In this work, we also employed redshifts measured with WFC3/G141 slitless (grism) spectroscopy from the \hsthyp \space Survey \citep{forrest_hst-hyperion_2025}. The \hsthyp \space Survey is a Cycle 29 program (PI: Lemaux, PID 16684) consisting of 50 orbits of direct imaging with HST/WFC3/F160W and slitless grism spectroscopy with HST/WFC3/G141 over 25 pointings in the COSMOS field. The pointing locations and position angles were chosen to target the main density peaks of Hyperion, as well as to be complementary to existing grism observations of the COSMOS field from 3D-HST \citep{brammer_3d-hst_2012,momcheva_3d-hst_2016}. Derived redshifts from this survey are a result of the combination of re-reduced 3D-HST data along with the new HST pointings with all sources passing multiple rounds of visual inspection (see \citet{forrest_hst-hyperion_2025} \space for a summary of the methodology). This work resulted in 5,629 reliable redshifts that we employ in this study.

Grism redshift quality flags (\qfgriz) were assigned for each source based on a visual classification scheme of redshifts and fits measured with \texttt{GRIZLI} version 1.9.5 \citep{brammer_gbrammergrizli_2021}. In this scheme, \qfgriz \space ranges from \qfgriz $= 0$ to \qfgriz $= 5$ in steps of 1 with \qfgriz = 3, 4, 5 being of reliable quality ($67.4$\%, $80.6$\%, and $93.2$\% reliable, respectively) and \qfgriz = 0, 1, 2 being unreliable quality redshifts. 

\section{Methods}
\label{sec:meth}
\subsection{Sample selection and spectroscopic completeness}
\label{sec:sampsel}

To create the final sample for this study, we selected sources from our extensive spectroscopic catalog,  \hsthyp, and COSMOS2020. We primarily selected sources from the spectroscopic and/or grism observations, whereas COSMOS2020 sources are supplementary and are included to increase the completion of our sample and to reduce biases from spectroscopic target selection. However, all sources selected from our spectroscopic catalog and \hsthyp \space have a confirmed match within COSMOS2020. For this, we only employed single match non-blended sources from \hsthyp \space (see \citet{forrest_hst-hyperion_2025}), and, for sources selected from our spectroscopic catalog, we used only single match or the highest-quality multi-match sources (Q3/Q4; see \citet{forrest_environmental_2024}). Thus, all sources in our final sample have at least a measured photometric redshift from COSMOS2020 with some having additional spectroscopic and/or grism redshifts (see Table \ref{tab:zsources} and Figure \ref{fig:zhists} for the redshift breakdowns). We did not employ any redshift cut in our sample selection due to the Monte Carlo (MC) process used in this work (see Section \ref{sec:zMC}).

However, we did make additional cuts for all sources using their IRAC channel 1 (ch1, 3.6 $\mu$m) and channel 2 (ch2, 4.5 $\mu$m) magnitudes at m$_{\rm IRAC1} < 24.8$ or m$_{\rm IRAC2} < 24.8 $ (roughly $0.03 \times M^*$ at $z\sim2.5$, see \citealt{weaver_cosmos2020_2023}). At the redshift of Hyperion ($z\sim2.5$), IRAC ch1 and ch2 correlate strongly with stellar mass by probing rest-frame Y and J bands, respectively, and are largely immune to contributions from dust or young stellar populations. We used both IRAC ch1 and IRAC ch2 (rather than just a single IRAC band) for this cut due to the poor angular resolution of IRAC. The IRCLEAN software \citep{hsieh_taiwan_2012} is used to extract photometry from those bands and uses the detection images in other bands as a prior for source extraction. This extraction can result in a non-detection in either IRAC band-pass (particularly for fainter sources), and, thus, to increase the completeness of our final sample we used both IRAC ch1 and ch2. If we were to use only IRAC ch1 for our sample selection, we would lose 10,378 sources or $\sim4.7$\% of our final sample. We chose our IRAC ch1 and ch2 magnitude cuts following the completeness limits for the COSMOS field (see \citealt{davidzon_cosmos2015_2017,weaver_cosmos2020_2022,weaver_cosmos2020_2023}) and we chose to make this magnitude cut rather than a stellar mass cut due to the MC process used to obtain our main results (see Section \ref{sec:comp frac}). This is done to create an overall sample that is stellar mass limited without being reliant on a fitted quantity, as well as to maximize the usefulness of the spectroscopic portion of our sample (i.e., galaxies measured with spectroscopic or grism redshifts in our sample fall off precipitously at IRAC $> 24.8$). We also found that our main results are invariant to our choice of an IRAC magnitude cut or a \texttt{LePhare} stellar mass cut. We employed the IRAC ch1 and IRAC ch2 values from COSMOS2020 for these cuts and galaxies without measured magnitudes in both IRAC channels by COSMOS2020 are not used. 

Additionally, we performed an astrometric cut to limit our sample to the sky region covered by Hyperion ($149.6^{\circ} \leq \alpha \leq 150.52^{\circ}$ and $1.74^{\circ} \leq \delta \leq 2.73^{\circ}$) utilizing COSMOS2020 astrometry. We also removed all sources designated as stars based on either the spectroscopic catalog or from the COSMOS2020 Classic Catalog \texttt{LePhare} fit \citep{ilbert_accurate_2006,arnouts_lephare_2011}. This results in a final sample of 220,356 unique sources. Table \ref{tab:zsources} gives the source breakdown of our final sample and Figure \ref{fig:zhists} provides the redshift distribution of our final sample. 

In Figure \ref{fig:iracmass}, we also provide for reference the stellar mass and the IRAC ch1 magnitude distribution of galaxies in our final sample between $2 < z < 3$ (i.e., the redshift range where we examine merger and interaction activity). Values for both the stellar mass and IRAC ch1 magnitude are drawn from COSMOS2020. With our imposed IRAC cuts, the resultant spectroscopic and photometric sample used to assess merger and interaction activity in Hyperion and the surrounding field is roughly 70\% complete to a stellar mass limit of log($M_{\ast}/M_{\odot}$)$\ga$9.3 (see also \citealt{weaver_cosmos2020_2022,weaver_cosmos2020_2023}).

\begin{figure}
    \centering
    \includegraphics[width=1\linewidth]{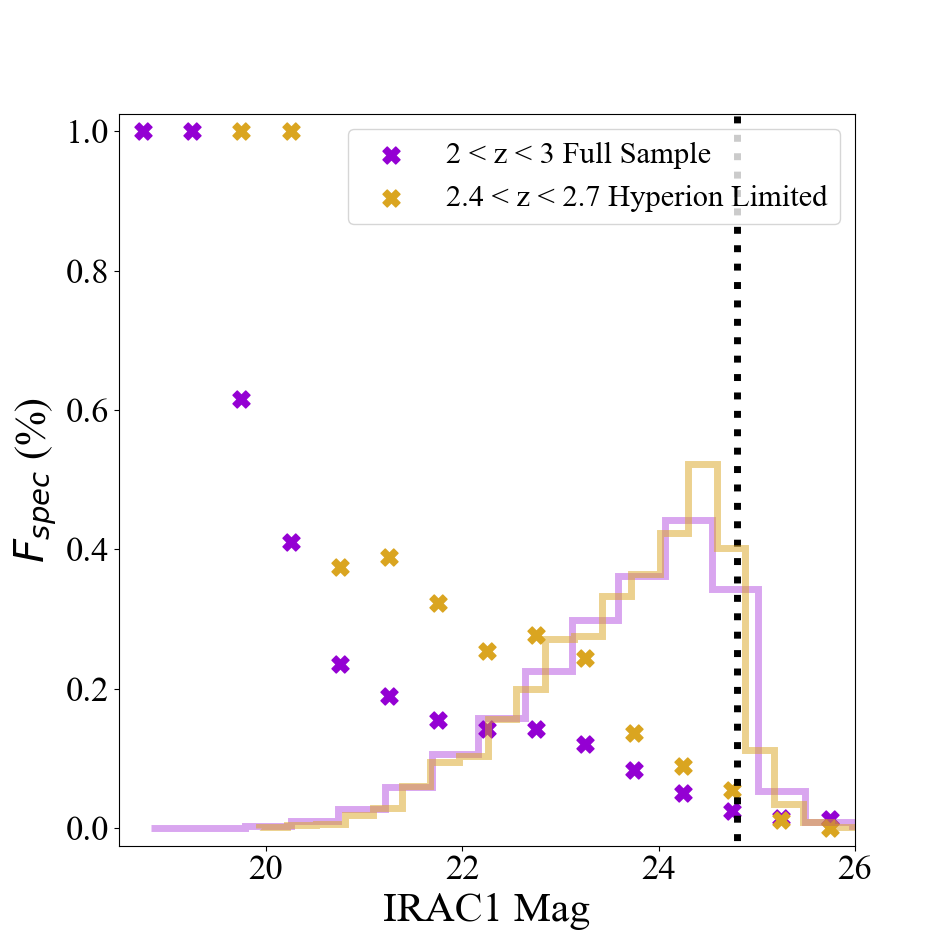}
    \caption{Spectroscopic completeness (or the fraction of sources with a spectroscopic and/or grism redshift, $F_{\text{spec}}$) as a function of IRAC ch1 magnitude for all sources from $2<z<3$ in our final sample (purple) and a sub-sample of sources limited to the redshift range of Hyperion with tighter astrometric cuts (gold). Also included for reference are the underlying normalized IRAC ch1 magnitude distributions for each population along with the location of the imposed IRAC ch1 (or ch2) magnitude cut (vertical dashed black line). Spectroscopic completeness for both samples declines sharply as a function of magnitude. However, there is notably higher completeness for galaxies limited more closely to the bounds of Hyperion.}
    \label{fig:speccomp}
\end{figure}

Though we do include over 22,000 sources with spectroscopic and/or grism redshifts in our final sample ($\sim2,900$ spectroscopic and/or grism sources from $2<z<3$), the majority of sources in our final sample of 220,356 galaxies are purely photometric sources ($\sim90$\%). From Figure \ref{fig:iracmass}, we see that the fainter (and low mass) end of our sample from $2<z<3$ in particular is dominated by photometric sources. This fact in combination with the variety of studies from which we draw spectroscopic redshifts -- along with the targeting strategy of \hsthyp \space -- suggests the spectroscopic and grism portion of our final sample is subject a complex selection function. Thus, it is important to explore the underlying spectroscopic completeness of our sample as a function of magnitude. 

In Figure \ref{fig:speccomp}, we plot spectroscopic completeness (or the fraction of sources with a spectroscopic and/or grism redshift, $F_{\text{spec}}$) as a function of IRAC ch1 magnitude for two distinct populations: the full sample of sources in our final sample from $2<z<3$ and a sample more specifically targeted to capture the bounds of Hyperion (i.e., the ``Hyperion limited'' sample). The $2<z<3$ full sample contains the same sources as shown in Figure \ref{fig:iracmass} and is selected as every source that has their ``best'' redshift lie within the given redshift range. The Hyperion limited sample contains 3,134 galaxies and is selected as all galaxies that have their ``best'' redshift from $2.4<z<2.7$ and has an additional astrometric cut of $149.9^{\circ} \leq \alpha \leq 150.42^{\circ}$ and $2^{\circ} \leq \delta \leq 2.5^{\circ}$ (to match the density contours of Hyperion, see Figure \ref{fig:szfsky}). We also include the underlying normalized IRAC ch1 magnitude distributions in Figure \ref{fig:speccomp} for reference. 

From Figure \ref{fig:speccomp}, we see a sharp decline in spectroscopic completeness as a function of magnitude for both the full sample and the Hyperion limited sample that matches the trends seen in Figure \ref{fig:iracmass}. However, we see that the spectroscopic completeness of the Hyperion limited sample is consistently higher than that of the full $2<z<3$ sample with the Hyperion limited sample seeing $\gtrsim10$\% more completeness until IRAC ch1 $\sim 23.5$ and $\sim3-5$\% more completeness until our imposed magnitude cut (i.e., IRAC ch1 $<24.8$). Due to the targeting strategies of the included C3VO observations and \hsthyp \space itself, this is expected. Our spectroscopic completeness for both samples is also best where we have lower source counts (IRAC ch1 $\lesssim23$) and noticeably lower where the bulk of our sample lies ($23 \lesssim \text{IRAC ch1} \lesssim 24.8$). We are sure to account for these varying levels of spectroscopic completeness when making our final close kinematic companion calculations (see Section \ref{sec:correct}). \\

\subsection{Environmental measures}
\label{sec:da env}

\begin{figure*}
    \centering
    \includegraphics[width=1\linewidth]{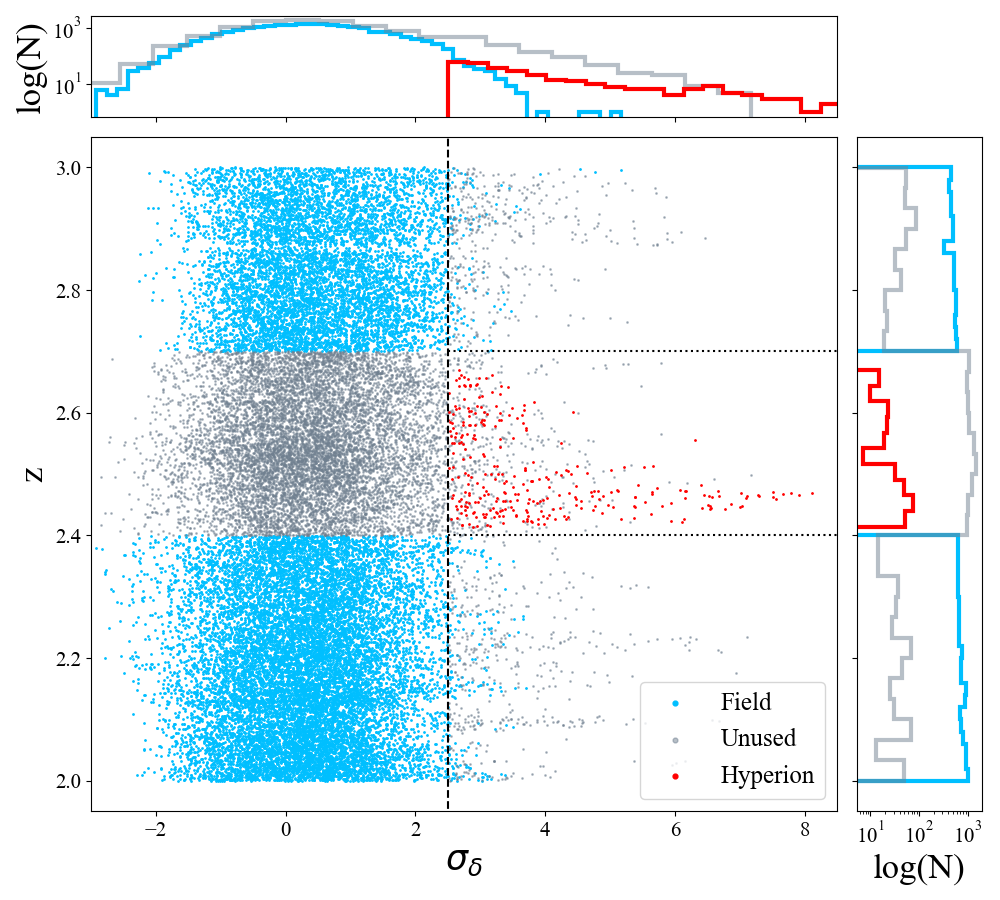}
    \caption{Example of the distribution of galaxies in redshift and overdensity ($\sigma_{\rm \delta}$) space for one MC iteration used in this work (MC \#82). Included are only those galaxies in the relevant redshift range for our fraction calculations (i.e., $2 < z < 3$). Galaxies associated with Hyperion for this iteration are marked in red and field galaxies for this iteration are marked in blue. Unused galaxies for this iteration are marked in gray (which includes galaxies in the redshift range of Hyperion that are not associated with overdensity and galaxies in the field sample redshift range that are associated with other large structure, see Section \ref{sec:deter env}). }
    \label{fig:mcsigmas}
\end{figure*}

\subsubsection{Local overdensity}
\label{sec:ldens}

In this work, we differentiate two distinct environments: Hyperion and the coeval field. This definition results from a determination of local overdensity (i.e., log(1+$\delta_{\rm gal}$)) drawn from Voronoi Monte Carlo (VMC) maps made for the Hyperion field (see \citealt{tomczak_glimpsing_2017,lemaux_chronos_2017, hung_establishing_2020, hung_optical_2021, lemaux_vimos_2022,hung_discovering_2025}). The Voronoi tessellation technique divides a 2D plane into a number of polygonal regions equal to the number of objects in that plane. A Voronoi cell for each object is then defined as the region closer to it than to any other object in the plane. Objects in the highest-density regions therefore have the smallest Voronoi cells, while objects in lower-density regions have larger cells. The inverse area of the cell sizes can thus be used to measure the local density at the position of the object bounded by the cell.

To obtain the 2D planes for VMC mapping, galaxies are partitioned into thin redshift slices with each slice spanning 7.5 proper $h_{70}^{-1}$ Mpc along the line of sight (approximately 3000 -- 4500 km s$^{-1}$ or $\Delta z \sim 0.03-0.05$ over the redshift range of this study, $2 < z < 3$), with a 90\% overlap between adjacent slices (see \cite{hung_establishing_2020} for a discussion on this choice). Each redshift slice is then projected onto a 2D Voronoi tessellation that is calculated using the positions of the galaxies within it. Local environmental density is defined as the inverse of the area of a Voronoi cell multiplied by the square of the angular diameter distance at the corresponding mean redshift of the slice. The Voronoi maps are then projected onto a 2D grid of pixels sized $75 \times 75$ proper kpc. The local environmental overdensity at pixel (i,j) is defined as $\log(1 + \delta_{\rm gal}) \space \equiv \space \log(1 + \frac{\Sigma_{\rm i,j} - \Tilde{\Sigma}}{\Tilde{\Sigma}})$ where $\Sigma_{\rm i,j}$ is the density at pixel $(i,j)$ and $\Tilde{\Sigma}$ is the median density of all the pixels where the map is reliable (i.e., the central 80\% of the slice).

A MC process is used to incorporate the redshift uncertainties. The above process of 2D planes partitioned by redshift slices is repeated 100 times, varying the redshifts of the member galaxies in each iteration. Within each iteration, galaxies with a spectroscopic redshift either have that redshift chosen or are cut based on the VUDS-like spectroscopic quality flag \citep{le_fevre_vimos_2015,lemaux_vimos_2022}. Both spectroscopic objects that fail to meet the quality flag cut and objects with no spectroscopic information have a new redshift assigned in each iteration that is sourced from sampling their photometric redshift PDFs in each trial. The voxel values used in final VMC maps are thus the median value of that voxel from these 100 iterations \citep{hung_establishing_2020, hung_optical_2021, lemaux_vimos_2022}. This technique has been employed for a variety of structures across a wide redshift range ($0.6 <  z < 4.6$; \citealt{darvish_comparative_2015,shen_properties_2017,shen_possible_2019,rumbaugh_x-ray-emitting_2017,pelliccia_searching_2019, hung_establishing_2020}), and is found to be strongly concordant with other density metrics and to trace known structures \citep{tomczak_glimpsing_2017,tomczak_conditional_2019,lemaux_persistence_2019,hung_discovering_2025}.

\subsubsection{Structure definition}
\label{sec:struc def}

Due to the extended nature of high-redshift protoclusters, the possibility exists that the entire field considered in this work may be overdense or underdense in a particular redshift slice, which could bias our overdensity calculations. As such, using the existing VMC maps, we fit a Gaussian to the distribution of overdensity values in each redshift slice, and, based on this fit, calculate the number of standard deviations of a galaxy's overdensity value above or below the fit mean (\sigdel; see \citealt{forrest_elentari_2023,forrest_environmental_2024,shah_identification_2024} for additional information on this calculation). This \sigdel \space encodes the 3D statistical significance in overdensity of each particular voxel within our VMC maps and galaxies are assigned overdensity values (and a \sigdel) based on the nearest Voronoi cell  to the galaxy’s coordinates (see Section \ref{sec:deter env}). Using the \sigdel \space values from our VMC maps, we define an overdense structure in a manner analogous to \citet{cucciati_progeny_2018,shen_implications_2021,shah_identification_2024,forrest_environmental_2024} by finding all contiguous voxels in our 3D maps with \sigdel $ > 2.5$ that include extremely overdense peak (\sigdel $ >4$, see \citet{sikorski25} for the motivation behind these thresholds). 

For this study, Hyperion is defined as the most massive overdense structure in our 3D VMC maps from $2<z<3$ that fits the aforementioned \sigdel \space criteria and has a mass of $\log(M/M_\odot) \sim 15.4$ and spans roughly $2.4 < z < 2.7$. This structure is similar to the most massive structure first characterized in \citet{cucciati_progeny_2018} and the definition of Hyperion used in this work is consistent with \citet{sikorski25} There exists other overdense structure from $2.4<z<2.7$ in our sample and VMC maps, but such structure is associated with separate overdensity peaks and are either unconnected to the structure defined as Hyperion in this work or are connected at low densities (i.e., \sigdel$<2.5$, see Figure \ref{fig:mcsigmas} and \citet{sikorski25} for further discussion). In contrast to our adopted definition of Hyperion and other overdense structure, the coeval field sample used in this work is primarily galaxies at similar redshifts ($2<z<3$) that are not associated with any overdense structure (see Section \ref{sec:deter env} and Figure \ref{fig:mcsigmas}).

\subsection{Companion fraction}
\label{sec:comp frac}

In order to assess potential merger and interaction activity in Hyperion and the coeval field, we developed a MC methodology to measure the fraction of galaxies with close kinematic companions (\fckc). In this methodology, we varied the redshifts of all galaxies in our final sample to find potential companion systems (i.e., galaxies with one or more close kinematic companions) and determined the local environment of such systems. We performed 100 MC iterations based on the following prescription:

\begin{enumerate}
    \item Vary the redshifts for all sources in our final sample (Section \ref{sec:zMC})
    \item Determine the environment of all galaxies in the relevant redshift range (Section \ref{sec:deter env})
    \item Identify galaxies with close kinematic companions (Section \ref{sec:cID})
    \item Calculate the fraction of galaxies with close kinematic companions (\fckc) for Hyperion and for the coeval field (Section \ref{sec:frac calc})
\end{enumerate}

Following these steps, we applied a correction factor to obtain our final \fckc \space values (see Section \ref{sec:correct}). The choice of this methodology was the result of extensive reliability testing on the feasibility of using photometric redshifts to identify galaxy pairs and galaxies with potential companions. We chose 100 MC iterations to strike a reasonable balance between obtaining a representative sampling of the redshift probability distributions (zPDFs) for our photometric sources and not over-weighing spectroscopic sources. We verified that our 100 MC iterations capture the uncertainty and secondary redshift peaks for a variety of zPDFs (see Appendix \ref{app:zpdfs} and Figure \ref{fig:ex_zPDFs}). The final \fckc \space values reported in this study are the medians of the 100 MC iterations. 

\subsubsection{Redshift Monte Carlo}
\label{sec:zMC}

We utilized the following procedure to MC the redshifts of our sample for 100 iterations with only galaxies that fall between $2<z<3$ for that MC iteration being used for further calculations. For each iteration, the redshift variation decision hinged on the type of redshift(s) available and, for sources with a spectroscopic redshift (spec-z) and/or grism redshift (grism-z), the quality flags associated with those redshifts. The 100 MC iterations of the redshifts of our final sample that are detailed in this study are drawn from those presented in \citet{sikorski25}, and are organized as follows:

\begin{itemize}
    \item Galaxies with a spec-z: We either used the available measured spectroscopic redshift or drew from the associated zPDF for that object from COSMOS2020 (v2.0 \texttt{LePhare} fit) for each iteration. To determine whether or not to keep the spec-z or draw from the zPDF, we used the corresponding spectroscopic redshift quality flags. Since we employ the VUDS/DEEP2 quality flag scheme, objects with the highest-quality redshifts (q$_{\rm flag} = X3, X4$) have their spec-z kept $\sim$99.3\% of the time and sources with other reliable spectroscopic redshifts (q$_{\rm flag} = X2, X9$) have their spec-z kept in $\sim$70\% of the iterations. Galaxies with low-quality spec-zs (q$_{\rm flag} = X0, X1$) have their redshifts drawn from the corresponding COSMOS2020 source zPDF each iteration.
    \item Galaxies with a grism-z: We used a similar method to that of spec-z sources, where we either drew from a tight Gaussian centered at the grism-z (with a width of $46/14100\times(1+z_{\text{grism}})$, see \citet{forrest_hst-hyperion_2025}), or draw from the source's COSMOS2020 zPDF for each iteration. For sources with reliable grism redshifts, we sampled from the tight grism-z Gaussian for 93.2\% (\qfgriz $ = 5$), 80.6\% (\qfgriz$ = 4$) and 67.4\% (\qfgriz$ = 3$) of our MC iterations, and draw from the COSMOS2020 zPDF for other iterations. Galaxies with low-quality grism-zs have their redshifts drawn from their COSMOS2020 zPDF each iteration.
    \item Galaxies with multiple reliable redshifts: (i.e., a reliable grism-z and a reliable spec-z) We used a hierarchy that prioritizes the most reliable redshifts first. As such, we preferentially used highest-quality spec-zs (99.3\% reliable), then the highest-quality grism-zs (93.2\% and 80.6\% reliable, respectively), then other reliable spec-zs (70\% reliable) and finally other reliable grism-zs (67.4\% reliable). With this, we employed a similar scheme as already described where, for these sources, we first attempted to take the existing most reliable spec-z or sample from the tight grism-z Gaussian based on the reliability. However, for these sources, if the most reliable spec-z or grism-z was not selected for that iteration, we then utilized the other reliable redshift at a rate that corresponds with its reliability. If neither redshift is chosen, we drew from the corresponding COSMOS2020 zPDF.
    \item Galaxies with only a photometric redshift: (photo-z) We drew from the COSMOS2020 zPDF for each iteration.
\end{itemize}

 Once  we determined all redshifts for each iteration, we used \texttt{LePhare} to fit the available COSMOS2020 photometry for all galaxies that land in the relevant redshift range (i.e., $2 < z < 3 $) for that iteration with the redshift fixed to the assigned redshift for that iteration. We used these fits to obtain the stellar masses for each relevant galaxy for each iteration (see \citet{sikorski25} for the methodology and substantial discussion). 

 \begin{figure*}
    \centering
    \includegraphics[width=0.49\textwidth]{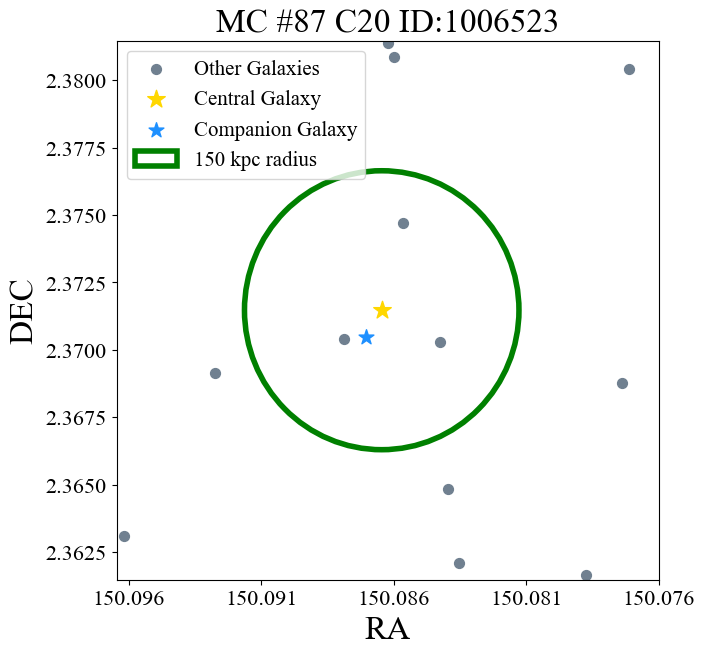}
    \includegraphics[width=0.49\textwidth]{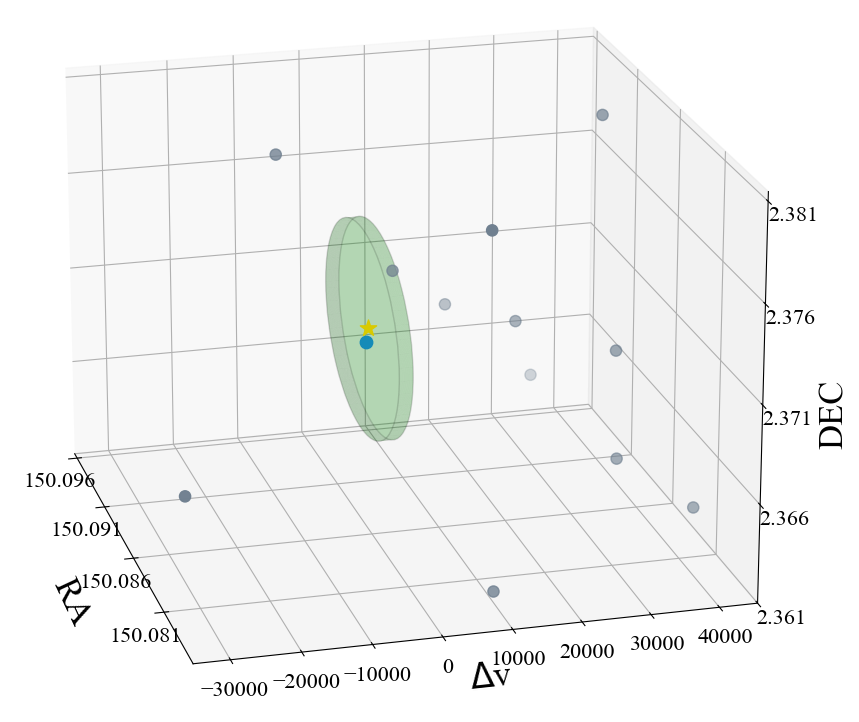}
    \caption{On-sky 2D view and 3D reconstruction of a potential galaxy companion for COSMOS2020 (C20) object \#1006523 found during our MC process (MC iteration \#87). The central galaxy (C20\#1006523) is marked in gold with our 150 kpc search radius centered on that galaxy shown in green. The companion galaxy that meets both the projected separation and LoS velocity difference criteria is denoted in blue. Other nearby galaxies are shown in gray. While multiple galaxies fall within the projected separation criteria for this galaxy, the 3D LoS velocity difference cut helps isolate the single companion from other projected companions.}
    \label{fig:postage}
\end{figure*}

 \subsubsection{Environmental determination}
\label{sec:deter env}

Once our redshifts and stellar masses had been determined for each iteration, we also needed to assign the relevant environment for that MC iteration. Thus, for each MC, we identified for each galaxy the \sigdel \space corresponding to the nearest voxel based on the galaxy's astrometric coordinates (COSMOS2020 RA/Dec) and redshift (from that iteration) from the 3D VMC maps that have been created for the COSMOS field (Section \ref{sec:da env}). From these overdensity values, we assigned either ``Hyperion'', ``field'', or ``unused'' for each galaxy in redshift range  of $2 < z < 3$ for that iteration. Galaxies are assigned to Hyperion based on the definition stated in Section \ref{sec:struc def} ($\Tilde{N}_{\text{Hyperion}}=309$ galaxies per iteration). All galaxies in the redshift range $2.4 < z < 2.7$ that are not assigned to Hyperion are designated ``unused'' to mitigate the effects of galaxies potentially infalling into the Hyperion structure (see \citealt{staab_protoclusters_2024}) or galaxies that are associated with other nearby overdense peaks \citep{cucciati_progeny_2018}. Field galaxies are thus all sources outside the redshift range of Hyperion ($2 < z < 2.4$ or $2.7 < z < 3$) that are not a part of any significantly massive overdense structure (i.e., do not belong to any \sigdel $> 2.5$ voxel that is contiguously connected to a \sigdel $> 4$ voxel with an overall structure mass of $\log(M_*/M_\odot) > 13$). This excludes from our field sample galaxies that belong to other known structures in the COSMOS field at relevant redshifts and astrometric coordinates \citep{diener_proto-groups_2013,cucciati_discovery_2014,lemaux_vimos_2014,yuan_keckmosfire_2014,casey_massive_2015,chiang_surveying_2015,franck_candidate_2016,lee_shadow_2016,wang_discovery_2016,cucciati_progeny_2018,lemaux_vimos_2018,darvish_spectroscopic_2020,koyama_planck-selected_2021,polletta_spectroscopic_2021,ata_predicted_2022,newman_population_2022,hung_discovering_2025}. This removal of galaxies in known structures from the field sample is done to help isolate the effect of the overdense Hyperion structure in comparison to field galaxies that are not associated with overdense environments. Unused sources for each MC iteration also include galaxies outside the redshift range considered in this work (i.e., $2 < z < 3$). In total, unused sources are galaxies outside the relevant z-range, galaxies within the z-range of Hyperion that are not associated with the massive overdense peak, or galaxies that are associated with other large structures in redshift ranges of our field sample. An example of the a full MC of galaxies from this work can be seen in Figure \ref{fig:mcsigmas}.

\subsubsection{Companion identification}
\label{sec:cID}

For each MC iteration, we identified all galaxies with potential companions based on their physical location for that iteration. For this, we followed the typical procedure within the literature and select potential companion galaxies based on on two primary factors: projected spatial separation and velocity difference of the member galaxies (see \citealt{lambas_galaxy_2003,lambas_galaxy_2012,robotham_galaxy_2014,ferreras_galaxy_2017,nottale_catalog_2018}). Similar to \citet{shah_investigating_2020,shah_investigating_2022}, we define companion galaxies as galaxies that are within $d_{\rm proj} < 150$ kpc and $\Delta v_{\rm LoS} < 1000$ km s$^{-1}$ of another galaxy. While other studies only consider maximum projected separations of companion galaxies to be in the range of 80--100 kpc \citep{patton_galaxy_2011,scudder_galaxy_2012,ellison_galaxy_2013-1}, we opt for a larger value of 150 kpc as there are studies that show that interacting galaxies begin to impact one another at such distances \citep{patton_galaxy_2013,shah_investigating_2020,shah_investigating_2022}, and, as we are interested in future merger and interaction activity in this work, galaxies in our redshift range of $2 < z < 3$ at distances of $\sim150$ kpc have ample time to merge by $z \sim 0$ (see Section \ref{sec:time merge}). Additionally, we found that our results are invariant with respect to the choice of a lower projected separation criteria, and that a $\gtrsim2\sigma$ difference in measured \fckc \space between the field and Hyperion persists at smaller $r_{\rm proj}$ (see Section \ref{sec:results}, as well as Section \ref{sec:time merge} and Table \ref{tab:timescales}).

Projected separations for companion galaxies were calculated by multiplying the angular separation between the two galaxies by the angular diameter distance of the average redshift of the two galaxies ($d_{\rm proj} \equiv \Theta_{\rm sep} * D_{\rm A})$. Line-of-sight (LoS) velocities are calculated based on the redshift of each galaxy for that MC iteration. At the redshift range of Hyperion, a $\Delta v_{\rm LoS} < 1000$ km s$^{-1}$ corresponds to a $\Delta z \lesssim 0.01$ between two galaxies and a $d_{\rm proj} < 150$ kpc corresponds to $\Theta_{\rm sep} \lesssim 18"$. An example of a galaxy with an identified companion based on our selection criteria can be seen in Figure \ref{fig:postage}. We chose to plot both a 2D and 3D representation of the same galaxy to highlight the importance of our LoS velocity criteria, as that cut is the most crucial in identifying projected companions versus potential true companions --- particularly in sky regions of higher galaxy number density. 

We did not make any cuts on the galaxy mass ratio during the companion identification process, but we did cut our final sample of companion systems used for all calculations to only consider major interactions or mergers. To be consistent with the literature, we adopted the definition of mass ratios less than 4 (4:1) to be major interactions and/or mergers and all others to be minor interactions and/or mergers (e.g., the ratio of the mass of the most massive galaxy to the mass of the least massive galaxy, see \citealt{ellison_galaxy_2013} and \citealt{mantha_major_2018}). The stellar masses for calculating these mass ratios are the result of the \texttt{LePhare} SED fitting described in \citet{sikorski25} and are based on the redshifts obtained in the MC process.

\subsubsection{Fraction calculation}
\label{sec:frac calc}

For each MC iteration, we calculated a fraction of close kinematic companions (\fckc) for both Hyperion and the coeval field considering only galaxies that have companions that would result in a major iteration and/or merger ($\leq$4:1 in mass ratio). This calculation is simply done by taking the unique number of galaxies with companions or that are companions (to mitigate galaxies with multiple potential companions; $\sim14.0\pm8.9$\% of Hyperion companion galaxies over our 100 MC iterations and $\sim3.6\pm0.8$\% of field companion galaxies) divided by the total number of galaxies in Hyperion or the field for that MC iteration ($f_{\rm ckc} = N_{\rm companions} / N_{\rm gals}$). This gives us our uncorrected companion fractions for each MC iteration. Typically, $N_{\rm companions} \sim43$ galaxies for Hyperion and $\sim1141$ galaxies for field. We then take the median \fckc \space over these 100 MC iterations as the final uncorrected companion fraction: $14\pm1.8$\% for Hyperion and $5\pm0.2$\% for the field (with the associated error on this measurement being the 1$\sigma$ spread of the 100 MC iterations, see Section \ref{sec:correct} for full error budget). 

\subsection{Simulated lightcone}
\label{sec:cone}

In order to validate and make any potential corrections to our companion methodology, we applied our MC methodology to simulated galaxy observations where the ``true'' \fckc \space is known. For this, we employed 
the GAlaxy Evolution and Assembly (GAEA; \citealt{hirschmann_galaxy_2016,de_lucia_tracing_2024}) semi-analytic model (SAM) to generate simulated galaxy catalogs. We used the GAEA model version described in \citet{xie_h2-based_2017}\footnote{This version includes an enhanced treatment of the partition of cold gas in atomic and molecular hydrogen along with an update to the model used for disc sizes.}, and applied it to the dark matter merger trees of the Millennium Simulation \citep{springel_simulations_2005}. We followed the procedure of \citet{zoldan_h_2017} to generate a lightcone from the output of the GAEA SAM. The simulated lightcone used in this work is identical to the one used in \citet{hung_discovering_2025}. 

More specifically, we used the $1\times1$ deg$^2$ field denoted ``Mock1'' from \citet{hung_discovering_2025} cut at IRAC/ch1 $< 24.8$. We chose to use this portion of the simulated lightcone as there are multiple versions of this region created with at different levels of spectroscopic redshift fraction (SzF). To create these different lightcone versions at varying SzF, \citet{hung_discovering_2025} broke down that region of the lightcone into bins of magnitude and redshift, and, using statistics that compared the spectroscopic and photometric redshifts within the COSMOS field (see \citealt{lemaux_vimos_2022}), generated ``observed'' spectroscopic catalogs using different SzF selection functions as a function of the magnitude and redshift from the COSMOS field. Spectroscopic redshifts were assigned randomly to galaxies in each given magnitude and redshift bin (the former of which were chosen to roughly contain an equal number of objects in each bin) and spectroscopic quality flags were assigned to mimic the distribution of quality flags in the real data. The base SzF level for this selection function (SzF 1.0) was modeled at the underlying level of the COSMOS field spectroscopic completeness at the time of the mock catalog creation ($\sim7$\%, \citealt{hung_discovering_2025}). The other SzF fractions can be found in Table \ref{tab:szffracs} representing the five different Mock1 simulated lightcone catalog versions. 

\begin{figure}
    \centering
    \includegraphics[width=1\linewidth]{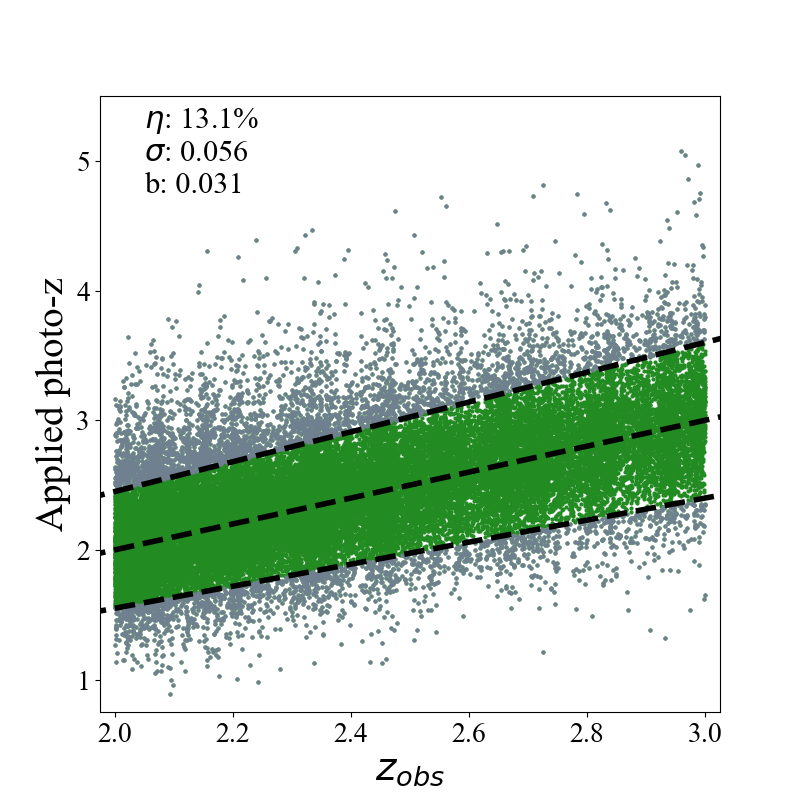}
    \caption{Applied photometric redshifts (photo-zs) versus true simulation redshifts ($z_{\rm obs}$) for all galaxies in the Mock1 region of the simulated lightcone between $ 2 < z_{\rm obs} < 3$. Galaxies that meet $|\Delta z|<0.15(1+z_{\rm obs})$ are marked in green \citep{hildebrandt_cfhtlens_2012}. The statistics for these galaxies are listed in the upper right hand corner (outlier fraction, scatter, and bias, respectively). Photo-zs are applied to all galaxies in our simulated lightcone based on the observed photo-z statistics in COSMOS2020 as a function of magnitude and redshift. The scatter in photo-zs creates difficulty in recovering companion systems (see Section \ref{sec:cone valid}).}
    \label{fig:lightconezs}
\end{figure}

All objects within each version of the Mock1 region were also assigned a photometric redshift regardless of whether or not they were assigned a spectroscopic redshift using the following formula: $z_{\rm phot} = z_{\rm obs} + B(1+z_{\rm obs}) + N \sigma_{\rm pz} (1 + z_{\rm obs}$), where B is the spectroscopic to photometric redshift bias in the galaxies magnitude and redshift bin, N is a value sampled from a normalized Gaussian distribution for each object, and $\sigma_{\rm pz}$ corresponds to the $\sigma_{\rm NMAD}$ of the photometric redshift scatter within that bin. Photometric redshift uncertainties (at $\pm1\sigma$) for each simulated galaxy were drawn from the PDF of the fractional photometric redshift error (i.e., $(z_{\rm phot,1\sigma,upper} - z_{\rm phot}) / (1+z_{\rm phot})$ and $(z_{\rm phot} - z_{\rm phot,1\sigma,lower}) / (1+z_{\rm phot})$) that was calculated through the statistics of objects in each magnitude bin in the real COSMOS data \citep{lemaux_vimos_2022}. Thus, at the end of this process each simulated galaxy had a photometric redshift with an associated $\pm1\sigma$ uncertainty. The photometric redshifts generated via this parametrization can be seen in Figure \ref{fig:lightconezs} for relevant galaxies from the Mock1 region of the simulated lightcone (i.e., from $2 < z_{\rm obs} < 3$).

\subsubsection{Lightcone validation}
\label{sec:cone valid}

We applied our MC companion methodology to each version of the Mock1 catalog (i.e., at each different SzF) using an identical method to that detailed in Section \ref{sec:comp frac} --- aside from the determination of environment --- in order to estimate how accurately our methodology can recover true galaxy companions based on our observational definition at different levels of spectroscopic completeness. For each Mock1 SzF catalog, we performed 100 MC iterations looking for galaxy companions in the redshift range $2 < z < 3$ (i.e., the same redshift range we used for our real observed data). While we tracked companions of all mass ratios, we again considered only companion systems that would potentially result in a major merger or interaction (maximum mass ratio of 4:1) to be consistent with the operational definition adopted for our observed data on what types of systems are included in our \fckc \space and other calculations. 

To assess our recovery of true companion galaxies, we used the true redshifts given from the simulation, $z_{\rm obs}$, a value that incorporates peculiar velocities, and find all galaxies with major companions based on our criteria given in Section \ref{sec:cID}. From this, we obtained the true \fckc \space  of 24.3\% for the simulated galaxies in the lightcone over $2 < z_{\rm obs} < 3$ (a value consistent with the literature, see Figure \ref{fig:money}). We then calculated the median purity and median completeness of the companion galaxies identified in our 100 MC iterations at each SzF based on the true companion galaxies in the lightcone. As expected, we find that both the median purity and median completeness of companion galaxies identified with our MC methodology increase with SzF as, at lower levels of spectroscopic completeness, larger numbers of galaxies are assigned ``observed'' photometric redshifts that obscure their true location (which are randomly scattered from their true redshift; see Figure \ref{fig:lightconezs}) which in turn greatly reduces the likelihood they are identified as companions in our methodology or causes them to be identified in false companion systems. The median purity and median completeness of the companion galaxies identified in our application of the MC methodology at each SzF can be found in Table \ref{tab:szffracs} and in Figure \ref{fig:recovery} (upper panel).

\begin{table}[h]
    \caption{\label{tab:szffracs} Mock lightcone and validation statistics}
    \centering
    \begin{tabular}{c|c|c|c|c}
         SzF& F$_{\text{spec}}$ & Purity & Completeness & Correction \\
         \hline
         \hline
         0.5 & 3.5\% & 24.2\% & 5.4\% & 4.47 \\
         1.0 & 7.0\% & 25.1\% & 5.6\% & 4.44\\
         1.5 & 10.5\% & 26.7\% & 6.2\% & 4.31\\
         2.0 & 14.0\% & 28.6\% & 6.8\% &4.19\\
         4.0 & 28.0\% & 38.4\% & 10.8\% &3.55 \\
        \hline
    \end{tabular}
    \tablefoot{SzF and corresponding spectroscopic fraction (i.e., fraction of galaxies with a quality spectroscopic redshift) are provided for each version of the Mock1 lightcone. Also given is the median purity and median completeness of identified companion galaxies at each SzF level using our MC companion methodology along with the derived correction factors (see Section \ref{sec:correct}). All fractions and values are for only objects brighter than $M_{\text{IRAC1}}<24.8$ in the simulated lightcone to match our catalog selection. }
\end{table}

\subsubsection{Correction factor}
\label{sec:correct}

\begin{figure}[t]
    \centering
    \includegraphics[width=1\linewidth]{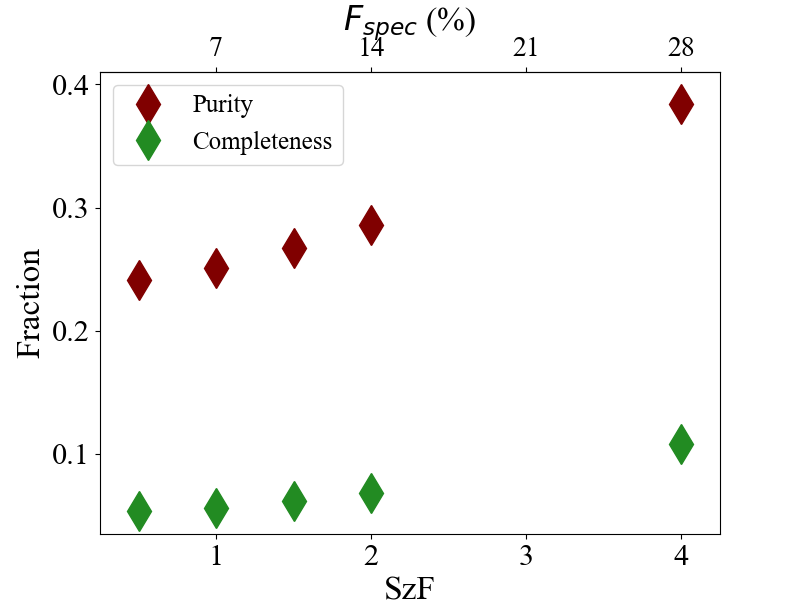}
    \includegraphics[width=1\linewidth]{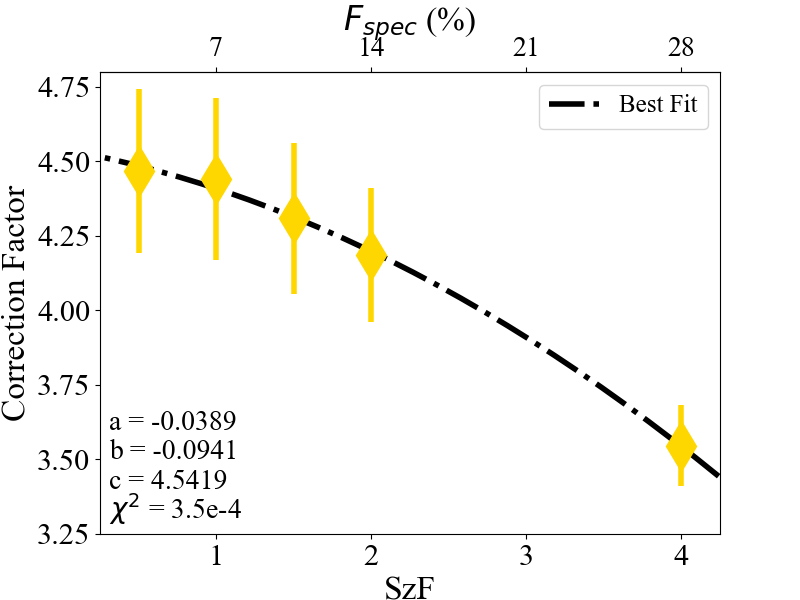}
    \caption{Top: Median purity and median  completeness of companion galaxies identified via the application of our MC methodology on the Mock1 lightcone at different levels of SzF. Bottom: Correction factor ($\equiv \frac{\text{purity}}{\text{completeness}}$) as a function of SzF. The best-fit second order polynomial used in the interpolation and determination of our final correction factors is shown in black with the parameters of the fit given in the lower left (along with the fit $\chi^2$). Though the median purity and median completeness of identified companion galaxies increases with SzF, a correction factor is needed to properly assess the true underlying \fckc.}
    \label{fig:recovery}
\end{figure}

With our MC companion methodology, we see an increase in both the purity and completeness of identified companion galaxies as a function of SzF. This is expected as in our methodology quality spec-zs (which are measured at $z_{\rm obs}$ for the simulated lightcone data) are often kept as the measured redshift value, which in turn results in more frequent recovery with our companion criteria (even a variation of $\Delta z = 0.05$ results in no companion) and thus higher completeness and increased purity. However, the purity and completeness achieved with our MC methodology does not approach unity -- even at 28\% spectroscopic completeness -- resulting in the derivation of a correction factor to adjust for the identification of false positives and the lack of true recovery of companion galaxies. We constructed this correction factor to be equivalent to the median purity divided by the median completeness of identified companion galaxies at a given SzF. These correction factors can be found in Table \ref{tab:szffracs} and Figure \ref{fig:recovery} (bottom panel). We used these calculated correction factors to inform the choice of our final correction factors for the Hyperion and the field. 

\begin{figure}[h!]
    \centering
    \includegraphics[width=1\linewidth]{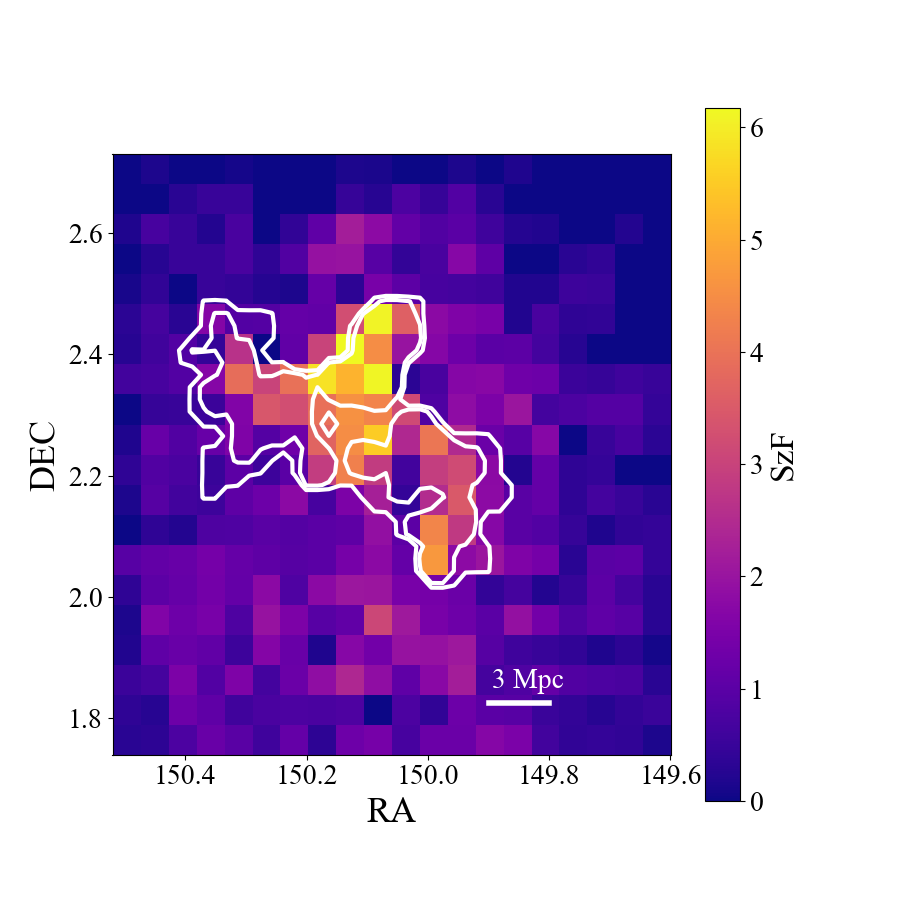}
    \caption{SzF for sources with a photo-z with $2 < z < 3$ in our final sample in $20\times20$ bins of even RA and Dec (each bin is approximately $\sim2.8'\times3.0'$). Over-plotted are the 2.5\sigdel \space and 4\sigdel \space contours of the massive structure defined as Hyperion in this work.  As expected from the underlying spec-z surveys used in this study, we see increased SzF in the regions covered by the Hyperion overdensity finding a median SzF of $\sim1.78$ for Hyperion and a median SzF of $\sim0.67$ for the field. These median SzFs are used to determine the final correction factors applied to our uncorrected \fckc \space values.}
    \label{fig:szfsky}
\end{figure}

However, SzF is not constant as a function of astrometric coordinates in our final sample, as the C3VO and \hsthyp \space observations are specifically targeted at the overdense peaks in the Hyperion structure which increases the SzF non-uniformly in RA/Dec space. To explore this effect, we plot SzF as a function of RA/Dec in Figure \ref{fig:szfsky} in $20\times20$ even bins of RA and Dec (approximately $\sim2.8'\times3.0'$) for all galaxies in our final sample with a photometric redshift in the range $2 < z < 3$. Plotted for reference are the $\sigma_{\rm \delta} > 2.5$ and $\sigma_{\rm \delta} > 4$ contours of the overdense structure that comprises Hyperion in this work compressed into two-dimensions (i.e., the regions used for the Hyperion \fckc). 

Using the $20\times20$ RA/Dec grid seen in Figure \ref{fig:szfsky}, we calculated a median SzF for pixels inside the Hyperion contours (SzF-Hyp) and a median SzF for pixels outside of the SzF contours (SzF-Field). From this, we obtain an SzF-Hyp $\sim1.78$ ($F_{\text{spec}} \sim12.4$\%) and an SzF-Field $\sim 0.67$ ($F_{\text{spec}} \sim4.7$\%). We translate these SzF values to the proper correction factors needed to correct the ``raw'' Hyperion \fckc \space and field \fckc \space by interpolating between the calculated correction factors via the second-order polynomial fit seen in Figure \ref{fig:recovery}). The correction factor estimates based on this fit and the median SzF of Hyperion and the field are $\sim4.25$ and $\sim4.46$, respectively, and are applied to our uncorrected \fckc \space values to obtain our final \fckc \space measurements. Due to the large difference already present in our uncorrected companion fractions (see Section \ref{sec:comp frac}), our results are immutable to our choice of correction factor over the range of SzF considered in this work. 

We carefully account for the uncertainties in our median SzF estimation for Hyperion and the field and its impact on the chosen correction factor, along with the variation in correction factor from applying our MC methodology to the lightcone data (i.e, the variation in the purity and completeness of our identified companion galaxies), into the final error budget of our \fckc \space measurements. To incorporate these uncertainties, we combined all three sources of error in quadrature: (1) the scatter in correction factor from the lightcone at a given SzF, (2) the variation in correction factor based on the $\sigma_{\text{NMAD}}$ of the SzF in Hyperion and the field, and (3) the median scatter in uncorrected \fckc \space for Hyperion and the field. The resulting error is then applied to our corrected \fckc \space values to determine the final uncertainties reported in Section \ref{sec:results}.

\section{Results}
\label{sec:results}

\begin{table*}[ht]
    \caption{\label{tab:fracs} Relevant merger rates and pair fractions in the literature}
    \centering
    \begin{tabular}{c|c|c|c|c|c|c}
         Structure & Type & f$_{\rm type}$ (\%) & M$_*$/Mag Range & Major? & Redshift Range & Reference \\
         \hline
         \hline
         SSA22 & Merger & 48$\pm10$ & 19.0 $\leq R_{\rm AB} \leq$ 25.5 mag& No & $z$ $\simeq$ 3.1 & 1 \\
         \hline
         Field & Merger & 31$\pm5$ & 19.0 $\leq R_{\rm AB} \leq$ 25.5 mag& No & $2.5 \leq z \leq 3.5$ & 1  \\
         \hline
         SSA22 & Merger & 38$^{+37}_{-20}$ & $R_{\rm AB} \leq$ 25.5 mag & No & $z$ $\simeq$ 3.1 & 2 \\
         \hline
         Field & Merger & 41$^{+11}_{-9}$ & $R_{\rm AB} \leq$ 25.5 mag & No & $ 2.9 \leq z \leq 3.3 $ & 2 \\
         \hline
         BOSS1244 &Pair & 22$\pm$5 & $\log(M_*/M_\odot) >$ 10.3 & Yes &  $z \simeq$ 2.24 & 3 \\
         \hline 
         BOSS1542 & Pair & 33$\pm$6 & $\log(M_*/M_\odot) >$ 10.3 & Yes &  $z \simeq$ 2.24 & 3 \\
         \hline 
         Field & Pair & 12$\pm$2 & $\log(M_*/M_\odot) >$ 10.3 & Yes &  2.1 $< z <$ 2.4 & 3 \\
         \hline
         $\bar{\delta} = 0.06$ & Merger & 20$\pm0.4$ & $m_{\rm UV} < 23.5$ & Yes & $z\sim2$& 4 \\
         \hline
         $\bar{\delta} = 0.76$ & Merger & 20.6$\pm0.9$ & $m_{\rm UV} < 23.5$ & Yes & $z\sim2$& 4 \\
         \hline
         $\bar{\delta} = 0.06$ & Merger & 20$\pm1.5$ & $m_{\rm UV} < 23.5$ & Yes & $z\sim3$& 4 \\
         \hline
         $\bar{\delta} = 0.94$ & Merger & 24.6$^{+2.5}_{-2.2}$ & $m_{\rm UV} < 23.5$ & Yes &  $z\sim3$& 4 \\
         \hline
         $\bar{\delta} = 0.22$ & Merger & 20$\pm0.5$ & $m_{\rm UV} < 23.5$ & Yes & $z\sim4-5$& 4 \\
         \hline
         $\bar{\delta} = 2.64$ & Merger & 27.7$^{+5.2}_{-4.4}$ & $m_{\rm UV} < 23.5$ & Yes & $z\sim4-5$& 4 \\
         \hline
         $\bar{\delta} = 5.11$ & Merger & 35.9$^{+14.2}_{-10.7}$ & $m_{\rm UV} < 23.5$ & Yes &  $z\sim4-5$& 4 \\
         \hline
         Hyperion & CKC & $59_{-10}^{+9}$ & 9.3 $\lesssim \log(M_*/M_\odot)$ & Yes &  2.4 $< z <$ 2.7 & This Work \\
         \hline
         Field & CKC & $23_{-1.8}^{+1.7}$ & 9.3 $\lesssim \log(M_*/M_\odot)$ & Yes &  $2< z < 2.4$ \& $2.7< z < 3$ & This Work \\
         \hline
         \hline
        
    \end{tabular}
    \tablefoot{All rates given are derived in similar fashions to our \fckc \space metric and are for studies that measure rates for high-z structures at relevant redshift ranges to that of Hyperion ($z\gtrsim2$). Though the assessment techniques and selection methods vary greatly for the limited studies currently available, merger and interaction activity is consistently found to be higher within more overdense regions and/or structures. For the fractions drawn from \citet{shibuya_galaxy_2025}, values are reported for different average levels of galaxy overdensity ($\delta$) rather than at a field versus structure level. Based on our definition of structure (see Section \ref{sec:struc def}), a $\delta \gtrsim 0.62 $ is the rough threshold we associate with potential structure. }
    \tablebib{(1) \citet{hine_enhanced_2016}; (2) \citet{monson_nature_2021}; (3) \citet{liu_what_2023}; (4) \citet{shibuya_galaxy_2025}}
\end{table*}

By applying the correction factors derived in Section \ref{sec:correct} to our uncorrected companion fractions from Section \ref{sec:comp frac}, we obtain a final \fckc \space for Hyperion and a final \fckc \space for the field. We find a $\gtrsim2.5\times$ enhancement in \fckc \space for galaxies in the overdense structure of Hyperion with over half of all Hyperion galaxies having a nearby companion, as we measure a corrected \fckc \space $=$ \frachyp \space for Hyperion and an \fckc$ =$ \fracfield \space for the field (a $\gtrsim3\sigma$ difference). Though an overdense region similar to Hyperion naturally suggests higher \fckc \space --- due to many galaxies existing in relatively close proximity --- this measurement properly validates that intuition. Additionally, protoclusters at the epoch of Hyperion are still diffuse systems that span many cMpc and sometimes wide redshift ranges ($\Delta z \sim 0.5$ in the case of Hyperion), so satisfying our stringent companion criteria is not easy; particularly in the LoS direction where galaxies are required to be tightly correlated in redshift space ($\Delta z \lesssim 0.01$). 

\begin{figure*}
    \centering
    \includegraphics[width=1\linewidth]{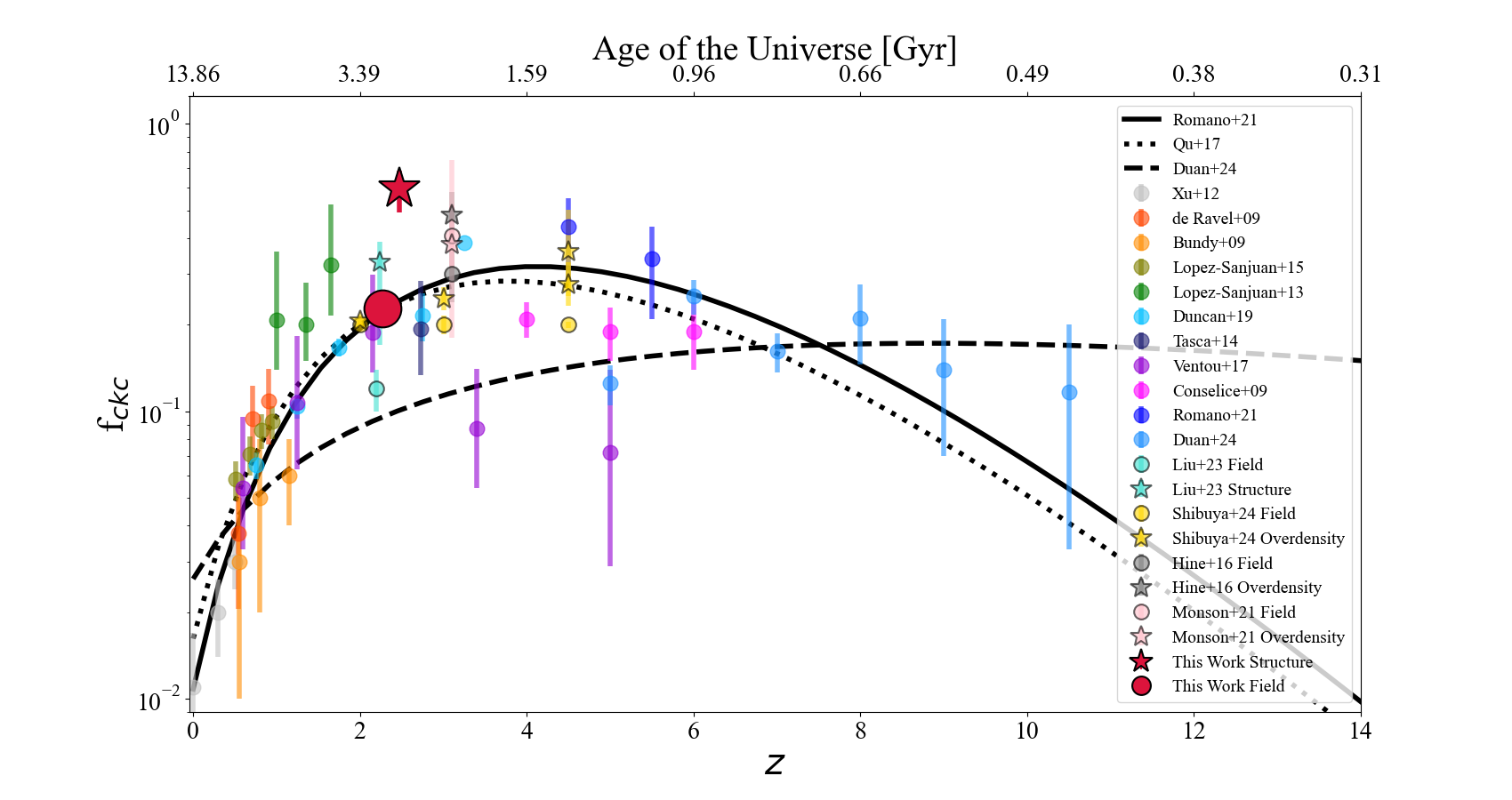}
    \caption{Evolution of \fckc \space and \fckc-like measurements as a function of redshift. The values measured in this work are plotted in red with black borders with the star denoting our structure measurement and the circle denoting our field measurement. Other studies that examine merger and interaction rates in relation to overdensity or structure at relevant epochs have their results plotted similarly (i.e., structure or overdensity fraction as a star and field fraction as a circle, \citealt{hine_enhanced_2016,monson_nature_2021,liu_what_2023,shibuya_galaxy_2025}) Various values from the literature that measure relevant major merger and pair fractions are also plotted for reference \citep{bundy_greater_2009,conselice_structures_2009,de_ravel_vimos_2009,xu_cosmic_2012,lopez-sanjuan_massiv_2013,tasca_evidence_2014,lopez-sanjuan_alhambra_2015,ventou_muse_2017,duncan_observational_2019,romano_alpine-alma_2021,duan_galaxy_2025}, along with simulation predictions from \citet{qu_chronicle_2017} and empirical fits from \citet{romano_alpine-alma_2021} and \citet{duan_galaxy_2025}. Overall, we find that \fckc \space evolves sharply at low redshift before peaking at $z\sim4$ with the measured \fckc \space for Hyperion, though at lower redshift ($z\sim2.5$), existing above the peak value. Though some scatter exists, the existing measured merger and interaction rates for structures and overdensities are higher than the corresponding field rates at similar redshifts. }
    \label{fig:money}
\end{figure*}

Though there are relatively few studies in the literature that examine quantities similar to \fckc \space at a relevant redshift range to this work ($2 < z < 3$), we note that the field \fckc \space derived in this study is in good agreement with other predictions for major merger and interaction fractions, demonstrating the validity of our applied technique (see Figure \ref{fig:money}). The dearth of other studies is often attributed to the lack of spectroscopic coverage to confirm galaxies with companions or to the lack of completeness in photometry which makes probabilistic methodologies more difficult. However, we do wish to acknowledge that the methodologies for investigating companions, interactions, and merger activity (i.e., close pair and morphological studies) in the literature are varied, and there is no consensus on the best approach, as data quality and availability varies greatly --- particularly for methods that attempt to include sources with photo-zs. We do our best to compare to values in the literature with as similar companion selection criteria to those employed for our \fckc \space methodology whenever possible. 

We also wish to highlight the only three other studies in the literature that explore merger and interaction activity in high-z ($z>2$) protoclusters from \citet{hine_enhanced_2016}, \citet{monson_nature_2021}, and \citet{liu_what_2023}. The most recent work, \citet{liu_what_2023}, finds a similar enhancement in merger and interaction activity in both of their structures (BOSS1244 and BOSS1542) in comparison to their coeval field sample, but their search is limited to only high-mass ($\log(M_*/M_\odot) \geq 10.3$) H$\alpha$ emission-line (HAE) galaxies. \citet{hine_enhanced_2016} and \citet{monson_nature_2021} both employ observations from the Hubble Space Telescope (HST) to study the SSA22 protocluster at $z=3.1$. However, these two studies obtain marginally discrepant results when assessing mergers within SSA22 with \citet{hine_enhanced_2016} finding an $\sim$1.5$\sigma$ enhancement in merger activity in the protocluster relative to the field and \citet{monson_nature_2021} finding statistically equivalent rates of mergers in the field and protocluster. As discussed in \citet{monson_nature_2021}, the potential difference in the two results is likely due to small number statistics in combination with the different rest-frame wavelength ranges probed by each study for their morphological merger classification (rest-frame UV from HST/ACS F814W imaging for \citet{hine_enhanced_2016} and rest-frame optical from HST/WFC3 F160W for \citet{monson_nature_2021}). Overall, these three studies offer the most relevant comparisons to this work and are statically in agreement with our results, though with large uncertainties in each case. Our study includes an order of magnitude more galaxies both in terms of overall sample and potential identified galaxies with companions resulting in much tighter uncertainty intervals. The fractions derived in \citet{hine_enhanced_2016}, \citet{monson_nature_2021}, and \citet{liu_what_2023} are included along with their field fractions in Table \ref{tab:fracs}. 

Also included in Table \ref{tab:fracs} is the results of \citet{shibuya_galaxy_2025}. This study measures the major merger fraction for galaxies in regions of different galaxy overdensity ($\delta =(N_{\rm gal} -\bar{N}_{\rm gal})/ \bar{N}_{\rm gal}$) in the $\sim300$ deg$^2$ area of the combined HSC Strategic Survey Program \citep{aihara_hyper_2018,aihara_first_2018,aihara_second_2019,aihara_third_2022} and CFHT Large Area U-band Survey \citep{sawicki_cfht_2019} and thus is of interesting comparison for this work. \citet{shibuya_galaxy_2025} find a roughly linear increase in the fraction of major mergers as a function of overdensity from $z \sim 2 - 5$, but the relative increase in merger fraction with increased density is marginal. This overall increase in the merger fraction with increased density aligns well will the results of this study, though the fractional increase in merger and interaction activity indicated by our \fckc \space measurements is much greater. However, a direct comparison in these terms between our work and \citet{shibuya_galaxy_2025} is difficult as they report their measured merger fractions at different average levels of galaxy density ($\bar{\delta}$) rather than at a structure versus field level (a $\delta \gtrsim 0.62$ roughly corresponds with the threshold we use to discern between structure and field populations based on our VMC mapping technique and definition of structure, see Section \ref{sec:struc def}). Additionally, \citet{shibuya_galaxy_2025} employ only photo-zs in their calculations which can create issues in separating out what galaxies belong to a field versus structure population \citep{hung_establishing_2020,hung_discovering_2025}.

To summarize our findings in comparison to both other environmental studies and to general merger and interaction studies in the field, we plot our measured \fckc \space values along with various major pair and major merger rates from the literature as a function of redshift in Figure \ref{fig:money}. For this figure, we use the studies (and selection criteria) chosen in \citet{romano_alpine-alma_2021} along with additional relevant selections from the literature to expand the redshift range and to display the other previously discussed studies that consider environment when calculating their pair and/or merger fractions. For included studies that do consider environmental implications, the assumptions and methods for their field measurements and structure measurements are the same, but the assumptions and applied methodology across studies --- regardless of environmental consideration --- can vary greatly. Thus, it can be difficult to make specific comparisons between studies though some general trends can be found. From newly obtained measurements of the early Universe ($z > 6$; \citealt{duan_galaxy_2025}), we see a slight evolution of the fraction of merging and interacting galaxies that eventually peaks at $z\sim4$ with a strong drop off in the local Universe after $z\sim2$. This evolution roughly maps the currently known cosmic star-formation history \citep{madau_cosmic_2014,harikane_comprehensive_2023,harikane_pure_2024,harikane_jwst_2025} and AGN history \citep{aird_evolution_2015,kulkarni_evolution_2019} of the Universe, though offset to slightly higher redshifts. However, the corrected \fckc \space value measured for Hyperion at $z\sim2.5$ is above the peak of merger and interaction activity at $z\sim4$, and significantly above similar values for the coeval field. In combination with the sparse other literature measurements that are also mostly enhanced compared to the field, this result suggests that merger and interaction activity is more prevalent in dense environments at higher redshifts ($z>2$) when structure is still forming and developing.

Given the wide breadth of companion selection criteria currently used in the literature, and to help facilitate future comparison, we also will report corrected \fckc \space values for Hyperion and the field cut at $r_{\rm proj} < 50$ kpc and at $r_{\rm proj} < 100$ kpc. To calculate these \fckc \space values, we take our same MC iterations for both the observed data and lightcone data, but cut all identified companion systems at larger projected separation criteria. We then perform the same SzF fitting as seen in Figure \ref{fig:recovery} for the recovered lightcone companion galaxies at this $r_{\rm proj}$ limit and interpolate to the median SzFs of Hyperion and the field given in Section \ref{sec:correct}. We apply these updated correction factors to the ``raw'' companion fractions from the observed data based on the smaller $r_{\rm proj}$ cuts with the same total error considerations to obtain our final corrected \fckc \space values and errors at reduced projected separations. We find that our result of enhanced merger and interaction activity in Hyperion is not sensitive to the choice of projected separation criteria, and calculate a \fckc$=48^{+9.6}_{-10}$\% for Hyperion and a \fckc$=16^{+2.0}_{-2.0}$\% for the field at $r_{\rm proj} < 100$ kpc. Similarly, we recover a \fckc$=40^{+10}_{-12}$\% for Hyperion and a \fckc$=9.0^{+2.2}_{-2.2}$ for the field at $r_{\rm proj} < 50$ kpc. While our field \fckc \space values begin to deviate from other measurements in the literature at smaller separations, there are additional assumptions we are unable to control for between all studies (namely sample selection and any correction or weighting scheme), and our trend of a significant ($\gtrsim$2$\sigma$) increase in merger activity in Hyperion relative to the coeval field still holds.

\begin{figure*}[h]
    \centering
    \includegraphics[width=1\linewidth]{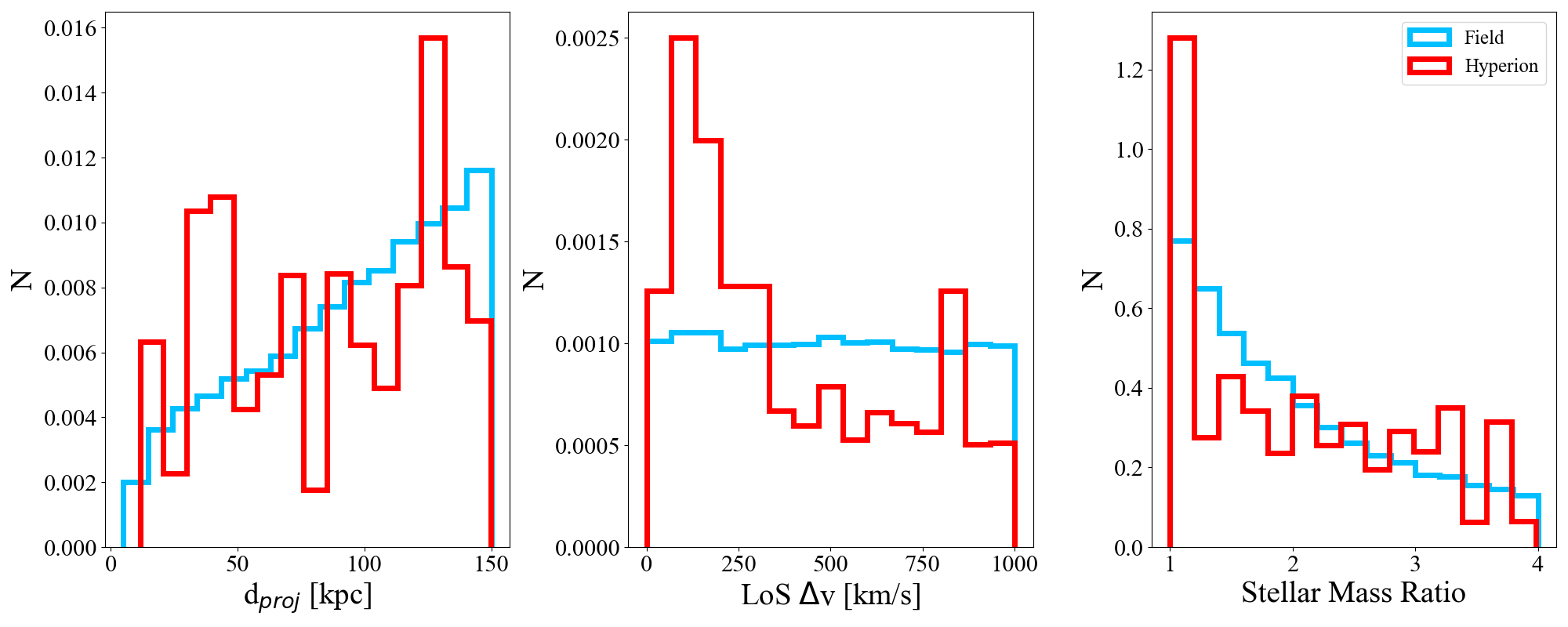}
    \caption{Normalized distributions of projected separation ($d_{\text{proj}}$; left), LoS velocity ($\Delta v_{LoS}$; center), and stellar mass ratio (right) between companion systems found in Hyperion (red) and the coeval field (blue). The plotted distributions contain all companion systems found over each of the 100 MC iterations used to identify close kinematic companions, so companion systems identified in multiple iterations are plotted multiple times. As opposed to the uniform and monotonic distributions seen for companion systems in the field, companion systems in Hyperion exhibit more complex distributions of $d_{\text{proj}}$, $\Delta v_{LoS}$, and stellar mass ratio. }
    \label{fig:pairprops}
\end{figure*}

We also include Figure \ref{fig:pairprops} to compare the distributions of projected separation ($d_{\text{proj}}$), LoS velocity ($\Delta v_{LoS}$), and stellar mass ratio between companion systems found in Hyperion and the coeval field. The plotted distributions are normalized for easier comparison and contain all companion systems found over each of our 100 MC iterations (i.e., companion systems identified in multiple iterations are included multiple times). Over our 100 MC iterations, we see a few trends emerging between companion systems found in Hyperion versus companion systems found in the field. Whereas the occurrence of close kinematic companions for the field monotonically increases with $d_{\text{proj}}$, we find increased numbers of companion systems at $d_{\text{proj}} \sim 40$ kpc and $d_{\text{proj}} \sim 125$ kpc for Hyperion. We also find that the underlying distribution of $\Delta v_{LoS}$ for companion systems in the field is roughly uniform while companion systems in Hyperion are more tightly correlated in redshift space with many companion systems existing at $\Delta v_{LoS} <250$ km/s. Lastly, companion systems in the field see a monotonic decrease with stellar mass ratio in contrast to the buildup of companion systems at $M_{*}$ ratio $\sim1$ for Hyperion. However, since Hyperion sees an increase in spectroscopic completeness in comparison to our field sample, some of the observed trends in the companion system parameters for Hyperion are likely the result of more consistent redshifts being used in our MC process which results the preservation of the same companion systems across MC iterations. Further implications of the differences in these underlying distributions for companion systems in Hyperion and the field are discussed in the following sections (i.e., Section \ref{sec:tides} and Section \ref{sec:time merge}).

\subsection{Tidal strength}
\label{sec:tides}

In order to quantify the impact of any potential interaction activity for companion galaxies found in this study, we utilize the tidal strength $Q$ that companion galaxies produces on a central (more massive) galaxy \citep{dahari_companions_1984,verley_amiga_2007,verley_amiga_2007-1,sabater_effect_2013,argudo-fernandez_amiga_2013,argudo-fernandez_effects_2014}. This tidal strength parameter attempts to quantify the relationship between the tidal forces exerted on the galaxy,

\begin{equation}
    F_{\text{tidal}} \simeq \frac{M_i\times D_p}{R_{\rm i,p}^3},
\end{equation}

\noindent and the internal binding force of galaxy, 

\begin{equation}
    F_{\text{bind}} = \frac{M_p}{D_p^2},
\end{equation}

\noindent making the tidal strength, $Q$, equivalent to the ratio of these forces, 

\begin{equation}
    Q_{\rm iP} \equiv \frac{F_{\text{tidal}}}{F_{\text{bind}}} \propto \frac{M_i}{M_P} \left( \frac{D_P}{R_{\rm iP}}\right)^3,
\end{equation}

\noindent where $M_i$ and $M_P$ are the stellar masses of the companion and primary galaxy, respectively. $D_P$ is the apparent diameter of the primary galaxy, and $R_{\rm iP}$ is the projected physical distance between companion and primary galaxy ($R_{\rm i,p} \equiv r_{\rm proj}$). The total tidal strength is thus the summed tidal strength exerted by all companions and is defined as

\begin{equation}
    Q = \log_{10} \left(\sum_i Q_{\rm iP}\right).
\end{equation}

\noindent This logarithm of the sum of the tidal strength created by all nearby companions is thus a dimensionless estimate of the gravitational interaction strength \citep{verley_amiga_2007}, and generally varies between -5 and +2 depending on the mass ratio and projected distance to companion galaxies. Based on this definition of $Q$, the tidal forces exerted by the companion galaxy on the primary galaxy is equivalent to the binding forces only when $Q \geq 0$. However, based on the results of numerical simulations \citep{athanassoula_spiral_1984,byrd_tidal_1992}, as well as observational results \citep{varela_properties_2004}, perturbations may be induced by external tidal forces from companion galaxies when those tidal forces amount to at least 1\% of the internal binding force (i.e., $Q \geq -2$). 

For each of our 100 MC iterations, we calculated the median tidal strength ($\Tilde{Q}$) for all galaxies with companions in the field and in Hyperion. If there are multiple companions in nearby proximity, we find the tidal strength on the most massive galaxy in every system of companions. For this calculation, we used the masses derived in Section \ref{sec:zMC} and the projected distances calculated in Section \ref{sec:cID}. For the apparent diameter of the primary galaxy ($D_P$), we use 2$\times r_{80}$ with $r_{80}$ being estimated from Equation 2 of \citet{mowla_mass-dependent_2019} using their best fit parameters from $2 < z < 2.5$ or $2.5 < z < 3$. From the individual $\Tilde{Q}$ values from each of the 100 MC iterations, we calculated the median $\Tilde{Q}$ over all 100 MC iterations ($\Tilde{Q}_{100}$) for both companion systems in Hyperion and companion systems in the field. We find a $\Tilde{Q}_{100} \sim -3.04\pm0.18$ for Hyperion companion systems and a $\Tilde{Q}_{100} \sim -3.43\pm0.03$ for the field companion systems. Thus, the tidal strength exerted by companion galaxies in Hyperion (via $\Tilde{Q}_{100}$) is on average enhanced by $\sim 0.4$ dex ($\sim2.5\times$) in comparison to the tidal strength exerted by companion galaxies in the field. This comports with the higher average fraction of galaxies with multiple companions in Hyperion ($\sim14.0\pm8.9$\% relative to $\sim3.6\pm0.8$\% for the field) along with the higher average fraction of companion galaxies within closer proximity in Hyperion ($d_{\text{proj}} < 50$ kpc; $\sim28.0\pm6.5$\% for Hyperion and $\sim17.6\pm1.4$\% for the field; also see Figure \ref{fig:pairprops}), as those are the key factors in determining the exerted tidal strength. 

\begin{figure}[t]
    \centering
    \includegraphics[width=1\linewidth]{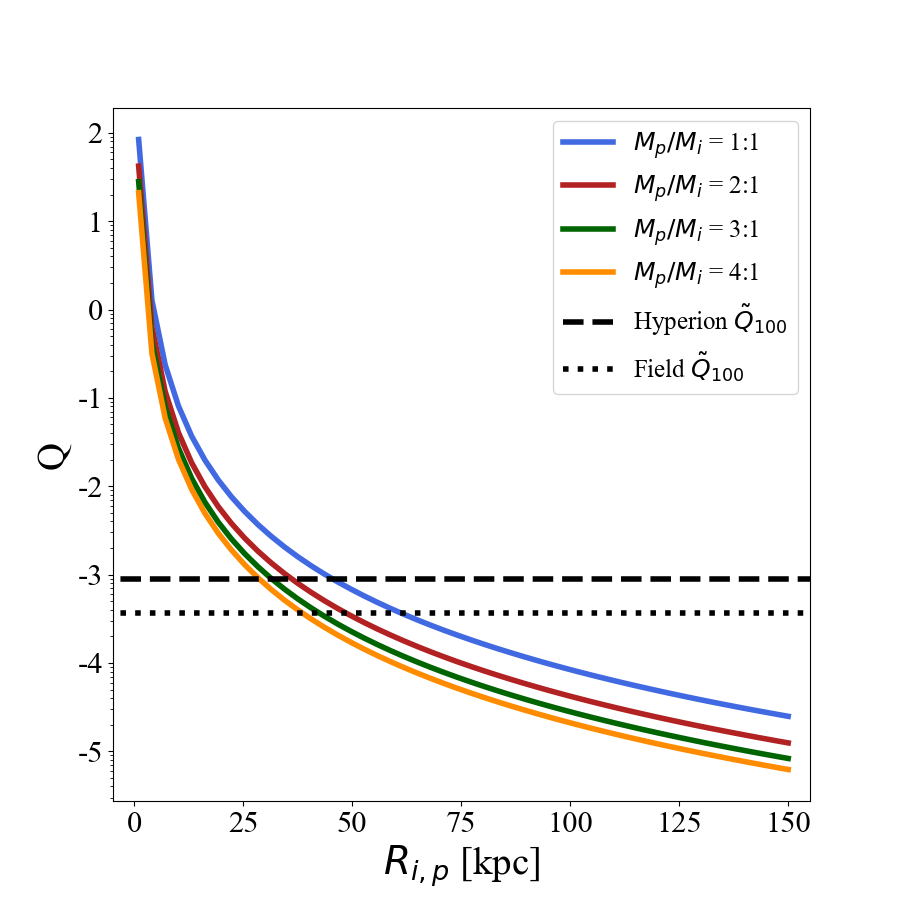}
    \caption{Dependence of the $Q$ (tidal strength) on $R_{\rm i,p}$ (the projected physical distance between companion and primary galaxy) for range of companion mass ratios considered in this work. The $Q$ value contours are estimated using the typical size of a star-forming galaxy at $z \sim 2.5$ from \citet{ribeiro_size_2016}. Also included for reference are the $\Tilde{Q}_{100}$ value for Hyperion companion systems and the $\Tilde{Q}_{100}$ value for field companion systems. The tidal forces induced by the companion galaxy do not dominate over the binding forces of the primary galaxy until $R_{\rm i,p} \lesssim 10$ kpc, but we do find increased tidal strengths in Hyperion companion systems due to increased percentages of multi-companion systems and systems at smaller projected separations.}
    \label{fig:q-param}
\end{figure}

In Figure \ref{fig:q-param}, we plot the dependence of the tidal strength (Q) on the projected physical distance between companion and primary galaxy ($R_{\rm i,p}$) for a variety of mass ratios considered in this work along with the $\Tilde{Q}_{100}$ values for Hyperion and the field. Though we consider a wide range of projected physical distances when selecting companions, the exerted tidal forces only outweigh the binding forces when $R_{\rm i,p} \lesssim 10$ kpc. 
The $\Tilde{Q}_{100}$ exerted by companion galaxies in Hyperion over our 100 MC iterations is indeed stronger than the $\Tilde{Q}_{100}$ exerted by companion galaxies in our field sample over 100 MC iterations, but neither value indicates a population of companion systems dominated by those which are necessarily inducing extreme tidal forces (i.e., $Q \geq 0$). Rather both the companion systems within Hyperion and within the field are primarily well below this threshold and typically have larger physical separations ($\bar{R_{\rm i,p}} > 10$ kpc). Therefore, we calculated the median fraction of companion systems where $Q \geq -2$ (i.e., the minimum threshold to induce perturbations) over our 100 MC iterations for Hyperion and for the field ($\Tilde{f}_{Q\geq-2}$) finding a $\Tilde{f}_{Q\geq-2} =  16.7\pm6.0$\% for Hyperion and a $\Tilde{f}_{Q\geq-2} = 9.7\pm1.2$\% for the field. This suggests that the companions within Hyperion are on average inducing more instabilities than their counterpart companion galaxies in the field. 

\begin{table*}[]
    \caption{\label{tab:timescales} Companion statistics for our identified systems }
    \centering
    \begin{tabular}{c|c|c|c|c}
        Sample & Value [Units] & $r_{\rm proj}\leq30$  [kpc]  & $30 < r_{\rm proj}\leq50$ [kpc] & $50 < r_{\rm proj}\leq 150$ [kpc]  \\
        \hline
        \hline
        & $\Delta\Tilde{v}$ [km/s]& 421 & 258 & 288 \\
        \cline{2-5}
        & $\Tilde{M_*}$ [$\log(M_*/M_\odot)$] & $9.83\pm0.48$ & $10.14\pm0.14$ & $9.84\pm0.13$ \\
        \cline{2-5}
        & $\Tilde{z}$ & $2.52\pm0.05$ & $2.47\pm0.01$& $2.47\pm0.01$\\
        \cline{2-5}
        Hyperion & $\Tilde{\sigma}_{\rm \delta}$ & $3.59\pm1.04$ & $4.93\pm0.59$ & $4.45\pm0.74$  \\
        \cline{2-5}
        & $\Tilde{N}_{\rm bin}$ & $2\pm1.3$ & $5\pm1.1$& $17\pm3.7$ \\
        \cline{2-5}
        & $\Tilde{f}_{\rm bin}$ & $7.7\pm5.2$\% & $20.3\pm5.1$\%& $72.0\pm6.5$\% \\
        \cline{2-5}
        & $\langle T_{\rm merge} \rangle$ [Gyr] & $3.54_{-1.09}^{+2.03}$ & $4.06_{-0.44}^{+0.53}$ & $9.32_{-1.20}^{+1.49}$ \\
        \hline
        & $\Delta\Tilde{v}$ [km/s] & 470 & 490 & 494 \\
        \cline{2-5}
        & $\Tilde{M_*}$ [$\log(M_*/M_\odot)$]& $9.60\pm0.06$ & $9.61\pm0.05$& $9.54\pm0.02$ \\
        \cline{2-5}
        & $\Tilde{z}$ & $2.31\pm0.11$ & $2.26\pm0.04$ & $2.27\pm0.02$ \\
        \cline{2-5}
        Field & $\Tilde{\sigma}_{\rm \delta}$ & $0.56\pm0.20$ & $0.55\pm0.17$ & $0.59\pm0.05$  \\
        \cline{2-5}
        & $\Tilde{N}_{\rm bin}$ & $44\pm6.7$ & $58\pm7.2$& $489\pm23$ \\
        \cline{2-5}
        & $\Tilde{f}_{\rm bin}$ & $7.5\pm1.0$\% & $9.7\pm1.2$\%& $82.4\pm1.4$\% \\ 
        \cline{2-5}
        & $\langle T_{\rm merge} \rangle$ [Gyr] & $4.10_{-0.28}^{+0.32}$ & $6.48_{-0.40}^{+0.44}$ & $13.10_{-0.31}^{+0.33}$ \\
        \hline
        \hline
    \end{tabular}
    \tablefoot{Provided are the median LoS velocity differences ($\Delta\Tilde{v}$), median stellar masses ($\Tilde{M}_*$), and median redshifts ($\Tilde{z}$) for each bin of projected separation for companion systems in Hyperion and the field. Also included are the $\langle T_{\rm merge} \rangle$ values and median $\sigma_\delta$ density ($\Tilde{\sigma}_{\delta}$) for each bin, as well as the median overall counts ($\Tilde{N}_{\rm bin}$) and the median fraction of companion systems ($\Tilde{f}_{\rm bin}$) at that projected separation. }
\end{table*}

\subsection{Potential merger timescales}
\label{sec:time merge}

As we select galaxies with close kinematic companions in our sample to estimate future merger activity --- rather than selecting galaxies currently undergoing mergers through morphology --- it is important that we attempt to quantify the potential timescales associated with any future mergers. We primarily focus on major mergers within this work, so this timescale gives an indication of the timescales by which these galaxies will grow significantly in stellar mass (either through induced star formation or upon coalescence). However, since we consider a wide range of projected physical distances and LoS velocity differences for our companion criteria, calculating a single merger timescale for all close kinematic companions found in our sample is too simplistic and does not accurately capture the full distribution of potential merger timescales. 

To calculate the associated merger timescales of the close kinematic companions identified within the 100 MC iterations of our sample, we used the formalism described in \citet{kitzbichler_calibration_2008}

\begin{equation}
    \langle T_{\rm merge} \rangle^{-(1/2)} = T_0^{-(1/2)}+ f_1 z + f_2 (\log M_* -10).
\end{equation}

\begin{figure}
    \centering
    \includegraphics[width=1.0\linewidth]{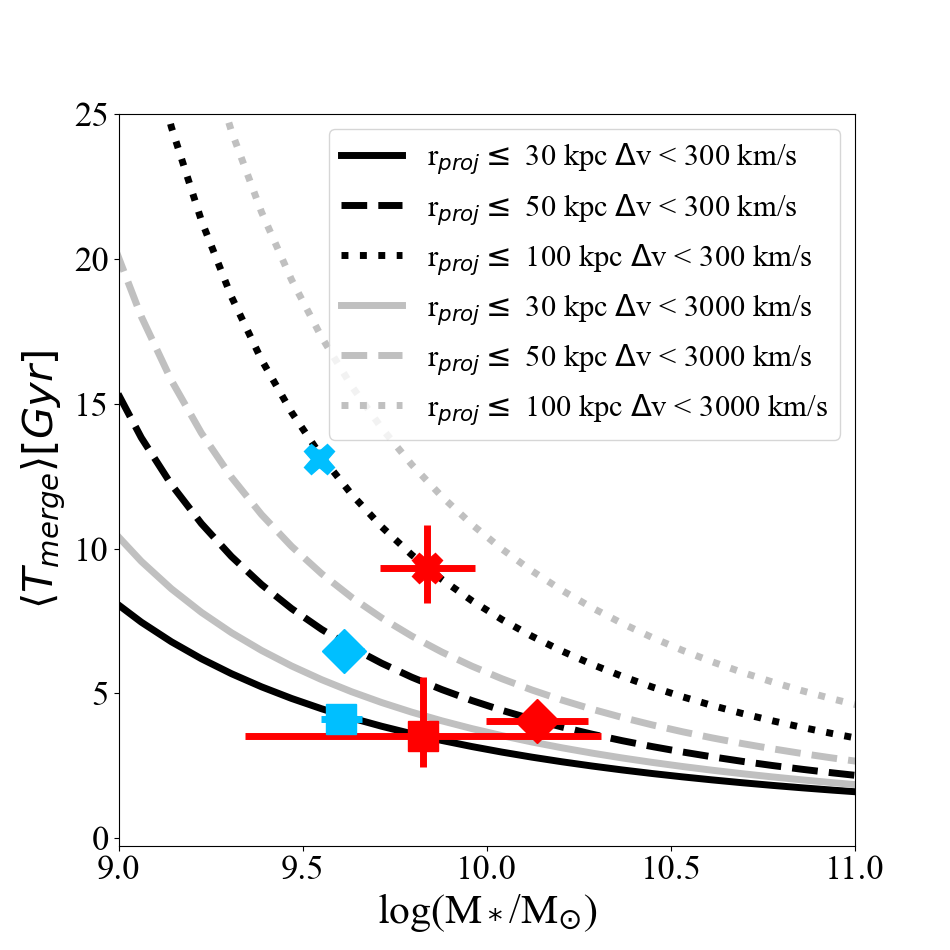}
    \caption{$\langle T_{\rm merge} \rangle$ as a function of stellar mass. Contours represent the six different companion selection criteria and corresponding fitting coefficients from \citet{kitzbichler_calibration_2008}. These contours are all plotted at the median redshift of all pairs over all MC iterations ($z \sim 2.48)$. We also include the calculated $\langle T_{\rm merge} \rangle$ values for Hyperion (red) and the field (blue) at each of projected separation criteria employed (i.e., $r_{\rm proj}\leq30$ kpc (squares), $30 <r_{\rm proj}\leq50$ kpc (diamonds),  and $50 < r_{\rm proj}\leq150$ kpc (crosses)). The $\langle T_{\rm merge} \rangle$ values calculated for Hyperion are shorter in each bin due to the larger stellar masses of galaxies in Hyperion companion systems.}
    \label{fig:timescale}
\end{figure}

For this equation, \citet{kitzbichler_calibration_2008} (hereafter KW$+$08) performed a two-dimensional linear regression fit on simulated galaxy pair catalogs from the Millennium Simulation to obtain $\langle T_{\rm merge} \rangle$ as a function of stellar mass threshold (for both galaxies) and redshift for major merger systems and provide six discrete sets of coefficients that vary based on the applied companion identification criteria (i.e., projected separation, $r_{\rm proj}$, and LoS velocity difference, $\Delta v_{\rm LoS}$). The selected $r_{\rm proj}$ and  $\Delta v_{\rm LoS}$ values that create these fitting coefficients are intended to capture a representative range of projected physical distances used in companion studies (i.e., maximal values of 30, 50 or 100 kpc h$^{-1}$) and for application to both spectroscopic redshift samples ($\Delta v_{\rm LoS} < 300$ km s$^{-1}$) and photometric redshift samples ($\Delta v_{\rm LoS} < 3000$ km s$^{-1}$). The employed companion criteria, and thus the fitting coefficients, have a notable impact on the measured values of $\langle T_{\rm merge} \rangle$ (see Figure \ref{fig:timescale}).

Though we used different criteria in selecting our sample of companion galaxies across our 100 MC iterations, we attempt to mimic the selection criteria of KW$+$08 when calculating $\langle T_{\rm merge} \rangle$ values. To do this, we split our identified companion systems into three bins of projected physical distance ($r_{\rm proj} < 30$ [kpc], $30 < r_{\rm proj} < 50$ [kpc], and $50 < r_{\rm proj} < 150$ [kpc]) for the companion systems in Hyperion and companion systems in the field for each of our 100 MC iterations (i.e., six samples per MC iteration). We then compute the median LoS velocity difference within each of those six bins over all 100 MC iterations, and find that the typical $\Delta v_{\rm LoS}$ for each sample is $\lesssim 500$ km s$^{-1}$ (see Table \ref{tab:timescales} and Figure \ref{fig:pairprops}). Based on these values, we decide to utilize the coefficients based on $\Delta v_{\rm LoS} < 300$ km s$^{-1}$ for calculating $\langle T_{\rm merge} \rangle$. We believe this choice is justified as, though we employ a sample of both spectroscopic and photometric redshifts, our MC methodology mitigates the effect of large photometric redshift errors and enforces a stringent $\Delta v_{\rm LoS}$ criteria that is in line with studies that only employ spectroscopic redshifts across similar redshift ranges (e.g., \citealt{shah_investigating_2020,shah_investigating_2022}). We also calculate the median galaxy stellar mass ($\Tilde{M_*}$) and the median redshift ($\Tilde{z}$) for all companion systems in our six bins across the 100 MC iterations. 

Using the appropriate $r_{\rm proj}$ and $\Delta v_{\rm LoS}$ coefficients per the mimicked companion selection criteria (i.e., $r_{\rm proj}\leq30$ kpc \& $\Delta v_{\rm LoS} < 300$ km s$^{-1}$, $r_{\rm proj}\leq50$ kpc \& $\Delta v_{\rm LoS} < 300$ km s$^{-1}$, $r_{\rm proj}\leq150$ kpc \& $\Delta v_{\rm LoS} < 300$ km s$^{-1}$) along with our measured $\Tilde{M_*}$ and $\Tilde{z}$ values, we calculated the $\langle T_{\rm merge} \rangle$ for each of our six bins (values reported in Table \ref{tab:timescales}, also see Figure \ref{fig:timescale}). As expected, we find that $\langle T_{\rm merge} \rangle$ decreases with decreasing $r_{\rm proj}$ for both companion systems in Hyperion and companion systems in the field with the companion systems of projected separations of less than 30 kpc having the shortest merger timescales. However, for every bin of projected separation, the average merger timescale for companion systems in Hyperion is shorter than the average merger timescale for the corresponding companion systems in the field. Though a slight offset exists in $\Tilde{z}$ between the Hyperion and field companion systems, we control for variations in the projected separation through our binning and thus the measured difference in $\langle T_{\rm merge} \rangle$ between our Hyperion and field companion systems is primarily due to the varying $\Tilde{M_*}$ between companion systems in Hyperion and the field. The increased stellar masses for systems in Hyperion results in faster merger timescales in comparison to the field and comports with the increased stellar masses seen in Hyperion \citep{sikorski25}. 

We also calculated the median fraction of companion systems within each bin ($\Tilde{f}_{\rm bin} \equiv$ number of systems at that projected separation divided by the number of total companion systems). This calculation gives us insight into what fraction of companion systems in Hyperion and in the field should be associated with a given merger timescale. Through this calculation, we find similar fractions of companion systems within our smallest projected separation bin ($\sim8$\% within $r_{\rm proj} \leq 30$ kpc), but diverging fractions of companion systems in our other projected separation bins. Most notably, in the intermediate projected separation bin ($30 < r_{\rm proj} < 50$ kpc), Hyperion companion systems that meet this criteria account for $\sim20$\% of companion systems across all 100 MC iterations whereas field companion systems in this bin only account for $\sim10$\% of companion systems. Due to the higher density of Hyperion, this is not necessarily surprising, but, since we do not observe a corresponding increase in companions systems within the smallest projected separation bin, this increase in companion systems at intermediate separations is likely physical and supports the increased merger and interaction activity measured by the \fckc \space values found in this work. In fact, the lack of a corresponding increase in companions systems within the smallest projected separation bin in Hyperion could be the result of the closest companion systems having already merged, as this process could accelerated compared to the similar field systems due to increased stellar mass. This is further corroborated by the median $\sigma_{\delta}$ ($\Tilde{\sigma}_{\delta}$) associated with our projected separation binned companion systems (see Table \ref{tab:timescales}; $\Tilde{\sigma}_{\delta} \sim 3.37$ for all Hyperion galaxies both with and without companions over the 100 MC iterations and $\Tilde{\sigma}_{\delta} \sim 0.34$ for all field galaxies). Whereas we find a relatively flat trend in $\Tilde{\sigma}_{\delta}$ for companion systems in the field, there is a notable jump in $\Tilde{\sigma}_{\delta}$ for companion systems with intermediate projected separations in Hyperion that suggests the most dense regions are sites of increased merger and interaction activity. 

In Figure \ref{fig:timescale}, we plot the evolution of $\langle T_{\rm merge} \rangle$ as a function of stellar mass. To further demonstrate the important contribution of the selected fitting coefficients, we include contours at all six permutations of companion separation criteria from KW$+$08. As is evident from these contours, the $r_{\rm proj}$ and $\Delta v_{\rm LoS}$ criteria that provide these coefficients have a significant impact on the resulting measurement of $\langle T_{\rm merge} \rangle$ --- especially at lower companion stellar masses. We also plot the location of our calculated $\langle T_{\rm merge} \rangle$ values for Hyperion (red) and for the field (blue). As discussed above, the increased stellar masses seen in the galaxies that comprise the Hyperion companion systems is primarily responsible for their faster merger timescales in comparison to the field. This increase in timescale is most notable for the largest projected separation bin, but the intermediate projected separation bin sees the largest difference in median stellar mass.

The range of $\langle T_{\rm merge} \rangle$ values calculated in this work follows the wide range of allowed projected separations (i.e., $r_{\rm proj} < 150$ kpc) used to identify companion systems within our MC methodology. Though the merger timescales associated with companion systems at the widest projected separations are $\sim10$ Gyr, the age of the Universe at the redshift range of this work is $\sim2.7$ Gyr (at $z\sim$2.5 via \citealt{wright_cosmology_2006}) giving many of our companion systems time to potentially merge by $z = 0$. Our companion systems identified at closer projected separations have much shorter $\langle T_{\rm merge} \rangle$ values and many will potentially merge by $z \sim 1$.

\section{Discussion}
\label{sec:diss}

\subsection{Structure development}\label{sec:struc dev}

Local clusters are the sites of incredible stellar mass buildup and some of their member galaxies are the most massive galaxies in the present day --- in particular brightest cluster galaxies (BCGs) that dominate cluster systems. Due to their close proximity in the local Universe, we have been able to study such systems to understand their underlying physical mechanisms that result from the high density of galaxies and the presence of the ICM. We find large velocity dispersions ($\gtrsim 1000$ km s$^{-1}$) in the densest regions of local virialized clusters, and such velocity dispersions are large enough that they tend to dissuade member galaxies from merging \citep{lin_where_2010}. Rather, many galaxies dynamically interact (i.e., galaxy harassment; \citealt{moore_galaxy_1996,moore_morphological_1998}) with one another before they continue on along their course through the dense regions of the cluster. We do see some merger activity toward the cluster outskirts \citep{moran_reflections_2007}, but those limited mergers and the happenstance interactions in the denser regions only partially help us to understand the current state of these systems, and do not fully explain the large buildup of mass in clusters nor their overall development. 

It is well established in the literature that beyond the local Universe ($z \geq$ 0.1) many of the galaxies that come to lie within cluster systems experience preprocessing as they move along the DM filaments or move from group environments to cluster environments \citep{zabludoff_environment_1996,fujita_pre-processing_2004,mcgee_accretion_2009,de_lucia_environmental_2012,bahe_disruption_2019,kraljic_galaxies_2019,salerno_filaments_2019,sarron_pre-processing_2019,reeves_gogreen_2021,baxter_gogreen_2022,werner_satellite_2022,baxter_when_2023} and that these regions oftentimes see more activity than the clusters themselves. Here, there is evidence of more galaxies with spiral morphologies, bluer colors, higher SF, as well as increased rates of mergers in comparison to the clusters that these galaxies are infalling toward \citep{lemaux_assembly_2012,strazzullo_galaxy_2013}. Therefore, this preprocessing at intermediate densities is often thought to a be dominant component in the development of the galaxies that will come to reside in the denser regions of clusters. 

In this work, we offer another significant contribution to the buildup of stellar mass in structure: the direct buildup of mass via enhanced rates of interactions and mergers within the very structures that will become massive clusters (i.e., protoclusters). This contribution appears reasonable as protoclusters are still diffuse in comparison to their cluster descendants with more moderate velocity dispersions (generally $\sim$ a few hundred kilometers per second and $\sim700$ km s$^{-1}$ for Hyperion \citealt{cucciati_progeny_2018}), though are much denser than their coeval surroundings, and thus their member galaxies exist at higher galaxy densities while no longer being subject to the large velocity dispersions that dissuade merger activity in cluster systems. Therefore, there is much less restriction on what galaxies will merge within protoclusters once they reach close enough distances to begin to gravitationally influence one another. There is also emerging evidence within protoclusters for stellar mass buildup as a result of this increased merger and interaction activity, as studied protoclusters are exhibiting increased stellar masses and more developed SMFs \citep{shimakawa_mahalo_2018,shimakawa_mahalo_2018-1,forrest_environmental_2024} --- including the stellar mass maturation seen in Hyperion itself \citep{sikorski25}. The increased stellar masses of galaxies in the overdense regions of protoclusters is consistent with a scenario where these protocluster galaxies are undergoing or have already undergone increased stellar mass buildup relative to field galaxies \citep{forrest_environmental_2024}. 

The observed trend of elevated SMFs in overdense regions relative to the field continues and strengthens at later epochs ($z \sim 1$; \citealt{tomczak_glimpsing_2017}) with the shape of the SMF depending strongly on local environment and increases in the relative numbers in high-mass to low-mass galaxies in denser environments. This observed increase of high-mass galaxies in dense environment implies that the local environment is destroying lower-mass galaxies via merging or stellar stripping and/or high-mass galaxies are experiencing more accelerated growth. In order to explain these potential avenues, many studies have found that star formation alone cannot be responsible and that galaxy-galaxy mergers are essential in the development of these populations and to recreate the SMFs seen in $z \sim 1$ clusters \citep{davidzon_vimos_2016,steinhardt_reconciling_2017,tomczak_glimpsing_2017}. In particular, \citet{tomczak_glimpsing_2017} found that to reproduce the SMFs of the highest-density regions considered in their study $\gtrsim80$\% of galaxies likely have undergone merger events and that a large majority of these mergers would have to occur in intermediate-density environments. 

The enhanced merger and interaction activity at higher densities suggested by the \fckc \space rates measured within this work along with other measurements of increased merger and interaction activity in denser environments at $z>2$ (e.g., \citealt{hine_enhanced_2016,monson_nature_2021,liu_what_2023,shibuya_galaxy_2025}) indicate a scenario where mergers and interactions are potentially a important component for the creation of observed $z \sim 1$ cluster populations and eventually $z \sim 0$ cluster populations. The timescales associated with potential mergers via the identified companion systems in this work ($\sim3 - 10$ Gyr) roughly correlate with the timescales for a high-z structures such as Hyperion to mature into their low-z descendants ($\sim3$ Gyr and $\sim10$ Gyr for $z = 1$ and $z = 0$ structure, respectively). Massive galaxies in structures, like a BCG, are predicted to have the growth of their accreted mass fractions be dominated by mergers \citep{de_lucia_hierarchical_2007,ascaso_violent_2014,lidman_evidence_2012,oleary_emerge_2021,shen_implications_2021} which again requires long timescales that are in agreement with the merging timescales found for the companion systems in Hyperion in this work. Additionally, increases in the rates of mergers and interactions are also seen in some $z \lesssim 2$ dense environments \citep{lin_where_2010,kampczyk_environmental_2013,lotz_caught_2013,coogan_merger-driven_2018,watson_galaxy_2019} which potentially exhibit moderate velocity dispersions that are in line with those observed for protocluster systems. Together, the high densities, lower velocity dispersions, elevated SMFs, and enhanced merger and interaction rates seen in protocluster systems, in combination with observations of lower-z cluster systems, all suggest an evolution where mergers and interactions are a crucial element within regions of high galaxy density at $z \gtrsim 2$. 

\subsection{Gas reservoirs in Hyperion}\label{sec:gas}

Though mergers and interactions may be a potential key mechanism for the transformation from protocluster populations to cluster populations, there are other modes which may be contributing to the observed mass buildup seen in high-z structures including the preferential enhancement of SFR in high-mass galaxies in overdense regions or an underlying bias for galaxies in overdense environments in comparison to those in the field \citep{ahad_environment-dependent_2024}. However, the increased number of massive and/or quiescent galaxies found within high-z clusters and protoclusters (e.g., \citealt{scoville_large_2007,muzzin_gemini_2012,nantais_stellar_2016,van_der_burg_stellar_2018,van_der_burg_gogreen_2020,mcconachie_spectroscopic_2022}) suggests that stellar mass buildup happened rapidly, with mergers or large reservoirs of molecular gas being the main candidates. 

While this work is primarily concerned with the merger and interaction activity of Hyperion, recent work by \citet{gururajan_gas_2025} has investigated the potential avenue of large molecular gas reservoirs within Hyperion for the buildup of stellar mass seen within this structure. \citet{gururajan_gas_2025} finds decreases in molecular gas fractions with increasing density along with increases in SFR and star-formation efficiency at higher densities within Hyperion. The enhanced star formation seen at higher densities in that work is in line with the observed reversal in the SFR-density at relevant epochs \citep{lemaux_vimos_2022}, but the lack of a corresponding increases in molecular gas content at higher densities in Hyperion implies that the molecular gas alone cannot be the trigger for the increased star formation and thus the increased stellar masses within Hyperion \citep{sikorski25}. Instead, other mechanisms outside of additional inflowing gas could be contributing --- such as the mergers and interactions signified by the \fckc \space values measured within this work. The combination of the internal star formation from previously existing molecular gas and external merger and interaction events are likely together driving the evolution of Hyperion and its member galaxies.

\section{Conclusions and future work}
\label{sec:conc}

In this work, we seek to measure the rate of galaxy interactions and mergers within the Hyperion proto-supercluster (z $\sim$ 2.5) in comparison to mergers and interactions within the coeval field in order to assess the importance and effect of merger and interaction activity on the mass assembly within large-scale structure at higher redshifts ($z \gtrsim 2$). For this, we employed a combined spectroscopically and photometrically selected galaxy sample drawn from variety of sources including COSMOS2020 \citep{weaver_cosmos2020_2022}, the C3VO Survey \citep{lemaux_vimos_2022}, and VUDS \citep{le_fevre_vimos_2015} among others. Our sample contains 220,356 galaxies spanning a wide redshift range ($ 0 < z < 7$) and the sky region of Hyperion within the COSMOS field ($149.6^{\circ} \leq \alpha \leq 150.52^{\circ}$ and $1.74^{\circ} \leq \delta \leq 2.73^{\circ}$).

From our sample, we calculated the fraction of galaxies with close kinematic companions (\fckc) via the methodology developed in this work. In short, this methodology takes 100 MC realizations of our entire galaxy sample, and, for galaxies that fall in the relevant redshift range ($2 < z < 3$) in a given iteration, we quantify the environmental density of each galaxy based on VMC maps generated for the COSMOS field and identify potential companions for each galaxy based on projected spatial separation and LoS velocity difference criteria: $d_{\rm proj} < 150$ kpc and $\Delta v_{\rm rec} < 1000$ km s$^{-1}$, respectively. From this combination of environmental information and identified galaxy companions, we are thus able to calculate a \fckc \space for Hyperion and for the field.

To validate the companion methodology constructed in this study, we applied the exact same approach, without the environmental determination, to a simulated lightcone from the GAEA SAM \citep{hirschmann_galaxy_2016,de_lucia_tracing_2024} at varying levels of spectroscopic completeness (SzF). We then compared the identified companion galaxies at each level of SzF to the true underlying companion galaxies based on the actual galaxy locations within the simulation. We find that a correction factor is needed to adjust our recovered \fckc \space for all levels of SzF due varying levels of purity and completeness in identified companion galaxies, though this correction factor weakens with increased levels of spectroscopic completeness. We primarily attribute the need for this correction factor to the relatively large redshift uncertainties for photometric sources that wash out potential true galaxy companions within our MC process.  

Based on our verified MC companion methodology, our main results are as follows:

\begin{itemize}
    \item We measure a $\gtrsim2.5\times$ increase in the fraction of galaxies with close kinematic companions in Hyperion relative to the field, finding a corrected \fckc \space $ =$ \frachyp \space for Hyperion and an \fckc \space $=$ \fracfield \space for the field (a difference of $\gtrsim3\sigma$). This suggests galaxies in Hyperion experience increased rates of merger and interaction activity in comparison to field galaxies at similar epochs. The rate of merger and interaction activity for field galaxies (via our field \fckc \space measurement) is well aligned with other measured and predicted fractions from the literature at comparable redshifts \citep{qu_chronicle_2017,romano_alpine-alma_2021}, and the heightened activity we find in the dense structure of Hyperion is similar to earlier works that have investigated the relation between environment and merger and interaction activity at similar epochs \citep{hine_enhanced_2016,monson_nature_2021,liu_what_2023,shibuya_galaxy_2025}.
    \item In measuring the tidal strengths exerted by companion galaxies identified in Hyperion and the field, we find a similar $\sim2.5\times$ increase in the median tidal strength exerted by Hyperion companion galaxies over our 100 MC iterations ($\Tilde{Q}_{100} \sim -3.04\pm0.18$ for Hyperion companion systems and a $\Tilde{Q}_{100} \sim -3.43\pm0.03$ for field companion systems). This increase in $\Tilde{Q}_{100}$ is marginal, but is likely due to the increased number of galaxies with multiple companions in Hyperion as well as the closer proximity of many of companions found within Hyperion. We also calculated the fraction of companion galaxies actively inducing perturbations (i.e., $Q \geq -2$) and find an increase for Hyperion companion systems as well ($\Tilde{f}_{Q\geq-2}=16.7\pm6.0$\% for Hyperion versus a $\Tilde{f}_{Q\geq-2}=9.7\pm1.2$\% for the field).
    \item After binning the companion systems in Hyperion and the field by projected separation, we estimate the potential merger timescales of our identified companions. As expected, the average merger timescale ($\langle T_{\rm merge} \rangle$) decreases at smaller projected separations for companion systems in both Hyperion and the field. However, in each bin of projected separation, the $\langle T_{\rm merge} \rangle$ of Hyperion companion systems is shorter --- primarily due to the increased masses of Hyperion galaxies in comparison to the field \citep{sikorski25}.
    \item In addition to the differences in $\langle T_{\rm merge} \rangle$, we find that companion systems in Hyperion have, on average, lower LoS velocity differences and increased fractions at smaller projected separations. These effects are most pronounced in our intermediate projected separation bin ($30 < r_{\rm proj} < 50$ kpc) where Hyperion companion systems exist at the highest overdensities. 
    \item The $\langle T_{\rm merge} \rangle$ values measured for Hyperion are short enough ($\sim3-10$ Gyr) to suggest that most of our identified Hyperion companion systems, if indeed merging, will merge by $z \sim 0$, with companion systems at $r_{\rm proj} < 50$ kpc potentially merging by $z\sim1$. This timescale is short enough to contribute to the buildup of stellar mass already seen in $z\sim1$ galaxy clusters \citep{tomczak_glimpsing_2017} and is in stark contrast the $\langle T_{\rm merge} \rangle$ values estimated for the field where only companion systems at the smallest projected separations (i.e., $r_{\rm proj} < 30$ kpc) are likely to have time to merge by $z \sim1$.
\end{itemize}

Overall, the confluence of increased merger and interaction activity in Hyperion (based on the measured enhancement of \fckc), the increased tidal strength of these increased interactions in Hyperion, and the shorter timescales associated with future mergers for companion systems in Hyperion all suggest that interactions and mergers within Hyperion may be a dominant and influential element in the assembly and buildup of large-scale structure and their constituent galaxies at higher redshifts (i.e., the regime of the protocluster at $z \gtrsim 2$). The increase in merger and interaction activity in high-z structure measured within this work and others \citep{hine_enhanced_2016,monson_nature_2021,liu_what_2023,shibuya_galaxy_2025} is an important component we need to consider as we continue to seek to fully understand the development of protoclusters into clusters; particularly in providing important contributions to the stellar mass buildup observed in high-z cluster and protocluster systems. Therefore, we plan to build upon this work and apply our developed methodology to a large sample protoclusters --- namely the 500$+$ protostructures identified by \citet{hung_discovering_2025} --- in the near future. This effort will help to fully disentangle the relation between merger and interaction activity and environment across a wide range of redshifts, dynamical states, and system masses for large-scale structure as they are still forming and maturing. This will provide a definitive insight into whether the currently observed enhancement of mergers and and interactions seen within protostructures is ubiquitous or if the level of enhancement varies with redshift or based on the properties of the overall system. With this larger sample of protoclusters, we will also be able to flesh out the merger and interaction activity at various different density levels (rather than in at a structure versus field level) to determine if similar trends to lower-redshift structure are still prevalent (i.e., ``preprocessing'' at lower density levels).

\begin{acknowledgements}
The authors thank the anonymous referee for the feedback and contributions which have improved the manuscript.
BF acknowledges support from JWST-GO-02913.001-A. DCB is supported by an NSF Astronomy and Astrophysics Postdoctoral Fellowship under award AST-2303800. DCB is also supported by the UC Chancellor's Postdoctoral Fellowship.
RA acknowledges financial support from project PID2023-147386NB-I00 and the State Agency for Research of the Spanish MCIU through 'Center of Excellence Severo Ochoa' award to the IAA-CSIC (SEV-2017-0709) and CEX2021-001131-S funded by MCIN/AEI/10.13039/501100011033.
This work is also based on observations collected at the European Southern Observatory under ESO programmes 175.A-0839, 179.A-2005, and 185.A-0791. 
This work supported by the National Science Foundation under Grant No. 1908422. 
This research is based on observations made with the NASA/ESA Hubble Space Telescope obtained from the Space Telescope Science Institute, which is operated by the Association of Universities for Research in Astronomy, Inc., under NASA contract NAS 5–26555. These observations are associated with program GO-16684.
Supported by the international Gemini Observatory, a program of NSF NOIRLab, which is managed by the Association of Universities for Research in Astronomy (AURA) under a cooperative agreement with the U.S. National Science Foundation, on behalf of the Gemini partnership of Argentina, Brazil, Canada, Chile, the Republic of Korea, and the United States of America.
This work is based in part on data products produced at Terapix available at the Canadian Astronomy Data Centre as part of the Canada-France-Hawaii Telescope Legacy Survey, a collaborative project of NRC and CNRS. This work is based, in part, on observations made with the Spitzer Space Telescope, which is operated by the Jet Propulsion Laboratory, California Institute of Technology under a contract with NASA.
Based on observations obtained with MegaPrime/MegaCam, a joint project of CFHT and CEA/IRFU, at the Canada-France-Hawaii Telescope (CFHT) which is operated by the National Research Council (NRC) of Canada, the Institut National des Science de l'Univers of the Centre National de la Recherche Scientifique (CNRS) of France, and the University of Hawaii. This work is based in part on data products produced at Terapix available at the Canadian Astronomy Data Centre as part of the Canada-France-Hawaii Telescope Legacy Survey, a collaborative project of NRC and CNRS.
This research is based in part on data collected at the Subaru Telescope, which is operated by the National Astronomical Observatory of Japan. We are honored and grateful for the opportunity of observing the Universe from Maunakea, which has the cultural, historical, and natural significance in Hawaii.
Some of the data presented herein were obtained at Keck Observatory, which is a private 501(c)3 non-profit organization operated as a scientific partnership among the California Institute of Technology, the University of California, and the National Aeronautics and Space Administration. The Observatory was made possible by the generous financial support of the W. M. Keck Foundation.
The authors wish to recognize and acknowledge the very significant cultural role and reverence that the summit of Maunakea has always had within the indigenous Hawaiian community.  We are most fortunate to have the opportunity to conduct observations from this mountain. 
\end{acknowledgements}

\textit{Facilities:}  HST(WFC3), Keck(MOSFIRE), Keck(DEIMOS), VLT(VIMOS)

\textit{Software:} \texttt{Astropy} \citep{astropy_collaboration_astropy_2013,astropy_collaboration_astropy_2018,astropy_collaboration_astropy_2022}, \texttt{Matplotlib} \citep{hunter_matplotlib_2007}, \texttt{NumPy} \citep{oliphant_python_2007,harris_array_2020}, \texttt{pandas} \citep{team_pandas-devpandas_2024}, \texttt{SciPy} \citep{virtanen_scipy_2020}

\bibliographystyle{aa}

\bibliography{main}

@ARTICLE{sikorski25,
       author = {{Sikorski}, Derek and {Forrest}, Ben and {Lemaux}, Brian C. and {Shen}, Lu and {Giddings}, Finn and {Gal}, Roy and {Cucciati}, Olga and {Golden-Marx}, Emmet and {Hu}, Weida and {Hung}, Denise and {Lubin}, Lori and {Ronayne}, Kaila and {Shah}, Ekta and {Bardelli}, Sandro and {Baxter}, Devontae C. and {Gururajan}, Gayathri and {Tresse}, Laurence and {Zamorani}, Giovanni and {Diamond}, Joel and {Guaita}, Lucia and {Hathi}, Nimish and {Zucca}, Elena},
        title = "{The HST-Hyperion Survey: Environmental Imprints on the Stellar-Mass Function at z=2.5}",
      journal = {arXiv e-prints},
         year = 2025,
        month = sep,
          eid = {arXiv:2509.02714},
        pages = {[arXiv:2509.02714]},
          doi = {10.48550/arXiv.2509.02714},
archivePrefix = {arXiv},
       eprint = {2509.02714},
 primaryClass = {astro-ph.GA},
       adsurl = {https://ui.adsabs.harvard.edu/abs/2025arXiv250902714S}
}

@article{ahad_environment-dependent_2024,
  title = {An Environment-Dependent Halo Mass Function as a Driver for the Early Quenching of z {$\geq$} 1.5 Cluster Galaxies},
  author = {Ahad, Syeda Lammim and Muzzin, Adam and Bah{\'e}, Yannick M. and Hoekstra, Henk},
  year = 2024,
  month = mar,
  journal = {MNRAS},
  volume = {528},
  pages = {6329--6339},
  issn = {0035-8711},
  doi = {10.1093/mnras/stae341},
  urldate = {2025-01-02}
}

@article{aihara_first_2018,
  title = {First Data Release of the {{Hyper Suprime-Cam Subaru Strategic Program}}},
  author = {Aihara, Hiroaki and Armstrong, Robert and Bickerton, Steven and Bosch, James and Coupon, Jean and Furusawa, Hisanori and Hayashi, Yusuke and Ikeda, Hiroyuki and Kamata, Yukiko and Karoji, Hiroshi and Kawanomoto, Satoshi and Koike, Michitaro and Komiyama, Yutaka and Lang, Dustin and Lupton, Robert H. and Mineo, Sogo and Miyatake, Hironao and Miyazaki, Satoshi and Morokuma, Tomoki and Obuchi, Yoshiyuki and Oishi, Yukie and Okura, Yuki and Price, Paul A. and Takata, Tadafumi and Tanaka, Manobu M. and Tanaka, Masayuki and Tanaka, Yoko and Uchida, Tomohisa and Uraguchi, Fumihiro and Utsumi, Yousuke and Wang, Shiang-Yu and Yamada, Yoshihiko and Yamanoi, Hitomi and Yasuda, Naoki and Arimoto, Nobuo and Chiba, Masashi and Finet, Francois and Fujimori, Hiroki and Fujimoto, Seiji and Furusawa, Junko and Goto, Tomotsugu and Goulding, Andy and Gunn, James E. and Harikane, Yuichi and Hattori, Takashi and Hayashi, Masao and He{\l}miniak, Krzysztof G. and Higuchi, Ryo and Hikage, Chiaki and Ho, Paul T. P. and Hsieh, Bau-Ching and Huang, Kuiyun and Huang, Song and Imanishi, Masatoshi and Iwata, Ikuru and Jaelani, Anton T. and Jian, Hung-Yu and Kashikawa, Nobunari and Katayama, Nobuhiko and Kojima, Takashi and Konno, Akira and Koshida, Shintaro and Kusakabe, Haruka and Leauthaud, Alexie and Lee, Chien-Hsiu and Lin, Lihwai and Lin, Yen-Ting and Mandelbaum, Rachel and Matsuoka, Yoshiki and Medezinski, Elinor and Miyama, Shoken and Momose, Rieko and More, Anupreeta and More, Surhud and Mukae, Shiro and Murata, Ryoma and Murayama, Hitoshi and Nagao, Tohru and Nakata, Fumiaki and Niida, Mana and Niikura, Hiroko and Nishizawa, Atsushi J. and Oguri, Masamune and Okabe, Nobuhiro and Ono, Yoshiaki and Onodera, Masato and Onoue, Masafusa and Ouchi, Masami and Pyo, Tae-Soo and Shibuya, Takatoshi and Shimasaku, Kazuhiro and Simet, Melanie and Speagle, Joshua and Spergel, David N. and Strauss, Michael A. and Sugahara, Yuma and Sugiyama, Naoshi and Suto, Yasushi and Suzuki, Nao and Tait, Philip J. and Takada, Masahiro and Terai, Tsuyoshi and Toba, Yoshiki and Turner, Edwin L. and Uchiyama, Hisakazu and Umetsu, Keiichi and Urata, Yuji and Usuda, Tomonori and Yeh, Sherry and Yuma, Suraphong},
  year = 2018,
  month = jan,
  journal = {PASJ},
  volume = {70},
  pages = {S8},
  issn = {0004-6264},
  doi = {10.1093/pasj/psx081},
  urldate = {2024-12-31}
}

@article{aihara_hyper_2018,
  title = {The {{Hyper Suprime-Cam SSP Survey}}: {{Overview}} and Survey Design},
  shorttitle = {The {{Hyper Suprime-Cam SSP Survey}}},
  author = {Aihara, Hiroaki and Arimoto, Nobuo and Armstrong, Robert and Arnouts, St{\'e}phane and Bahcall, Neta A. and Bickerton, Steven and Bosch, James and Bundy, Kevin and Capak, Peter L. and Chan, James H. H. and Chiba, Masashi and Coupon, Jean and Egami, Eiichi and Enoki, Motohiro and Finet, Francois and Fujimori, Hiroki and Fujimoto, Seiji and Furusawa, Hisanori and Furusawa, Junko and Goto, Tomotsugu and Goulding, Andy and Greco, Johnny P. and Greene, Jenny E. and Gunn, James E. and Hamana, Takashi and Harikane, Yuichi and Hashimoto, Yasuhiro and Hattori, Takashi and Hayashi, Masao and Hayashi, Yusuke and He{\l}miniak, Krzysztof G. and Higuchi, Ryo and Hikage, Chiaki and Ho, Paul T. P. and Hsieh, Bau-Ching and Huang, Kuiyun and Huang, Song and Ikeda, Hiroyuki and Imanishi, Masatoshi and Inoue, Akio K. and Iwasawa, Kazushi and Iwata, Ikuru and Jaelani, Anton T. and Jian, Hung-Yu and Kamata, Yukiko and Karoji, Hiroshi and Kashikawa, Nobunari and Katayama, Nobuhiko and Kawanomoto, Satoshi and Kayo, Issha and Koda, Jin and Koike, Michitaro and Kojima, Takashi and Komiyama, Yutaka and Konno, Akira and Koshida, Shintaro and Koyama, Yusei and Kusakabe, Haruka and Leauthaud, Alexie and Lee, Chien-Hsiu and Lin, Lihwai and Lin, Yen-Ting and Lupton, Robert H. and Mandelbaum, Rachel and Matsuoka, Yoshiki and Medezinski, Elinor and Mineo, Sogo and Miyama, Shoken and Miyatake, Hironao and Miyazaki, Satoshi and Momose, Rieko and More, Anupreeta and More, Surhud and Moritani, Yuki and Moriya, Takashi J. and Morokuma, Tomoki and Mukae, Shiro and Murata, Ryoma and Murayama, Hitoshi and Nagao, Tohru and Nakata, Fumiaki and Niida, Mana and Niikura, Hiroko and Nishizawa, Atsushi J. and Obuchi, Yoshiyuki and Oguri, Masamune and Oishi, Yukie and Okabe, Nobuhiro and Okamoto, Sakurako and Okura, Yuki and Ono, Yoshiaki and Onodera, Masato and Onoue, Masafusa and Osato, Ken and Ouchi, Masami and Price, Paul A. and Pyo, Tae-Soo and Sako, Masao and Sawicki, Marcin and Shibuya, Takatoshi and Shimasaku, Kazuhiro and Shimono, Atsushi and Shirasaki, Masato and Silverman, John D. and Simet, Melanie and Speagle, Joshua and Spergel, David N. and Strauss, Michael A. and Sugahara, Yuma and Sugiyama, Naoshi and Suto, Yasushi and Suyu, Sherry H. and Suzuki, Nao and Tait, Philip J. and Takada, Masahiro and Takata, Tadafumi and Tamura, Naoyuki and Tanaka, Manobu M. and Tanaka, Masaomi and Tanaka, Masayuki and Tanaka, Yoko and Terai, Tsuyoshi and Terashima, Yuichi and Toba, Yoshiki and Tominaga, Nozomu and Toshikawa, Jun and Turner, Edwin L. and Uchida, Tomohisa and Uchiyama, Hisakazu and Umetsu, Keiichi and Uraguchi, Fumihiro and Urata, Yuji and Usuda, Tomonori and Utsumi, Yousuke and Wang, Shiang-Yu and Wang, Wei-Hao and Wong, Kenneth C. and Yabe, Kiyoto and Yamada, Yoshihiko and Yamanoi, Hitomi and Yasuda, Naoki and Yeh, Sherry and Yonehara, Atsunori and Yuma, Suraphong},
  year = 2018,
  month = jan,
  journal = {PASJ},
  volume = {70},
  pages = {S4},
  issn = {0004-6264},
  doi = {10.1093/pasj/psx066},
  urldate = {2024-12-31}
}

@article{aihara_second_2019,
  title = {Second Data Release of the {{Hyper Suprime-Cam Subaru Strategic Program}}},
  author = {Aihara, Hiroaki and AlSayyad, Yusra and Ando, Makoto and Armstrong, Robert and Bosch, James and Egami, Eiichi and Furusawa, Hisanori and Furusawa, Junko and Goulding, Andy and Harikane, Yuichi and Hikage, Chiaki and Ho, Paul T. P. and Hsieh, Bau-Ching and Huang, Song and Ikeda, Hiroyuki and Imanishi, Masatoshi and Ito, Kei and Iwata, Ikuru and Jaelani, Anton T. and Kakuma, Ryota and Kawana, Kojiro and Kikuta, Satoshi and Kobayashi, Umi and Koike, Michitaro and Komiyama, Yutaka and Li, Xiangchong and Liang, Yongming and Lin, Yen-Ting and Luo, Wentao and Lupton, Robert and Lust, Nate B. and MacArthur, Lauren A. and Matsuoka, Yoshiki and Mineo, Sogo and Miyatake, Hironao and Miyazaki, Satoshi and More, Surhud and Murata, Ryoma and Namiki, Shigeru V. and Nishizawa, Atsushi J. and Oguri, Masamune and Okabe, Nobuhiro and Okamoto, Sakurako and Okura, Yuki and Ono, Yoshiaki and Onodera, Masato and Onoue, Masafusa and Osato, Ken and Ouchi, Masami and Shibuya, Takatoshi and Strauss, Michael A. and Sugiyama, Naoshi and Suto, Yasushi and Takada, Masahiro and Takagi, Yuhei and Takata, Tadafumi and Takita, Satoshi and Tanaka, Masayuki and Terai, Tsuyoshi and Toba, Yoshiki and Uchiyama, Hisakazu and Utsumi, Yousuke and Wang, Shiang-Yu and Wang, Wenting and Yamada, Yoshihiko},
  year = 2019,
  month = dec,
  journal = {PASJ},
  volume = {71},
  pages = {114},
  issn = {0004-6264},
  doi = {10.1093/pasj/psz103},
  urldate = {2023-08-19}
}

@article{aihara_third_2022,
  title = {Third Data Release of the {{Hyper Suprime-Cam Subaru Strategic Program}}},
  author = {Aihara, Hiroaki and AlSayyad, Yusra and Ando, Makoto and Armstrong, Robert and Bosch, James and Egami, Eiichi and Furusawa, Hisanori and Furusawa, Junko and Harasawa, Sumiko and Harikane, Yuichi and Hsieh, Bau-Ching and Ikeda, Hiroyuki and Ito, Kei and Iwata, Ikuru and Kodama, Tadayuki and Koike, Michitaro and Kokubo, Mitsuru and Komiyama, Yutaka and Li, Xiangchong and Liang, Yongming and Lin, Yen-Ting and Lupton, Robert H. and Lust, Nate B. and MacArthur, Lauren A. and Mawatari, Ken and Mineo, Sogo and Miyatake, Hironao and Miyazaki, Satoshi and More, Surhud and Morishima, Takahiro and Murayama, Hitoshi and Nakajima, Kimihiko and Nakata, Fumiaki and Nishizawa, Atsushi J. and Oguri, Masamune and Okabe, Nobuhiro and Okura, Yuki and Ono, Yoshiaki and Osato, Ken and Ouchi, Masami and Pan, Yen-Chen and Plazas Malag{\'o}n, Andr{\'e}s A. and Price, Paul A. and Reed, Sophie L. and Rykoff, Eli S. and Shibuya, Takatoshi and Simunovic, Mirko and Strauss, Michael A. and Sugimori, Kanako and Suto, Yasushi and Suzuki, Nao and Takada, Masahiro and Takagi, Yuhei and Takata, Tadafumi and Takita, Satoshi and Tanaka, Masayuki and Tang, Shenli and Taranu, Dan S. and Terai, Tsuyoshi and Toba, Yoshiki and Turner, Edwin L. and Uchiyama, Hisakazu and Vijarnwannaluk, Bovornpratch and Waters, Christopher Z. and Yamada, Yoshihiko and Yamamoto, Naoaki and Yamashita, Takuji},
  year = 2022,
  month = apr,
  journal = {PASJ},
  volume = {74},
  pages = {247--272},
  issn = {0004-6264},
  doi = {10.1093/pasj/psab122},
  urldate = {2024-12-31}
}

@article{aird_evolution_2015,
  title = {The Evolution of the {{X-ray}} Luminosity Functions of Unabsorbed and Absorbed {{AGNs}} out to Z{$\sim$} 5},
  author = {Aird, J. and Coil, A. L. and Georgakakis, A. and Nandra, K. and Barro, G. and {P{\'e}rez-Gonz{\'a}lez}, P. G.},
  year = 2015,
  month = aug,
  journal = {MNRAS},
  volume = {451},
  pages = {1892--1927},
  issn = {0035-8711},
  doi = {10.1093/mnras/stv1062},
  urldate = {2024-11-05}
}

@article{alonso_active_2007,
  title = {Active Galactic Nuclei and Galaxy Interactions},
  author = {Alonso, M. Sol and Lambas, Diego G. and Tissera, Patricia and Coldwell, Georgina},
  year = 2007,
  month = mar,
  journal = {MNRAS},
  volume = {375},
  pages = {1017--1024},
  issn = {0035-8711},
  doi = {10.1111/j.1365-2966.2007.11367.x},
  urldate = {2023-05-04}
}

@article{alonso_galaxy_2004,
  title = {Galaxy Pairs in the {{2dF}} Survey - {{II}}. {{Effects}} of Interactions on Star Formation in Groups and Clusters},
  author = {Alonso, M. Sol and Tissera, Patricia B. and Coldwell, Georgina and Lambas, Diego G.},
  year = 2004,
  month = aug,
  journal = {MNRAS},
  volume = {352},
  pages = {1081--1088},
  issn = {0035-8711},
  doi = {10.1111/j.1365-2966.2004.08002.x},
  urldate = {2023-05-04}
}

@article{argudo-fernandez_amiga_2013,
  title = {The {{AMIGA}} Sample of Isolated Galaxies. {{XII}}. {{Revision}} of the Isolation Degree for {{AMIGA}} Galaxies Using the {{SDSS}}},
  author = {{Argudo-Fern{\'a}ndez}, M. and Verley, S. and Bergond, G. and Sulentic, J. and Sabater, J. and Fern{\'a}ndez Lorenzo, M. and Leon, S. and Espada, D. and {Verdes-Montenegro}, L. and {Santander-Vela}, J. D. and Ruiz, J. E. and {S{\'a}nchez-Exp{\'o}sito}, S.},
  year = 2013,
  month = dec,
  journal = {A\&A},
  volume = {560},
  pages = {A9},
  issn = {0004-6361},
  doi = {10.1051/0004-6361/201321326},
  urldate = {2024-11-05}
}

@article{argudo-fernandez_effects_2014,
  title = {Effects of the Environment on Galaxies in the {{Catalogue}} of {{Isolated Galaxies}}: Physical Satellites and Large Scale Structure},
  shorttitle = {Effects of the Environment on Galaxies in the {{Catalogue}} of {{Isolated Galaxies}}},
  author = {{Argudo-Fern{\'a}ndez}, M. and Verley, S. and Bergond, G. and Sulentic, J. and Sabater, J. and Fern{\'a}ndez Lorenzo, M. and Espada, D. and Leon, S. and {S{\'a}nchez-Exp{\'o}sito}, S. and {Santander-Vela}, J. D. and {Verdes-Montenegro}, L.},
  year = 2014,
  month = apr,
  journal = {A\&A},
  volume = {564},
  pages = {A94},
  issn = {0004-6361},
  doi = {10.1051/0004-6361/201322498},
  urldate = {2024-10-26}
}

@article{arnouts_lephare_2011,
  title = {{{LePHARE}}: {{Photometric Analysis}} for {{Redshift Estimate}}},
  shorttitle = {{{LePHARE}}},
  author = {Arnouts, S. and Ilbert, O.},
  year = 2011,
  month = aug,
  journal = {Astrophysics Source Code Library},
  pages = {[record ascl:1108.009]},
  urldate = {2023-05-05}
}

@article{ascaso_violent_2014,
  title = {The Violent Youth of Bright and Massive Cluster Galaxies and Their Maturation over 7 Billion Years},
  author = {Ascaso, B. and Lemaux, B. C. and Lubin, L. M. and Gal, R. R. and Kocevski, D. D. and Rumbaugh, N. and Squires, G.},
  year = 2014,
  month = jul,
  journal = {MNRAS},
  volume = {442},
  number = {1},
  pages = {589--615},
  issn = {0035-8711},
  doi = {10.1093/mnras/stu877},
  urldate = {2025-01-03},
  langid = {english}
}

@article{astropy_collaboration_astropy_2013,
  title = {Astropy: {{A}} Community {{Python}} Package for Astronomy},
  shorttitle = {Astropy},
  author = {{Astropy Collaboration} and Robitaille, Thomas P. and Tollerud, Erik J. and Greenfield, Perry and Droettboom, Michael and Bray, Erik and Aldcroft, Tom and Davis, Matt and Ginsburg, Adam and {Price-Whelan}, Adrian M. and Kerzendorf, Wolfgang E. and Conley, Alexander and Crighton, Neil and Barbary, Kyle and Muna, Demitri and Ferguson, Henry and Grollier, Fr{\'e}d{\'e}ric and Parikh, Madhura M. and Nair, Prasanth H. and Unther, Hans M. and Deil, Christoph and Woillez, Julien and Conseil, Simon and Kramer, Roban and Turner, James E. H. and Singer, Leo and Fox, Ryan and Weaver, Benjamin A. and Zabalza, Victor and Edwards, Zachary I. and Azalee Bostroem, K. and Burke, D. J. and Casey, Andrew R. and Crawford, Steven M. and Dencheva, Nadia and Ely, Justin and Jenness, Tim and Labrie, Kathleen and Lim, Pey Lian and Pierfederici, Francesco and Pontzen, Andrew and Ptak, Andy and Refsdal, Brian and Servillat, Mathieu and Streicher, Ole},
  year = 2013,
  month = oct,
  journal = {A\&A},
  volume = {558},
  pages = {A33},
  issn = {0004-6361},
  doi = {10.1051/0004-6361/201322068},
  urldate = {2025-02-11}
}

@article{astropy_collaboration_astropy_2018,
  title = {The {{Astropy Project}}: {{Building}} an {{Open-science Project}} and {{Status}} of the v2.0 {{Core Package}}},
  shorttitle = {The {{Astropy Project}}},
  author = {{Astropy Collaboration} and {Price-Whelan}, A. M. and Sip{\H o}cz, B. M. and G{\"u}nther, H. M. and Lim, P. L. and Crawford, S. M. and Conseil, S. and Shupe, D. L. and Craig, M. W. and Dencheva, N. and Ginsburg, A. and VanderPlas, J. T. and Bradley, L. D. and {P{\'e}rez-Su{\'a}rez}, D. and {de Val-Borro}, M. and Aldcroft, T. L. and Cruz, K. L. and Robitaille, T. P. and Tollerud, E. J. and Ardelean, C. and Babej, T. and Bach, Y. P. and Bachetti, M. and Bakanov, A. V. and Bamford, S. P. and Barentsen, G. and Barmby, P. and Baumbach, A. and Berry, K. L. and Biscani, F. and Boquien, M. and Bostroem, K. A. and Bouma, L. G. and Brammer, G. B. and Bray, E. M. and Breytenbach, H. and Buddelmeijer, H. and Burke, D. J. and Calderone, G. and Cano Rodr{\'i}guez, J. L. and Cara, M. and Cardoso, J. V. M. and Cheedella, S. and Copin, Y. and Corrales, L. and Crichton, D. and D'Avella, D. and Deil, C. and Depagne, {\'E}. and Dietrich, J. P. and Donath, A. and Droettboom, M. and Earl, N. and Erben, T. and Fabbro, S. and Ferreira, L. A. and Finethy, T. and Fox, R. T. and Garrison, L. H. and Gibbons, S. L. J. and Goldstein, D. A. and Gommers, R. and Greco, J. P. and Greenfield, P. and Groener, A. M. and Grollier, F. and Hagen, A. and Hirst, P. and Homeier, D. and Horton, A. J. and Hosseinzadeh, G. and Hu, L. and Hunkeler, J. S. and Ivezi{\'c}, {\v Z}. and Jain, A. and Jenness, T. and Kanarek, G. and Kendrew, S. and Kern, N. S. and Kerzendorf, W. E. and Khvalko, A. and King, J. and Kirkby, D. and Kulkarni, A. M. and Kumar, A. and Lee, A. and Lenz, D. and Littlefair, S. P. and Ma, Z. and Macleod, D. M. and Mastropietro, M. and McCully, C. and Montagnac, S. and Morris, B. M. and Mueller, M. and Mumford, S. J. and Muna, D. and Murphy, N. A. and Nelson, S. and Nguyen, G. H. and Ninan, J. P. and N{\"o}the, M. and Ogaz, S. and Oh, S. and Parejko, J. K. and Parley, N. and Pascual, S. and Patil, R. and Patil, A. A. and Plunkett, A. L. and Prochaska, J. X. and Rastogi, T. and Reddy Janga, V. and Sabater, J. and Sakurikar, P. and Seifert, M. and Sherbert, L. E. and {Sherwood-Taylor}, H. and Shih, A. Y. and Sick, J. and Silbiger, M. T. and Singanamalla, S. and Singer, L. P. and Sladen, P. H. and Sooley, K. A. and Sornarajah, S. and Streicher, O. and Teuben, P. and Thomas, S. W. and Tremblay, G. R. and Turner, J. E. H. and Terr{\'o}n, V. and {van Kerkwijk}, M. H. and {de la Vega}, A. and Watkins, L. L. and Weaver, B. A. and Whitmore, J. B. and Woillez, J. and Zabalza, V. and {Astropy Contributors}},
  year = 2018,
  month = sep,
  journal = {AJ},
  volume = {156},
  pages = {123},
  issn = {0004-6256},
  doi = {10.3847/1538-3881/aabc4f},
  urldate = {2025-02-11}
}

@article{astropy_collaboration_astropy_2022,
  title = {The {{Astropy Project}}: {{Sustaining}} and {{Growing}} a {{Community-oriented Open-source Project}} and the {{Latest Major Release}} (v5.0) of the {{Core Package}}},
  shorttitle = {The {{Astropy Project}}},
  author = {{Astropy Collaboration} and {Price-Whelan}, Adrian M. and Lim, Pey Lian and Earl, Nicholas and Starkman, Nathaniel and Bradley, Larry and Shupe, David L. and Patil, Aarya A. and Corrales, Lia and Brasseur, C. E. and N{\"o}the, Maximilian and Donath, Axel and Tollerud, Erik and Morris, Brett M. and Ginsburg, Adam and Vaher, Eero and Weaver, Benjamin A. and Tocknell, James and Jamieson, William and {van Kerkwijk}, Marten H. and Robitaille, Thomas P. and Merry, Bruce and Bachetti, Matteo and G{\"u}nther, H. Moritz and Aldcroft, Thomas L. and {Alvarado-Montes}, Jaime A. and Archibald, Anne M. and B{\'o}di, Attila and Bapat, Shreyas and Barentsen, Geert and Baz{\'a}n, Juanjo and Biswas, Manish and Boquien, M{\'e}d{\'e}ric and Burke, D. J. and Cara, Daria and Cara, Mihai and Conroy, Kyle E. and Conseil, Simon and Craig, Matthew W. and Cross, Robert M. and Cruz, Kelle L. and D'Eugenio, Francesco and Dencheva, Nadia and Devillepoix, Hadrien A. R. and Dietrich, J{\"o}rg P. and Eigenbrot, Arthur Davis and Erben, Thomas and Ferreira, Leonardo and {Foreman-Mackey}, Daniel and Fox, Ryan and Freij, Nabil and Garg, Suyog and Geda, Robel and Glattly, Lauren and Gondhalekar, Yash and Gordon, Karl D. and Grant, David and Greenfield, Perry and Groener, Austen M. and Guest, Steve and Gurovich, Sebastian and Handberg, Rasmus and Hart, Akeem and {Hatfield-Dodds}, Zac and Homeier, Derek and Hosseinzadeh, Griffin and Jenness, Tim and Jones, Craig K. and Joseph, Prajwel and Kalmbach, J. Bryce and Karamehmetoglu, Emir and Ka{\l}uszy{\'n}ski, Miko{\l}aj and Kelley, Michael S. P. and Kern, Nicholas and Kerzendorf, Wolfgang E. and Koch, Eric W. and Kulumani, Shankar and Lee, Antony and Ly, Chun and Ma, Zhiyuan and MacBride, Conor and Maljaars, Jakob M. and Muna, Demitri and Murphy, N. A. and Norman, Henrik and O'Steen, Richard and Oman, Kyle A. and Pacifici, Camilla and Pascual, Sergio and {Pascual-Granado}, J. and Patil, Rohit R. and Perren, Gabriel I. and Pickering, Timothy E. and Rastogi, Tanuj and Roulston, Benjamin R. and Ryan, Daniel F. and Rykoff, Eli S. and Sabater, Jose and Sakurikar, Parikshit and Salgado, Jes{\'u}s and Sanghi, Aniket and Saunders, Nicholas and Savchenko, Volodymyr and Schwardt, Ludwig and {Seifert-Eckert}, Michael and Shih, Albert Y. and Jain, Anany Shrey and Shukla, Gyanendra and Sick, Jonathan and Simpson, Chris and Singanamalla, Sudheesh and Singer, Leo P. and Singhal, Jaladh and Sinha, Manodeep and Sip{\H o}cz, Brigitta M. and Spitler, Lee R. and Stansby, David and Streicher, Ole and {\v S}umak, Jani and Swinbank, John D. and Taranu, Dan S. and Tewary, Nikita and Tremblay, Grant R. and {de Val-Borro}, Miguel and Van Kooten, Samuel J. and Vasovi{\'c}, Zlatan and Verma, Shresth and {de Miranda Cardoso}, Jos{\'e} Vin{\'i}cius and Williams, Peter K. G. and Wilson, Tom J. and Winkel, Benjamin and {Wood-Vasey}, W. M. and Xue, Rui and Yoachim, Peter and Zhang, Chen and Zonca, Andrea and {Astropy Project Contributors}},
  year = 2022,
  month = aug,
  journal = {ApJ},
  volume = {935},
  pages = {167},
  issn = {0004-637X},
  doi = {10.3847/1538-4357/ac7c74},
  urldate = {2025-02-11}
}

@article{ata_predicted_2022,
  title = {Predicted Future Fate of {{COSMOS}} Galaxy Protoclusters over 11 {{Gyr}} with Constrained Simulations},
  author = {Ata, Metin and Lee, Khee-Gan and Vecchia, Claudio Dalla and Kitaura, Francisco-Shu and Cucciati, Olga and Lemaux, Brian C. and Kashino, Daichi and M{\"u}ller, Thomas},
  year = 2022,
  month = jun,
  journal = {Nat. Astron.},
  volume = {6},
  pages = {857--865},
  issn = {2397-3366},
  doi = {10.1038/s41550-022-01693-0},
  urldate = {2025-02-08}
}

@article{athanassoula_spiral_1984,
  title = {The Spiral Structure of Galaxies.},
  author = {Athanassoula, E.},
  year = 1984,
  month = jan,
  journal = {Phys. Rep.},
  volume = {114},
  pages = {321--403},
  issn = {0370-1573},
  urldate = {2025-01-06}
}

@article{bahe_disruption_2019,
  title = {Disruption of Satellite Galaxies in Simulated Groups and Clusters: The Roles of Accretion Time, Baryons, and Pre-Processing},
  shorttitle = {Disruption of Satellite Galaxies in Simulated Groups and Clusters},
  author = {Bah{\'e}, Yannick M. and Schaye, Joop and Barnes, David J. and Dalla Vecchia, Claudio and Kay, Scott T. and Bower, Richard G. and Hoekstra, Henk and McGee, Sean L. and Theuns, Tom},
  year = 2019,
  month = may,
  journal = {MNRAS},
  volume = {485},
  pages = {2287--2311},
  issn = {0035-8711},
  doi = {10.1093/mnras/stz361},
  urldate = {2025-01-02}
}

@article{balogh_evidence_2016,
  title = {Evidence for a Change in the Dominant Satellite Galaxy Quenching Mechanism at z = 1},
  author = {Balogh, Michael L. and McGee, Sean L. and Mok, Angus and Muzzin, Adam and {van der Burg}, Remco F. J. and Bower, Richard G. and Finoguenov, Alexis and Hoekstra, Henk and Lidman, Chris and Mulchaey, John S. and Noble, Allison and Parker, Laura C. and Tanaka, Masayuki and Wilman, David J. and Webb, Tracy and Wilson, Gillian and Yee, Howard K. C.},
  year = 2016,
  month = mar,
  journal = {MNRAS},
  volume = {456},
  pages = {4364--4376},
  issn = {0035-8711},
  doi = {10.1093/mnras/stv2949},
  urldate = {2021-06-10}
}

@article{baxter_gogreen_2022,
  title = {The {{GOGREEN}} Survey: Constraining the Satellite Quenching Time-Scale in Massive Clusters at z {$\greaterequivlnt$} 1},
  shorttitle = {The {{GOGREEN}} Survey},
  author = {Baxter, Devontae C. and Cooper, M. C. and Balogh, Michael L. and Carleton, Timothy and Cerulo, Pierluigi and De Lucia, Gabriella and Demarco, Ricardo and McGee, Sean and Muzzin, Adam and Nantais, Julie and {Pintos-Castro}, Irene and Reeves, Andrew M. M. and Rudnick, Gregory H. and Sarron, Florian and {van der Burg}, Remco F. J. and Vulcani, Benedetta and Wilson, Gillian and Zaritsky, Dennis},
  year = 2022,
  month = oct,
  journal = {MNRAS},
  volume = {515},
  pages = {5479--5494},
  issn = {0035-8711},
  doi = {10.1093/mnras/stac2149},
  urldate = {2025-03-01}
}

@article{baxter_when_2023,
  title = {When the Well Runs Dry: Modelling Environmental Quenching of High-Mass Satellites in Massive Clusters at z {$\greaterequivlnt$} 1},
  shorttitle = {When the Well Runs Dry},
  author = {Baxter, Devontae C. and Cooper, M. C. and Balogh, Michael L. and Rudnick, Gregory H. and De Lucia, Gabriella and Demarco, Ricardo and Finoguenov, Alexis and Forrest, Ben and Muzzin, Adam and Reeves, Andrew M. M. and Sarron, Florian and Vulcani, Benedetta and Wilson, Gillian and Zaritsky, Dennis},
  year = 2023,
  month = dec,
  journal = {MNRAS},
  volume = {526},
  pages = {3716--3729},
  issn = {0035-8711},
  doi = {10.1093/mnras/stad2995},
  urldate = {2025-03-01}
}

@article{bertin_sextractor_1996,
  title = {{{SExtractor}}: {{Software}} for Source Extraction.},
  shorttitle = {{{SExtractor}}},
  author = {Bertin, E. and Arnouts, S.},
  year = 1996,
  month = jun,
  journal = {A\&AS},
  volume = {117},
  pages = {393--404},
  issn = {0365-01380004-6361},
  doi = {10.1051/aas:1996164},
  urldate = {2023-08-21}
}

@article{bond_how_1996,
  title = {How Filaments of Galaxies Are Woven into the Cosmic Web},
  author = {Bond, J. Richard and Kofman, Lev and Pogosyan, Dmitry},
  year = 1996,
  month = apr,
  journal = {Nature},
  volume = {380},
  pages = {603--606},
  issn = {0028-0836},
  doi = {10.1038/380603a0},
  urldate = {2023-05-04}
}

@article{boselli_effect_2009,
  title = {The Effect of Ram-Pressure Stripping and Starvation on the Star Formation Properties of Cluster Galaxies},
  author = {Boselli, A. and Boissier, S. and Cortese, L. and Gavazzi, G.},
  year = 2009,
  month = dec,
  journal = {Astron. Nachr.},
  volume = {330},
  pages = {904},
  issn = {0004-6337},
  doi = {10.1002/asna.200911259},
  urldate = {2021-06-10}
}

@article{boselli_quenching_2016,
  title = {Quenching of the Star Formation Activity in Cluster Galaxies},
  author = {Boselli, A. and Roehlly, Y. and Fossati, M. and Buat, V. and Boissier, S. and Boquien, M. and Burgarella, D. and Ciesla, L. and Gavazzi, G. and Serra, P.},
  year = 2016,
  month = nov,
  journal = {A\&A},
  volume = {596},
  pages = {A11},
  issn = {0004-6361},
  doi = {10.1051/0004-6361/201629221},
  urldate = {2022-12-12},
  langid = {english}
}

@article{boulade_megacam_2003,
  title = {{{MegaCam}}: The New {{Canada-France-Hawaii Telescope}} Wide-Field Imaging Camera},
  shorttitle = {{{MegaCam}}},
  author = {Boulade, Olivier and Charlot, Xavier and Abbon, P. and Aune, Stephan and Borgeaud, Pierre and Carton, Pierre-Henri and Carty, M. and Da Costa, J. and Deschamps, H. and Desforge, D. and Eppell{\'e}, Dominique and Gallais, Pascal and Gosset, L. and Granelli, Remy and Gros, Michel and {de Kat}, Jean and Loiseau, Denis and Ritou, J. -. and Rouss{\'e}, Jean Y. and Starzynski, Pierre and Vignal, Nicolas and Vigroux, Laurent G.},
  year = 2003,
  month = mar,
  journal={in Instrument Design and Performance for Optical/Infrared Ground-based Telescopes, eds. M. Iye, \& A. F. M. Moorwood, Proc. SPIE},
  volume = {4841},
  pages = {72--81},
  doi = {10.1117/12.459890},
  urldate = {2023-08-20}
}

@article{brammer_3d-hst_2012,
  title = {{{3D-HST}}: {{A Wide-field Grism Spectroscopic Survey}} with the {{Hubble Space Telescope}}},
  shorttitle = {{{3D-HST}}},
  author = {Brammer, Gabriel B. and {van Dokkum}, Pieter G. and Franx, Marijn and Fumagalli, Mattia and Patel, Shannon and Rix, Hans-Walter and Skelton, Rosalind E. and Kriek, Mariska and Nelson, Erica and Schmidt, Kasper B. and Bezanson, Rachel and {da Cunha}, Elisabete and Erb, Dawn K. and Fan, Xiaohui and F{\"o}rster Schreiber, Natascha and Illingworth, Garth D. and Labb{\'e}, Ivo and Leja, Joel and Lundgren, Britt and Magee, Dan and Marchesini, Danilo and McCarthy, Patrick and Momcheva, Ivelina and Muzzin, Adam and Quadri, Ryan and Steidel, Charles C. and Tal, Tomer and Wake, David and Whitaker, Katherine E. and Williams, Anna},
  year = 2012,
  month = jun,
  journal = {ApJS},
  volume = {200},
  pages = {13},
  issn = {0067-0049},
  doi = {10.1088/0067-0049/200/2/13},
  urldate = {2022-11-22}
}

@article{brammer_gbrammergrizli_2021,
       author = {{Brammer}, Gabe and {Matharu}, Jasleen},
        title = "{gbrammer/grizli: Release 2021}",
         year = 2021,
        month = jun,
        journal = {gbrammer/grizli: Release 2021, Version 1.3.2, Zenodo},
          eid = {10.5281/zenodo.5012699},
          pages = {doi:10.5281/zenodo.5012699},
      version = {1.3.2},
    publisher = {Zenodo},
       adsurl = {https://ui.adsabs.harvard.edu/abs/2021zndo...5012699B}
}

@article{bundy_greater_2009,
  title = {The {{Greater Impact}} of {{Mergers}} on the {{Growth}} of {{Massive Galaxies}}: {{Implications}} for {{Mass Assembly}} and {{Evolution}} since z Sime 1},
  shorttitle = {The {{Greater Impact}} of {{Mergers}} on the {{Growth}} of {{Massive Galaxies}}},
  author = {Bundy, Kevin and Fukugita, Masataka and Ellis, Richard S. and Targett, Thomas A. and Belli, Sirio and Kodama, Tadayuki},
  year = 2009,
  month = jun,
  journal = {ApJ},
  volume = {697},
  pages = {1369--1383},
  issn = {0004-637X},
  doi = {10.1088/0004-637X/697/2/1369},
  urldate = {2025-01-28}
}

@article{butcher_evolution_1978,
  title = {The Evolution of Galaxies in Clusters. {{II}}. {{The}} Galaxy Content of Nearby Clusters.},
  author = {Butcher, H. and Oemler, Jr., A.},
  year = 1978,
  month = dec,
  journal = {ApJ},
  volume = {226},
  pages = {559--565},
  issn = {0004-637X},
  doi = {10.1086/156640},
  urldate = {2022-11-21}
}

@article{byrd_tidal_1992,
  title = {Tidal {{Arms}} Are {{Ubiquitous}} in {{Spiral Galaxies}}},
  author = {Byrd, Gene G. and Howard, Sethanne},
  year = 1992,
  month = apr,
  journal = {AJ},
  volume = {103},
  pages = {1089},
  issn = {0004-6256},
  doi = {10.1086/116128},
  urldate = {2025-01-06}
}

@article{calvi_probing_2021,
  title = {Probing the Existence of a Rich Galaxy Overdensity at z = 5.2},
  author = {Calvi, Rosa and Dannerbauer, Helmut and Arrabal Haro, Pablo and Rodr{\'i}guez Espinosa, Jos{\'e} M. and {Mu{\~n}oz-Tu{\~n}{\'o}n}, Casiana and P{\'e}rez Gonz{\'a}lez, Pablo G. and Geier, Stefan},
  year = 2021,
  month = apr,
  journal = {MNRAS},
  volume = {502},
  pages = {4558--4575},
  issn = {0035-8711},
  doi = {10.1093/mnras/staa4037},
  urldate = {2022-12-12}
}

@article{capak_first_2007,
  title = {The {{First Release COSMOS Optical}} and {{Near-IR Data}} and {{Catalog}}},
  author = {Capak, P. and Aussel, H. and Ajiki, M. and McCracken, H. J. and Mobasher, B. and Scoville, N. and Shopbell, P. and Taniguchi, Y. and Thompson, D. and Tribiano, S. and Sasaki, S. and Blain, A. W. and Brusa, M. and Carilli, C. and Comastri, A. and Carollo, C. M. and Cassata, P. and Colbert, J. and Ellis, R. S. and Elvis, M. and Giavalisco, M. and Green, W. and Guzzo, L. and Hasinger, G. and Ilbert, O. and Impey, C. and Jahnke, K. and Kartaltepe, J. and Kneib, J. -P. and Koda, J. and Koekemoer, A. and Komiyama, Y. and Leauthaud, A. and Le Fevre, O. and Lilly, S. and Liu, C. and Massey, R. and Miyazaki, S. and Murayama, T. and Nagao, T. and Peacock, J. A. and Pickles, A. and Porciani, C. and Renzini, A. and Rhodes, J. and Rich, M. and Salvato, M. and Sanders, D. B. and Scarlata, C. and Schiminovich, D. and Schinnerer, E. and Scodeggio, M. and Sheth, K. and Shioya, Y. and Tasca, L. A. M. and Taylor, J. E. and Yan, L. and Zamorani, G.},
  year = 2007,
  month = sep,
  journal = {ApJS},
  volume = {172},
  pages = {99--116},
  issn = {0067-0049},
  doi = {10.1086/519081},
  urldate = {2023-08-18}
}

@article{casey_massive_2015,
  title = {A {{Massive}}, {{Distant Proto-cluster}} at z = 2.47 {{Caught}} in a {{Phase}} of {{Rapid Formation}}?},
  author = {Casey, C. M. and Cooray, A. and Capak, P. and Fu, H. and Kovac, K. and Lilly, S. and Sanders, D. B. and Scoville, N. Z. and Treister, E.},
  year = 2015,
  month = aug,
  journal = {ApJ},
  volume = {808},
  pages = {L33},
  issn = {0004-637X},
  doi = {10.1088/2041-8205/808/2/L33},
  urldate = {2025-02-08}
}

@article{chiang_ancient_2013,
  title = {Ancient {{Light}} from {{Young Cosmic Cities}}: {{Physical}} and {{Observational Signatures}} of {{Galaxy Proto-clusters}}},
  shorttitle = {Ancient {{Light}} from {{Young Cosmic Cities}}},
  author = {Chiang, Yi-Kuan and Overzier, Roderik and Gebhardt, Karl},
  year = 2013,
  month = dec,
  journal = {ApJ},
  volume = {779},
  pages = {127},
  issn = {0004-637X},
  doi = {10.1088/0004-637X/779/2/127},
  urldate = {2023-04-06}
}

@article{chiang_galaxy_2017,
  title = {Galaxy {{Protoclusters}} as {{Drivers}} of {{Cosmic Star Formation History}} in the {{First}} 2 {{Gyr}}},
  author = {Chiang, Yi-Kuan and Overzier, Roderik A. and Gebhardt, Karl and Henriques, Bruno},
  year = 2017,
  month = aug,
  journal = {ApJ},
  volume = {844},
  pages = {L23},
  issn = {0004-637X},
  doi = {10.3847/2041-8213/aa7e7b},
  urldate = {2022-11-21}
}

@article{chiang_surveying_2015,
  title = {Surveying {{Galaxy Proto-clusters}} in {{Emission}}: {{A Large-scale Structure}} at z = 2.44 and the {{Outlook}} for {{HETDEX}}},
  shorttitle = {Surveying {{Galaxy Proto-clusters}} in {{Emission}}},
  author = {Chiang, Yi-Kuan and Overzier, Roderik A. and Gebhardt, Karl and Finkelstein, Steven L. and Chiang, Chi-Ting and Hill, Gary J. and Blanc, Guillermo A. and Drory, Niv and Chonis, Taylor S. and Zeimann, Gregory R. and Hagen, Alex and Schneider, Donald P. and Jogee, Shardha and Ciardullo, Robin and Gronwall, Caryl},
  year = 2015,
  month = jul,
  journal = {ApJ},
  volume = {808},
  pages = {37},
  issn = {0004-637X},
  doi = {10.1088/0004-637X/808/1/37},
  urldate = {2025-02-08}
}

@article{christlein_disentangling_2005,
  title = {Disentangling {{Morphology}}, {{Star Formation}}, {{Stellar Mass}}, and {{Environment}} in {{Galaxy Evolution}}},
  author = {Christlein, Daniel and Zabludoff, Ann I.},
  year = 2005,
  month = mar,
  journal = {ApJ},
  volume = {621},
  pages = {201--214},
  issn = {0004-637X},
  doi = {10.1086/427427},
  urldate = {2022-11-21}
}

@article{colberg_clustering_2000,
  title = {Clustering of Galaxy Clusters in Cold Dark Matter Universes},
  author = {Colberg, J. M. and White, S. D. M. and Yoshida, N. and MacFarland, T. J. and Jenkins, A. and Frenk, C. S. and Pearce, F. R. and Evrard, A. E. and Couchman, H. M. P. and Efstathiou, G. and Peacock, J. A. and Thomas, P. A. and {Virgo Consortium}},
  year = 2000,
  month = nov,
  journal = {MNRAS},
  volume = {319},
  pages = {209--214},
  issn = {0035-8711},
  doi = {10.1046/j.1365-8711.2000.03832.x},
  urldate = {2021-06-10}
}

@article{colless_2df_2001,
  title = {The {{2dF Galaxy Redshift Survey}}: Spectra and Redshifts},
  shorttitle = {The {{2dF Galaxy Redshift Survey}}},
  author = {Colless, Matthew and Dalton, Gavin and Maddox, Steve and Sutherland, Will and Norberg, Peder and Cole, Shaun and {Bland-Hawthorn}, Joss and Bridges, Terry and Cannon, Russell and Collins, Chris and Couch, Warrick and Cross, Nicholas and Deeley, Kathryn and De Propris, Roberto and Driver, Simon P. and Efstathiou, George and Ellis, Richard S. and Frenk, Carlos S. and Glazebrook, Karl and Jackson, Carole and Lahav, Ofer and Lewis, Ian and Lumsden, Stuart and Madgwick, Darren and Peacock, John A. and Peterson, Bruce A. and Price, Ian and Seaborne, Mark and Taylor, Keith},
  year = 2001,
  month = dec,
  journal = {MNRAS},
  volume = {328},
  pages = {1039--1063},
  issn = {0035-8711},
  doi = {10.1046/j.1365-8711.2001.04902.x},
  urldate = {2021-06-10}
}

@article{conselice_structures_2009,
  title = {The Structures of Distant Galaxies - {{III}}. {{The}} Merger History of over 20000 Massive Galaxies at z {$<$} 1.2},
  author = {Conselice, Christopher J. and Yang, Cui and Bluck, Asa F. L.},
  year = 2009,
  month = apr,
  journal = {MNRAS},
  volume = {394},
  pages = {1956--1972},
  issn = {0035-8711},
  doi = {10.1111/j.1365-2966.2009.14396.x},
  urldate = {2024-11-13}
}

@article{contini_semi-analytic_2016,
  title = {Semi-Analytic Model Predictions of the Galaxy Population in Protoclusters},
  author = {Contini, E. and De Lucia, G. and Hatch, N. and Borgani, S. and Kang, X.},
  year = 2016,
  month = feb,
  journal = {MNRAS},
  volume = {456},
  pages = {1924--1935},
  issn = {0035-8711},
  doi = {10.1093/mnras/stv2852},
  urldate = {2025-02-23}
}

@article{coogan_merger-driven_2018,
  title = {Merger-Driven Star Formation Activity in {{Cl J1449}}+0856 at z = 1.99 as Seen by {{ALMA}} and {{JVLA}}},
  author = {Coogan, R. T. and Daddi, E. and Sargent, M. T. and Strazzullo, V. and Valentino, F. and Gobat, R. and Magdis, G. and Bethermin, M. and Pannella, M. and Onodera, M. and Liu, D. and Cimatti, A. and Dannerbauer, H. and Carollo, M. and Renzini, A. and Tremou, E.},
  year = 2018,
  month = sep,
  journal = {MNRAS},
  volume = {479},
  pages = {703--729},
  issn = {0035-8711},
  doi = {10.1093/mnras/sty1446},
  urldate = {2025-03-04}
}

@article{cooper_deep2_2008,
  title = {The {{DEEP2 Galaxy Redshift Survey}}: The Role of Galaxy Environment in the Cosmic Star Formation History},
  shorttitle = {The {{DEEP2 Galaxy Redshift Survey}}},
  author = {Cooper, Michael C. and Newman, Jeffrey A. and Weiner, Benjamin J. and Yan, Renbin and Willmer, Christopher N. A. and Bundy, Kevin and Coil, Alison L. and Conselice, Christopher J. and Davis, Marc and Faber, S. M. and Gerke, Brian F. and Guhathakurta, Puragra and Koo, David C. and Noeske, Kai G.},
  year = 2008,
  month = jan,
  journal = {MNRAS},
  volume = {383},
  pages = {1058--1078},
  issn = {0035-8711},
  doi = {10.1111/j.1365-2966.2007.12613.x},
  urldate = {2021-06-06}
}

@article{coupon_bright-star_2018,
  title = {The Bright-Star Masks for the {{HSC-SSP}} Survey},
  author = {Coupon, Jean and Czakon, Nicole and Bosch, James and Komiyama, Yutaka and Medezinski, Elinor and Miyazaki, Satoshi and Oguri, Masamune},
  year = 2018,
  month = jan,
  journal = {PASJ},
  volume = {70},
  pages = {S7},
  issn = {0004-6264},
  doi = {10.1093/pasj/psx047},
  urldate = {2023-08-21}
}

@article{cucciati_discovery_2014,
  title = {Discovery of a Rich Proto-Cluster at z = 2.9 and Associated Diffuse Cold Gas in the {{VIMOS Ultra-Deep Survey}} ({{VUDS}})},
  author = {Cucciati, O. and Zamorani, G. and Lemaux, B. C. and Bardelli, S. and Cimatti, A. and Le F{\`e}vre, O. and Cassata, P. and Garilli, B. and Le Brun, V. and Maccagni, D. and Pentericci, L. and Tasca, L. a. M. and Thomas, R. and Vanzella, E. and Zucca, E. and Amorin, R. and Capak, P. and Cassar{\`a}, L. P. and Castellano, M. and Cuby, J. G. and {de la Torre}, S. and Durkalec, A. and Fontana, A. and Giavalisco, M. and Grazian, A. and Hathi, N. P. and Ilbert, O. and Moreau, C. and Paltani, S. and Ribeiro, B. and Salvato, M. and Schaerer, D. and Scodeggio, M. and Sommariva, V. and Talia, M. and Taniguchi, Y. and Tresse, L. and Vergani, D. and Wang, P. W. and Charlot, S. and Contini, T. and Fotopoulou, S. and {L{\'o}pez-Sanjuan}, C. and Mellier, Y. and Scoville, N.},
  year = 2014,
  month = oct,
  journal = {A\&A},
  volume = {570},
  pages = {A16},
  issn = {0004-6361},
  doi = {10.1051/0004-6361/201423811},
  urldate = {2022-12-12},
  langid = {english}
}

@article{cucciati_progeny_2018,
  title = {The Progeny of a Cosmic Titan: A Massive Multi-Component Proto-Supercluster in Formation at z = 2.45 in {{VUDS}}},
  shorttitle = {The Progeny of a Cosmic Titan},
  author = {Cucciati, O. and Lemaux, B. C. and Zamorani, G. and Le F{\`e}vre, O. and Tasca, L. A. M. and Hathi, N. P. and Lee, K. -G. and Bardelli, S. and Cassata, P. and Garilli, B. and Le Brun, V. and Maccagni, D. and Pentericci, L. and Thomas, R. and Vanzella, E. and Zucca, E. and Lubin, L. M. and Amorin, R. and Cassar{\`a}, L. P. and Cimatti, A. and Talia, M. and Vergani, D. and Koekemoer, A. and Pforr, J. and Salvato, M.},
  year = 2018,
  month = nov,
  journal = {A\&A},
  volume = {619},
  pages = {A49},
  issn = {0004-6361},
  doi = {10.1051/0004-6361/201833655},
  urldate = {2023-08-20}
}

@article{daddi_new_2004,
  title = {A {{New Photometric Technique}} for the {{Joint Selection}} of {{Star-forming}} and {{Passive Galaxies}} at 1.4 {$<$}\textasciitilde{} z {$<$}\textasciitilde{} 2.5},
  author = {Daddi, E. and Cimatti, A. and Renzini, A. and Fontana, A. and Mignoli, M. and Pozzetti, L. and Tozzi, P. and Zamorani, G.},
  year = 2004,
  month = dec,
  journal = {ApJ},
  volume = {617},
  pages = {746--764},
  issn = {0004-637X},
  doi = {10.1086/425569},
  urldate = {2023-08-18}
}

@article{dahari_companions_1984,
  title = {Companions of {{Seyfert}} Galaxies : A Statistical Survey.},
  shorttitle = {Companions of {{Seyfert}} Galaxies},
  author = {Dahari, O.},
  year = 1984,
  month = jul,
  journal = {AJ},
  volume = {89},
  pages = {966--974},
  issn = {0004-6256},
  doi = {10.1086/113591},
  urldate = {2025-02-23}
}

@article{darvish_comparative_2015,
  title = {A {{Comparative Study}} of {{Density Field Estimation}} for {{Galaxies}}: {{New Insights}} into the {{Evolution}} of {{Galaxies}} with {{Environment}} in {{COSMOS}} out to Z{$\sim$}3},
  shorttitle = {A {{Comparative Study}} of {{Density Field Estimation}} for {{Galaxies}}},
  author = {Darvish, Behnam and Mobasher, Bahram and Sobral, David and Scoville, Nicholas and {Aragon-Calvo}, Miguel},
  year = 2015,
  month = jun,
  journal = {ApJ},
  volume = {805},
  pages = {121},
  issn = {0004-637X},
  doi = {10.1088/0004-637X/805/2/121},
  urldate = {2021-06-11}
}

@article{darvish_effects_2016,
  title = {The {{Effects}} of the {{Local Environment}} and {{Stellar Mass}} on {{Galaxy Quenching}} to z {$\sim$} 3},
  author = {Darvish, Behnam and Mobasher, Bahram and Sobral, David and Rettura, Alessandro and Scoville, Nick and Faisst, Andreas and Capak, Peter},
  year = 2016,
  month = jul,
  journal = {ApJ},
  volume = {825},
  pages = {113},
  issn = {0004-637X},
  doi = {10.3847/0004-637X/825/2/113},
  urldate = {2021-06-10}
}

@article{darvish_spectroscopic_2020,
  title = {Spectroscopic {{Confirmation}} of a {{Coma Cluster Progenitor}} at z {$\sim$} 2.2},
  author = {Darvish, Behnam and Scoville, Nick Z. and Martin, Christopher and Sobral, David and Mobasher, Bahram and Rettura, Alessandro and Matthee, Jorryt and Capak, Peter and Chartab, Nima and Hemmati, Shoubaneh and Masters, Daniel and Nayyeri, Hooshang and O'Sullivan, Donal and {Paulino-Afonso}, Ana and Sattari, Zahra and Shahidi, Abtin and Salvato, Mara and Lemaux, Brian C. and F{\`e}vre, Olivier Le and Cucciati, Olga},
  year = 2020,
  month = mar,
  journal = {ApJ},
  volume = {892},
  pages = {8},
  issn = {0004-637X},
  doi = {10.3847/1538-4357/ab75c3},
  urldate = {2025-02-08}
}

@article{das_galaxy_2021,
  title = {Galaxy Interactions in Different Environments: {{An}} Analysis of Galaxy Pairs from the {{SDSS}}},
  shorttitle = {Galaxy Interactions in Different Environments},
  author = {Das, Apashanka and Pandey, Biswajit and Sarkar, Suman and Dutta, Arunima},
  year = 2021,
  month = aug,
  journal = {arXiv e-prints},
  pages = {[arXiv:2108.05874]},
  doi = {10.48550/arXiv.2108.05874},
  urldate = {2023-05-04}
}

@article{davidzon_cosmos2015_2017,
  title = {The {{COSMOS2015}} Galaxy Stellar Mass Function . {{Thirteen}} Billion Years of Stellar Mass Assembly in Ten Snapshots},
  author = {Davidzon, I. and Ilbert, O. and Laigle, C. and Coupon, J. and McCracken, H. J. and Delvecchio, I. and Masters, D. and Capak, P. and Hsieh, B. C. and Le F{\`e}vre, O. and Tresse, L. and Bethermin, M. and Chang, Y. -Y. and Faisst, A. L. and Le Floc'h, E. and Steinhardt, C. and Toft, S. and Aussel, H. and Dubois, C. and Hasinger, G. and Salvato, M. and Sanders, D. B. and Scoville, N. and Silverman, J. D.},
  year = 2017,
  month = sep,
  journal = {A\&A},
  volume = {605},
  pages = {A70},
  issn = {0004-6361},
  doi = {10.1051/0004-6361/201730419},
  urldate = {2023-08-18}
}

@article{davidzon_vimos_2016,
  title = {The {{VIMOS Public Extragalactic Redshift Survey}} ({{VIPERS}}). {{Environmental}} Effects Shaping the Galaxy Stellar Mass Function},
  author = {Davidzon, I. and Cucciati, O. and Bolzonella, M. and De Lucia, G. and Zamorani, G. and Arnouts, S. and Moutard, T. and Ilbert, O. and Garilli, B. and Scodeggio, M. and Guzzo, L. and Abbas, U. and Adami, C. and Bel, J. and Bottini, D. and Branchini, E. and Cappi, A. and Coupon, J. and {de la Torre}, S. and Di Porto, C. and Fritz, A. and Franzetti, P. and Fumana, M. and Granett, B. R. and Guennou, L. and Iovino, A. and Krywult, J. and Le Brun, V. and Le F{\`e}vre, O. and Maccagni, D. and Ma{\l}ek, K. and Marulli, F. and McCracken, H. J. and Mellier, Y. and Moscardini, L. and Polletta, M. and Pollo, A. and Tasca, L. A. M. and Tojeiro, R. and Vergani, D. and Zanichelli, A.},
  year = 2016,
  month = feb,
  journal = {A\&A},
  volume = {586},
  pages = {A23},
  issn = {0004-6361},
  doi = {10.1051/0004-6361/201527129},
  urldate = {2025-01-02}
}

@article{de_lucia_environmental_2012,
  title = {The Environmental History of Group and Cluster Galaxies in a {{$\Lambda$}} Cold Dark Matter Universe},
  author = {De Lucia, Gabriella and Weinmann, Simone and Poggianti, Bianca M. and {Arag{\'o}n-Salamanca}, Alfonso and Zaritsky, Dennis},
  year = 2012,
  month = jun,
  journal = {MNRAS},
  volume = {423},
  number = {2},
  pages = {1277--1292},
  issn = {0035-8711},
  doi = {10.1111/j.1365-2966.2012.20983.x},
  urldate = {2025-01-02},
  langid = {english}
}

@article{de_lucia_hierarchical_2007,
  title = {The Hierarchical Formation of the Brightest Cluster Galaxies},
  author = {De Lucia, Gabriella and Blaizot, J{\'e}r{\'e}my},
  year = 2007,
  month = feb,
  journal = {MNRAS},
  volume = {375},
  number = {1},
  pages = {2--14},
  issn = {0035-8711},
  doi = {10.1111/j.1365-2966.2006.11287.x},
  urldate = {2025-01-03},
  langid = {english}
}

@article{de_lucia_tracing_2024,
  title = {Tracing the Quenching Journey across Cosmic Time},
  author = {De Lucia, Gabriella and Fontanot, Fabio and Xie, Lizhi and Hirschmann, Michaela},
  year = 2024,
  month = jul,
  journal = {A\&A},
  volume = {687},
  pages = {A68},
  issn = {0004-6361},
  doi = {10.1051/0004-6361/202349045},
  urldate = {2024-09-29}
}

@article{de_ravel_vimos_2009,
  title = {The {{VIMOS VLT Deep Survey}}. {{Evolution}} of the Major Merger Rate since z \textasciitilde{} 1 from Spectroscopically Confirmed Galaxy Pairs},
  author = {{de Ravel}, L. and Le F{\`e}vre, O. and Tresse, L. and Bottini, D. and Garilli, B. and Le Brun, V. and Maccagni, D. and Scaramella, R. and Scodeggio, M. and Vettolani, G. and Zanichelli, A. and Adami, C. and Arnouts, S. and Bardelli, S. and Bolzonella, M. and Cappi, A. and Charlot, S. and Ciliegi, P. and Contini, T. and Foucaud, S. and Franzetti, P. and Gavignaud, I. and Guzzo, L. and Ilbert, O. and Iovino, A. and Lamareille, F. and McCracken, H. J. and Marano, B. and Marinoni, C. and Mazure, A. and Meneux, B. and Merighi, R. and Paltani, S. and Pell{\`o}, R. and Pollo, A. and Pozzetti, L. and Radovich, M. and Vergani, D. and Zamorani, G. and Zucca, E. and Bondi, M. and Bongiorno, A. and Brinchmann, J. and Cucciati, O. and {de La Torre}, S. and Gregorini, L. and Memeo, P. and {Perez-Montero}, E. and Mellier, Y. and Merluzzi, P. and Temporin, S.},
  year = 2009,
  month = may,
  journal = {A\&A},
  volume = {498},
  pages = {379--397},
  issn = {0004-6361},
  doi = {10.1051/0004-6361/200810569},
  urldate = {2024-11-13}
}

@article{diener_proto-groups_2013,
  title = {Proto-Groups at 1.8 {$<$} z {$<$} 3 in the {{zCOSMOS-deep Sample}}},
  author = {Diener, C. and Lilly, S. J. and Knobel, C. and Zamorani, G. and Lemson, G. and Kampczyk, P. and Scoville, N. and Carollo, C. M. and Contini, T. and Kneib, J. -P. and Le Fevre, O. and Mainieri, V. and Renzini, A. and Scodeggio, M. and Bardelli, S. and Bolzonella, M. and Bongiorno, A. and Caputi, K. and Cucciati, O. and {de la Torre}, S. and {de Ravel}, L. and Franzetti, P. and Garilli, B. and Iovino, A. and Kova{\v c}, K. and Lamareille, F. and Le Borgne, J. -F. and Le Brun, V. and Maier, C. and Mignoli, M. and Pello, R. and Peng, Y. and Perez Montero, E. and Presotto, V. and Silverman, J. and Tanaka, M. and Tasca, L. and Tresse, L. and Vergani, D. and Zucca, E. and Bordoloi, R. and Cappi, A. and Cimatti, A. and Coppa, G. and Koekemoer, A. M. and {L{\'o}pez-Sanjuan}, C. and McCracken, H. J. and Moresco, M. and Nair, P. and Pozzetti, L. and Welikala, N.},
  year = 2013,
  month = mar,
  journal = {ApJ},
  volume = {765},
  pages = {109},
  issn = {0004-637X},
  doi = {10.1088/0004-637X/765/2/109},
  urldate = {2022-11-21}
}

@article{diener_protocluster_2015,
  title = {A {{Protocluster}} at z = 2.45},
  author = {Diener, C. and Lilly, S. J. and Ledoux, C. and Zamorani, G. and Bolzonella, M. and Murphy, D. N. A. and Capak, P. and Ilbert, O. and McCracken, H.},
  year = 2015,
  month = mar,
  journal = {ApJ},
  volume = {802},
  pages = {31},
  issn = {0004-637X},
  doi = {10.1088/0004-637X/802/1/31},
  urldate = {2022-11-21}
}

@article{dolag_simulating_2006,
  title = {Simulating Large-Scale Structure Formation with Magnetic Fields},
  author = {Dolag, K.},
  year = 2006,
  month = jun,
  journal = {Astron. Nachr.},
  volume = {327},
  pages = {575},
  issn = {0004-6337},
  doi = {10.1002/asna.200610595},
  urldate = {2021-06-10}
}

@article{dressler_evolution_1984,
  title = {The {{Evolution}} of {{Galaxies}} in {{Clusters}}},
  author = {Dressler, A.},
  year = 1984,
  journal = {ARA\&A},
  volume = {22},
  pages = {185},
  issn = {0066-4146},
  doi = {10.1146/annurev.astro.22.1.185},
  urldate = {2022-11-21},
  langid = {english}
}

@article{dressler_galaxy_1980,
  title = {Galaxy Morphology in Rich Clusters: Implications for the Formation and Evolution of Galaxies.},
  shorttitle = {Galaxy Morphology in Rich Clusters},
  author = {Dressler, A.},
  year = 1980,
  month = mar,
  journal = {ApJ},
  volume = {236},
  pages = {351--365},
  issn = {0004-637X},
  doi = {10.1086/157753},
  urldate = {2022-11-22}
}

@ARTICLE{duan_galaxy_2025,
       author = {{Duan}, Qiao and {Conselice}, Christopher J. and {Li}, Qiong and {Austin}, Duncan and {Harvey}, Thomas and {Adams}, Nathan J. and {Duncan}, Kenneth J. and {Trussler}, James and {Ferreira}, Leonardo and {Westcott}, Lewi and {Harris}, Honor and {Windhorst}, Rogier A. and {Holwerda}, Benne W. and {Broadhurst}, Thomas J. and {Coe}, Dan and {Cohen}, Seth H. and {Du}, Xiaojing and {Driver}, Simon P. and {Frye}, Brenda and {Grogin}, Norman A. and {Hathi}, Nimish P. and {Jansen}, Rolf A. and {Koekemoer}, Anton M. and {Marshall}, Madeline A. and {Nonino}, Mario and {Ortiz}, III, Rafael and {Pirzkal}, Nor and {Robotham}, Aaron and {Ryan}, Russell E. and {Summers}, Jake and {D'Silva}, Jordan C.~J. and {Willmer}, Christopher N.~A. and {Yan}, Haojing},
        title = "{Galaxy mergers in the epoch of reionization {\textendash} I. A JWST study of pair fractions, merger rates, and stellar mass accretion rates at z = 4.5{\textendash}11.5}",
      journal = {\mnras},
         year = 2025,
        month = jun,
       volume = {540},
       number = {1},
        pages = {774-805},
          doi = {10.1093/mnras/staf638},
archivePrefix = {arXiv},
       eprint = {2407.09472},
 primaryClass = {astro-ph.GA},
       adsurl = {https://ui.adsabs.harvard.edu/abs/2025MNRAS.540..774D}
}

@article{duncan_observational_2019,
  title = {Observational {{Constraints}} on the {{Merger History}} of {{Galaxies}} since z {$\approx$} 6: {{Probabilistic Galaxy Pair Counts}} in the {{CANDELS Fields}}},
  shorttitle = {Observational {{Constraints}} on the {{Merger History}} of {{Galaxies}} since z {$\approx$} 6},
  author = {Duncan, Kenneth and Conselice, Christopher J. and Mundy, Carl and Bell, Eric and Donley, Jennifer and Galametz, Audrey and Guo, Yicheng and Grogin, Norman A. and Hathi, Nimish and Kartaltepe, Jeyhan and Kocevski, Dale and Koekemoer, Anton M. and {P{\'e}rez-Gonz{\'a}lez}, Pablo G. and Mantha, Kameswara B. and Snyder, Gregory F. and Stefanon, Mauro},
  year = 2019,
  month = may,
  journal = {ApJ},
  volume = {876},
  pages = {110},
  issn = {0004-637X},
  doi = {10.3847/1538-4357/ab148a},
  urldate = {2023-08-08}
}

@article{einasto_120-mpc_1997,
  title = {A 120-{{Mpc}} Periodicity in the Three-Dimensional Distribution of Galaxy Superclusters},
  author = {Einasto, J. and Einasto, M. and Gottl{\"o}ber, S. and M{\"u}ller, V. and Saar, V. and Starobinsky, A. A. and Tago, E. and Tucker, D. and Andernach, H. and Frisch, P.},
  year = 1997,
  month = jan,
  journal = {Nature},
  volume = {385},
  pages = {139--141},
  issn = {0028-0836},
  doi = {10.1038/385139a0},
  urldate = {2022-12-09}
}

@article{eke_evolution_1998,
  title = {The {{Evolution}} of {{X-Ray Clusters}} in a {{Low-Density Universe}}},
  author = {Eke, Vincent R. and Navarro, Julio F. and Frenk, Carlos S.},
  year = 1998,
  month = aug,
  journal = {ApJ},
  volume = {503},
  pages = {569--592},
  issn = {0004-637X},
  doi = {10.1086/306008},
  urldate = {2021-06-10}
}

@article{elbaz_reversal_2007,
  title = {The Reversal of the Star Formation-Density Relation in the Distant Universe},
  author = {Elbaz, D. and Daddi, E. and Le Borgne, D. and Dickinson, M. and Alexander, D. M. and Chary, R.-R. and Starck, J.-L. and Brandt, W. N. and Kitzbichler, M. and MacDonald, E. and Nonino, M. and Popesso, P. and Stern, D. and Vanzella, E.},
  year = 2007,
  month = jun,
  journal = {A\&A},
  volume = {468},
  number = {1},
  pages = {33--48},
  issn = {0004-6361},
  doi = {10.1051/0004-6361:20077525},
  urldate = {2022-12-11},
  langid = {english}
}

@article{ellison_galaxy_2010,
  title = {Galaxy Pairs in the {{Sloan Digital Sky Survey}} - {{II}}. {{The}} Effect of Environment on Interactions},
  author = {Ellison, Sara L. and Patton, David R. and Simard, Luc and McConnachie, Alan W. and Baldry, Ivan K. and Mendel, J. Trevor},
  year = 2010,
  month = sep,
  journal = {MNRAS},
  volume = {407},
  pages = {1514--1528},
  issn = {0035-8711},
  doi = {10.1111/j.1365-2966.2010.17076.x},
  urldate = {2023-05-04}
}

@article{ellison_galaxy_2013,
  title = {Galaxy Pairs in the {{Sloan Digital Sky Survey}} - {{VIII}}. {{The}} Observational Properties of Post-Merger Galaxies},
  author = {Ellison, Sara L. and Mendel, J. Trevor and Patton, David R. and Scudder, Jillian M.},
  year = 2013,
  month = nov,
  journal = {MNRAS},
  volume = {435},
  pages = {3627--3638},
  issn = {0035-8711},
  doi = {10.1093/mnras/stt1562},
  urldate = {2023-05-05}
}

@article{ellison_galaxy_2013-1,
  title = {Galaxy Pairs in the {{Sloan Digital Sky Survey}} - {{VII}}. {{The}} Merger-Luminous Infrared Galaxy Connection},
  author = {Ellison, Sara L. and Mendel, J. Trevor and Scudder, Jillian M. and Patton, David R. and Palmer, Michael J. D.},
  year = 2013,
  month = apr,
  journal = {MNRAS},
  volume = {430},
  pages = {3128--3141},
  issn = {0035-8711},
  doi = {10.1093/mnras/sts546},
  urldate = {2023-05-05}
}

@article{ellison_galaxy_2022,
  title = {Galaxy Mergers Can Rapidly Shut down Star Formation},
  author = {Ellison, Sara L. and Wilkinson, Scott and Woo, Joanna and Leung, Ho-Hin and Wild, Vivienne and Bickley, Robert W. and Patton, David R. and Quai, Salvatore and Gwyn, Stephen},
  year = 2022,
  month = nov,
  journal = {MNRAS},
  volume = {517},
  pages = {L92--L96},
  issn = {0035-8711},
  doi = {10.1093/mnrasl/slac109},
  urldate = {2022-11-16}
}

@article{euclid_collaboration_euclid_2022,
  title = {Euclid Preparation. {{XVII}}. {{Cosmic Dawn Survey}}: {{Spitzer Space Telescope}} Observations of the {{Euclid}} Deep Fields and Calibration Fields},
  shorttitle = {Euclid Preparation. {{XVII}}. {{Cosmic Dawn Survey}}},
  author = {{Euclid Collaboration} and Moneti, A. and McCracken, H. J. and Shuntov, M. and Kauffmann, O. B. and Capak, P. and Davidzon, I. and Ilbert, O. and Scarlata, C. and Toft, S. and Weaver, J. and Chary, R. and Cuby, J. and Faisst, A. L. and Masters, D. C. and McPartland, C. and Mobasher, B. and Sanders, D. B. and Scaramella, R. and Stern, D. and Szapudi, I. and Teplitz, H. and Zalesky, L. and Amara, A. and Auricchio, N. and Bodendorf, C. and Bonino, D. and Branchini, E. and {Brau-Nogue}, S. and Brescia, M. and Brinchmann, J. and Capobianco, V. and Carbone, C. and Carretero, J. and Castander, F. J. and Castellano, M. and Cavuoti, S. and Cimatti, A. and Cledassou, R. and Congedo, G. and Conselice, C. J. and Conversi, L. and Copin, Y. and Corcione, L. and Costille, A. and Cropper, M. and Da Silva, A. and Degaudenzi, H. and Douspis, M. and Dubath, F. and Duncan, C. A. J. and Dupac, X. and Dusini, S. and Farrens, S. and Ferriol, S. and Fosalba, P. and Frailis, M. and Franceschi, E. and Fumana, M. and Garilli, B. and Gillis, B. and Giocoli, C. and Granett, B. R. and Grazian, A. and Grupp, F. and Haugan, S. V. H. and Hoekstra, H. and Holmes, W. and Hormuth, F. and Hudelot, P. and Jahnke, K. and Kermiche, S. and Kiessling, A. and Kilbinger, M. and Kitching, T. and Kohley, R. and K{\"u}mmel, M. and Kunz, M. and {Kurki-Suonio}, H. and Ligori, S. and Lilje, P. B. and Lloro, I. and Maiorano, E. and Mansutti, O. and Marggraf, O. and Markovic, K. and Marulli, F. and Massey, R. and Maurogordato, S. and Meneghetti, M. and Merlin, E. and Meylan, G. and Moresco, M. and Moscardini, L. and Munari, E. and Niemi, S. M. and Padilla, C. and Paltani, S. and Pasian, F. and Pedersen, K. and Pires, S. and Poncet, M. and Popa, L. and Pozzetti, L. and Raison, F. and Rebolo, R. and Rhodes, J. and Rix, H. and Roncarelli, M. and Rossetti, E. and Saglia, R. and Schneider, P. and Secroun, A. and Seidel, G. and Serrano, S. and Sirignano, C. and Sirri, G. and Stanco, L. and {Tallada-Cresp{\'i}}, P. and Taylor, A. N. and Tereno, I. and {Toledo-Moreo}, R. and Torradeflot, F. and Wang, Y. and Welikala, N. and Weller, J. and Zamorani, G. and Zoubian, J. and Andreon, S. and Bardelli, S. and Camera, S. and {Graci{\'a}-Carpio}, J. and Medinaceli, E. and Mei, S. and Polenta, G. and Romelli, E. and Sureau, F. and Tenti, M. and Vassallo, T. and Zacchei, A. and Zucca, E. and Baccigalupi, C. and {Balaguera-Antol{\'i}nez}, A. and Bernardeau, F. and Biviano, A. and Bolzonella, M. and Bozzo, E. and Burigana, C. and Cabanac, R. and Cappi, A. and Carvalho, C. S. and Casas, S. and Castignani, G. and {Colodro-Conde}, C. and Coupon, J. and Courtois, H. M. and Di Ferdinando, D. and Farina, M. and Finelli, F. and {Flose-Reimberg}, P. and Fotopoulou, S. and Galeotta, S. and Ganga, K. and {Garcia-Bellido}, J. and Gaztanaga, E. and Gozaliasl, G. and Hook, I. and Joachimi, B. and Kansal, V. and Keihanen, E. and Kirkpatrick, C. C. and Lindholm, V. and Mainetti, G. and Maino, D. and Maoli, R. and Martinelli, M. and Martinet, N. and Maturi, M. and Metcalf, R. B. and Morgante, G. and Morisset, N. and Nucita, A. and Patrizii, L. and Potter, D. and Renzi, A. and Riccio, G. and S{\'a}nchez, A. G. and Sapone, D. and Schirmer, M. and Schultheis, M. and Scottez, V. and Sefusatti, E. and Teyssier, R. and Tubio, O. and Tutusaus, I. and Valiviita, J. and Viel, M. and Hildebrandt, H.},
  year = 2022,
  month = feb,
  journal = {A\&A},
  volume = {658},
  pages = {A126},
  issn = {0004-6361},
  doi = {10.1051/0004-6361/202142361},
  urldate = {2024-10-31}
}

@article{evrard_galaxy_2002,
  title = {Galaxy {{Clusters}} in {{Hubble Volume Simulations}}: {{Cosmological Constraints}} from {{Sky Survey Populations}}},
  shorttitle = {Galaxy {{Clusters}} in {{Hubble Volume Simulations}}},
  author = {Evrard, A. E. and MacFarland, T. J. and Couchman, H. M. P. and Colberg, J. M. and Yoshida, N. and White, S. D. M. and Jenkins, A. and Frenk, C. S. and Pearce, F. R. and Peacock, J. A. and Thomas, P. A.},
  year = 2002,
  month = jul,
  journal = {ApJ},
  volume = {573},
  pages = {7--36},
  issn = {0004-637X},
  doi = {10.1086/340551},
  urldate = {2021-06-10}
}

@article{faber_deimos_2003,
       author = {{Faber}, Sandra M. and {Phillips}, Andrew C. and {Kibrick}, Robert I. and {Alcott}, Barry and {Allen}, Steven L. and {Burrous}, Jim and {Cantrall}, T. and {Clarke}, De and {Coil}, Alison L. and {Cowley}, David J. and {Davis}, Marc and {Deich}, William T.~S. and {Dietsch}, Ken and {Gilmore}, David K. and {Harper}, Carol A. and {Hilyard}, David F. and {Lewis}, Jeffrey P. and {McVeigh}, Molly and {Newman}, Jeffrey and {Osborne}, Jack and {Schiavon}, Ricardo and {Stover}, Richard J. and {Tucker}, Dean and {Wallace}, Vernon and {Wei}, Mingzhi and {Wirth}, Gregory and {Wright}, Christopher A.},
        title = "{The DEIMOS spectrograph for the Keck II Telescope: integration and testing}",
    journal = {in Instrument Design and Performance for Optical/Infrared Ground-based Telescopes, eds. M. Iye, \& A. F. M. Moorwood, Proc. SPIE,},
         year = 2003,
       editor = {{Iye}, Masanori and {Moorwood}, Alan F.~M.},
       series = {Society of Photo-Optical Instrumentation Engineers (SPIE) Conference Series},
       volume = {4841},
        month = mar,
        pages = {1657-1669},
          doi = {10.1117/12.460346},
       adsurl = {https://ui.adsabs.harvard.edu/abs/2003SPIE.4841.1657F}
}

@article{fazio_infrared_2004,
       author = {{Fazio}, G.~G. and {Hora}, J.~L. and {Allen}, L.~E. and {Ashby}, M.~L.~N. and {Barmby}, P. and {Deutsch}, L.~K. and {Huang}, J.-S. and {Kleiner}, S. and {Marengo}, M. and {Megeath}, S.~T. and {Melnick}, G.~J. and {Pahre}, M.~A. and {Patten}, B.~M. and {Polizotti}, J. and {Smith}, H.~A. and {Taylor}, R.~S. and {Wang}, Z. and {Willner}, S.~P. and {Hoffmann}, W.~F. and {Pipher}, J.~L. and {Forrest}, W.~J. and {McMurty}, C.~W. and {McCreight}, C.~R. and {McKelvey}, M.~E. and {McMurray}, R.~E. and {Koch}, D.~G. and {Moseley}, S.~H. and {Arendt}, R.~G. and {Mentzell}, J.~E. and {Marx}, C.~T. and {Losch}, P. and {Mayman}, P. and {Eichhorn}, W. and {Krebs}, D. and {Jhabvala}, M. and {Gezari}, D.~Y. and {Fixsen}, D.~J. and {Flores}, J. and {Shakoorzadeh}, K. and {Jungo}, R. and {Hakun}, C. and {Workman}, L. and {Karpati}, G. and {Kichak}, R. and {Whitley}, R. and {Mann}, S. and {Tollestrup}, E.~V. and {Eisenhardt}, P. and {Stern}, D. and {Gorjian}, V. and {Bhattacharya}, B. and {Carey}, S. and {Nelson}, B.~O. and {Glaccum}, W.~J. and {Lacy}, M. and {Lowrance}, P.~J. and {Laine}, S. and {Reach}, W.~T. and {Stauffer}, J.~A. and {Surace}, J.~A. and {Wilson}, G. and {Wright}, E.~L. and {Hoffman}, A. and {Domingo}, G. and {Cohen}, M.},
        title = "{The Infrared Array Camera (IRAC) for the Spitzer Space Telescope}",
      journal = {ApJS},
         year = 2004,
        month = sep,
       volume = {154},
       number = {1},
        pages = {10-17},
          doi = {10.1086/422843},
archivePrefix = {arXiv},
       eprint = {astro-ph/0405616},
 primaryClass = {astro-ph},
       adsurl = {https://ui.adsabs.harvard.edu/abs/2004ApJS..154...10F}
}

@article{ferreras_galaxy_2017,
  title = {Galaxy and {{Mass Assembly}} ({{GAMA}}): Probing the Merger Histories of Massive Galaxies via Stellar Populations},
  shorttitle = {Galaxy and {{Mass Assembly}} ({{GAMA}})},
  author = {Ferreras, I. and Hopkins, A. M. and Gunawardhana, M. L. P. and Sansom, A. E. and Owers, M. S. and Driver, S. and Davies, L. and Robotham, A. and Taylor, E. N. and Konstantopoulos, I. and Brough, S. and Norberg, P. and Croom, S. and Loveday, J. and Wang, L. and Bremer, M.},
  year = 2017,
  month = jun,
  journal = {MNRAS},
  volume = {468},
  pages = {607--619},
  issn = {0035-8711},
  doi = {10.1093/mnras/stx503},
  urldate = {2023-05-05}
}

@article{foltz_evolution_2018,
  title = {The {{Evolution}} of {{Environmental Quenching Timescales}} to z {$\sim$} 1.6: {{Evidence}} for {{Dynamically Driven Quenching}} of the {{Cluster Galaxy Population}}},
  shorttitle = {The {{Evolution}} of {{Environmental Quenching Timescales}} to z {$\sim$} 1.6},
  author = {Foltz, R. and Wilson, G. and Muzzin, A. and Cooper, M. C. and Nantais, J. and {van der Burg}, R. F. J. and Cerulo, P. and Chan, J. and Fillingham, S. P. and Surace, J. and Webb, T. and Noble, A. and Lacy, M. and McDonald, M. and Rudnick, G. and Lidman, C. and Demarco, R. and {Hlavacek-Larrondo}, J. and Yee, H. K. C. and Perlmutter, S. and Hayden, B.},
  year = 2018,
  month = oct,
  journal = {ApJ},
  volume = {866},
  pages = {136},
  issn = {0004-637X},
  doi = {10.3847/1538-4357/aad80d},
  urldate = {2022-12-12}
}

@article{forrest_elentari_2023,
       author = {{Forrest}, Ben and {Lemaux}, Brian C. and {Shah}, Ekta and {Staab}, Priti and {McConachie}, Ian and {Cucciati}, Olga and {Gal}, Roy R. and {Hung}, Denise and {Lubin}, Lori M. and {Cassar{\`a}}, Letizia P. and {Cassata}, Paolo and {Chang}, Wenjun and {Cooper}, M.~C. and {Decarli}, Roberto and {Gomez}, Percy and {Gururajan}, Gayathri and {Hathi}, Nimish and {Kashino}, Daichi and {Marchesini}, Danilo and {Marsan}, Z. Cemile and {McDonald}, Michael and {Muzzin}, Adam and {Shen}, Lu and {Urbano Stawinski}, Stephanie and {Talia}, Margherita and {Vergani}, Daniela and {Wilson}, Gillian and {Zamorani}, Giovanni},
        title = "{Elent{\'a}ri:a massive proto-supercluster at z {\ensuremath{\sim}} 3.3 in the COSMOS field}",
      journal = {\mnras},
         year = 2023,
        month = nov,
       volume = {526},
       number = {1},
        pages = {L56-L62},
          doi = {10.1093/mnrasl/slad114},
archivePrefix = {arXiv},
       eprint = {2307.15113},
 primaryClass = {astro-ph.GA},
       adsurl = {https://ui.adsabs.harvard.edu/abs/2023MNRAS.526L..56F}
}

@article{forrest_environmental_2024,
  title = {Environmental {{Effects}} on the {{Stellar Mass Function}} in a z {$\sim$} 3.3 {{Overdensity}} of {{Galaxies}} in the {{COSMOS Field}}},
  author = {Forrest, Ben and Lemaux, Brian C. and Shah, Ekta A. and Staab, Priti and Gal, Roy R. and Lubin, Lori M. and Cooper, M. C. and Cucciati, Olga and Hung, Denise and McConachie, Ian and Muzzin, Adam and Wilson, Gillian and Bardelli, Sandro and Cassar{\`a}, Letizia P. and Chang, Wenjun and Giddings, Finn and {Golden-Marx}, Emmet and Hathi, Nimish and Urbano Stawinski, Stephanie M. and Zucca, Elena},
  year = 2024,
  month = aug,
  journal = {ApJ},
  volume = {971},
  pages = {169},
  issn = {0004-637X},
  doi = {10.3847/1538-4357/ad5e78},
  urldate = {2024-10-31}
}

@article{forrest_hst-hyperion_2025,
  title = {The {{HST-Hyperion Survey}}: {{Grism Observations}} of a z {$\sim$} 2.5 {{Protosupercluster}}},
  shorttitle = {The {{HST-Hyperion Survey}}},
  author = {Forrest, Ben and Shen, Lu and Lemaux, Brian C. and Shah, Ekta and Cucciati, Olga and Gal, Roy R. and Giddings, Finn and {Golden-Marx}, Emmet and Hu, Weida and Ronayne, Kaila and Sikorski, Derek and Staab, Priti and Amor{\'i}n, Ricardo O. and Bardelli, Sandro and Garilli, Bianca and Hathi, Nimish and Hung, Denise and Lubin, Lori and Pelliccia, Debora and Ryan, Russell E. and Zamorani, Gianni and Zucca, Elena},
  year = 2025,
  month = may,
  journal = {ApJ},
  volume = {985},
  pages = {61},
  issn = {0004-637X},
  doi = {10.3847/1538-4357/adc252},
  urldate = {2025-06-18}
}

@article{forrest_massive_2020,
  title = {The {{Massive Ancient Galaxies}} at z {$>$} 3 {{NEar-infrared}} ({{MAGAZ3NE}}) {{Survey}}: {{Confirmation}} of {{Extremely Rapid Star Formation}} and {{Quenching Timescales}} for {{Massive Galaxies}} in the {{Early Universe}}},
  shorttitle = {The {{Massive Ancient Galaxies}} at z {$>$} 3 {{NEar-infrared}} ({{MAGAZ3NE}}) {{Survey}}},
  author = {Forrest, Ben and Marsan, Z. Cemile and Annunziatella, Marianna and Wilson, Gillian and Muzzin, Adam and Marchesini, Danilo and Cooper, M. C. and Chan, Jeffrey C. C. and McConachie, Ian and Gomez, Percy and {Kado-Fong}, Erin and La Barbera, Francesco and {Lange-Vagle}, Daniel and Nantais, Julie and Nonino, Mario and Saracco, Paolo and Stefanon, Mauro and {van der Burg}, Remco F. J.},
  year = 2020,
  month = nov,
  journal = {ApJ},
  volume = {903},
  pages = {47},
  issn = {0004-637X},
  doi = {10.3847/1538-4357/abb819},
  urldate = {2023-08-18}
}

@article{franck_candidate_2016,
  title = {The {{Candidate Cluster}} and {{Protocluster Catalog}} ({{CCPC}}) {{II}}. {{Spectroscopically Identified Structures Spanning}} 2{$<$} z {$<$} 6.6},
  author = {Franck, J. R. and McGaugh, S. S.},
  year = 2016,
  month = dec,
  journal = {ApJ},
  volume = {833},
  pages = {15},
  issn = {0004-637X},
  doi = {10.3847/0004-637X/833/1/15},
  urldate = {2025-02-08}
}

@article{frenk_galaxy_1996,
  title = {Galaxy {{Dynamics}} in {{Clusters}}},
  author = {Frenk, Carlos S. and Evrard, August E. and White, Simon D. M. and Summers, F. J.},
  year = 1996,
  month = dec,
  journal = {ApJ},
  volume = {472},
  pages = {460},
  issn = {0004-637X},
  doi = {10.1086/178079},
  urldate = {2021-06-10}
}

@article{fujita_pre-processing_2004,
  title = {Pre-{{Processing}} of {{Galaxies}} before {{Entering}} a {{Cluster}}},
  author = {Fujita, Yutaka},
  year = 2004,
  month = feb,
  journal = {PASJ},
  volume = {56},
  pages = {29--43},
  issn = {0004-6264},
  doi = {10.1093/pasj/56.1.29},
  urldate = {2024-12-19}
}

@article{gaia_collaboration_gaia_2016,
  title = {Gaia {{Data Release}} 1. {{Summary}} of the Astrometric, Photometric, and Survey Properties},
  author = {{Gaia Collaboration} and Brown, A. G. A. and Vallenari, A. and Prusti, T. and {de Bruijne}, J. H. J. and Mignard, F. and Drimmel, R. and Babusiaux, C. and {Bailer-Jones}, C. A. L. and Bastian, U. and Biermann, M. and Evans, D. W. and Eyer, L. and Jansen, F. and Jordi, C. and Katz, D. and Klioner, S. A. and Lammers, U. and Lindegren, L. and Luri, X. and O'Mullane, W. and Panem, C. and Pourbaix, D. and Randich, S. and Sartoretti, P. and Siddiqui, H. I. and Soubiran, C. and Valette, V. and {van Leeuwen}, F. and Walton, N. A. and Aerts, C. and Arenou, F. and Cropper, M. and H{\o}g, E. and Lattanzi, M. G. and Grebel, E. K. and Holland, A. D. and Huc, C. and Passot, X. and Perryman, M. and Bramante, L. and Cacciari, C. and Casta{\~n}eda, J. and Chaoul, L. and Cheek, N. and De Angeli, F. and Fabricius, C. and Guerra, R. and Hern{\'a}ndez, J. and {Jean-Antoine-Piccolo}, A. and Masana, E. and Messineo, R. and Mowlavi, N. and Nienartowicz, K. and {Ord{\'o}{\~n}ez-Blanco}, D. and Panuzzo, P. and Portell, J. and Richards, P. J. and Riello, M. and Seabroke, G. M. and Tanga, P. and Th{\'e}venin, F. and Torra, J. and Els, S. G. and {Gracia-Abril}, G. and Comoretto, G. and {Garcia-Reinaldos}, M. and Lock, T. and Mercier, E. and Altmann, M. and Andrae, R. and Astraatmadja, T. L. and {Bellas-Velidis}, I. and Benson, K. and Berthier, J. and Blomme, R. and Busso, G. and Carry, B. and Cellino, A. and Clementini, G. and Cowell, S. and Creevey, O. and Cuypers, J. and Davidson, M. and De Ridder, J. and {de Torres}, A. and Delchambre, L. and Dell'Oro, A. and Ducourant, C. and Fr{\'e}mat, Y. and {Garc{\'i}a-Torres}, M. and Gosset, E. and Halbwachs, J. -L. and Hambly, N. C. and Harrison, D. L. and Hauser, M. and Hestroffer, D. and Hodgkin, S. T. and Huckle, H. E. and Hutton, A. and Jasniewicz, G. and Jordan, S. and Kontizas, M. and Korn, A. J. and Lanzafame, A. C. and Manteiga, M. and Moitinho, A. and Muinonen, K. and Osinde, J. and Pancino, E. and Pauwels, T. and Petit, J. -M. and {Recio-Blanco}, A. and Robin, A. C. and Sarro, L. M. and Siopis, C. and Smith, M. and Smith, K. W. and Sozzetti, A. and Thuillot, W. and {van Reeven}, W. and Viala, Y. and Abbas, U. and Abreu Aramburu, A. and Accart, S. and Aguado, J. J. and Allan, P. M. and Allasia, W. and Altavilla, G. and {\'A}lvarez, M. A. and Alves, J. and Anderson, R. I. and Andrei, A. H. and Anglada Varela, E. and Antiche, E. and Antoja, T. and Ant{\'o}n, S. and Arcay, B. and Bach, N. and Baker, S. G. and {Balaguer-N{\'u}{\~n}ez}, L. and Barache, C. and Barata, C. and Barbier, A. and Barblan, F. and {Barrado y Navascu{\'e}s}, D. and Barros, M. and Barstow, M. A. and Becciani, U. and Bellazzini, M. and Bello Garc{\'i}a, A. and Belokurov, V. and Bendjoya, P. and Berihuete, A. and Bianchi, L. and Bienaym{\'e}, O. and Billebaud, F. and Blagorodnova, N. and {Blanco-Cuaresma}, S. and Boch, T. and Bombrun, A. and Borrachero, R. and Bouquillon, S. and Bourda, G. and Bouy, H. and Bragaglia, A. and Breddels, M. A. and Brouillet, N. and Br{\"u}semeister, T. and Bucciarelli, B. and Burgess, P. and Burgon, R. and Burlacu, A. and Busonero, D. and Buzzi, R. and Caffau, E. and Cambras, J. and Campbell, H. and Cancelliere, R. and {Cantat-Gaudin}, T. and Carlucci, T. and Carrasco, J. M. and Castellani, M. and Charlot, P. and Charnas, J. and Chiavassa, A. and Clotet, M. and Cocozza, G. and Collins, R. S. and Costigan, G. and Crifo, F. and Cross, N. J. G. and Crosta, M. and Crowley, C. and Dafonte, C. and Damerdji, Y. and Dapergolas, A. and David, P. and David, M. and De Cat, P. and {de Felice}, F. and {de Laverny}, P. and De Luise, F. and De March, R. and {de Martino}, D. and {de Souza}, R. and Debosscher, J. and {del Pozo}, E. and Delbo, M. and Delgado, A. and Delgado, H. E. and Di Matteo, P. and Diakite, S. and Distefano, E. and Dolding, C. and Dos Anjos, S. and Drazinos, P. and Duran, J. and Dzigan, Y. and Edvardsson, B. and Enke, H. and Evans, N. W. and Eynard Bontemps, G. and Fabre, C. and Fabrizio, M. and Faigler, S. and Falc{\~a}o, A. J. and Farr{\`a}s Casas, M. and Federici, L. and Fedorets, G. and {Fern{\'a}ndez-Hern{\'a}ndez}, J. and Fernique, P. and Fienga, A. and Figueras, F. and Filippi, F. and Findeisen, K. and Fonti, A. and Fouesneau, M. and Fraile, E. and Fraser, M. and Fuchs, J. and Gai, M. and Galleti, S. and Galluccio, L. and Garabato, D. and {Garc{\'i}a-Sedano}, F. and Garofalo, A. and Garralda, N. and Gavras, P. and Gerssen, J. and Geyer, R. and Gilmore, G. and Girona, S. and Giuffrida, G. and Gomes, M. and {Gonz{\'a}lez-Marcos}, A. and {Gonz{\'a}lez-N{\'u}{\~n}ez}, J. and {Gonz{\'a}lez-Vidal}, J. J. and Granvik, M. and Guerrier, A. and Guillout, P. and Guiraud, J. and G{\'u}rpide, A. and {Guti{\'e}rrez-S{\'a}nchez}, R. and Guy, L. P. and Haigron, R. and Hatzidimitriou, D. and Haywood, M. and Heiter, U. and Helmi, A. and Hobbs, D. and Hofmann, W. and Holl, B. and Holland, G. and Hunt, J. A. S. and Hypki, A. and Icardi, V. and Irwin, M. and {Jevardat de Fombelle}, G. and Jofr{\'e}, P. and Jonker, P. G. and Jorissen, A. and Julbe, F. and Karampelas, A. and Kochoska, A. and Kohley, R. and Kolenberg, K. and Kontizas, E. and Koposov, S. E. and Kordopatis, G. and Koubsky, P. and {Krone-Martins}, A. and Kudryashova, M. and Kull, I. and Bachchan, R. K. and {Lacoste-Seris}, F. and Lanza, A. F. and Lavigne, J. -B. and {Le Poncin-Lafitte}, C. and Lebreton, Y. and Lebzelter, T. and Leccia, S. and Leclerc, N. and {Lecoeur-Taibi}, I. and Lemaitre, V. and Lenhardt, H. and Leroux, F. and Liao, S. and Licata, E. and Lindstr{\o}m, H. E. P. and Lister, T. A. and Livanou, E. and Lobel, A. and L{\"o}ffler, W. and L{\'o}pez, M. and Lorenz, D. and MacDonald, I. and Magalh{\~a}es Fernandes, T. and Managau, S. and Mann, R. G. and Mantelet, G. and Marchal, O. and Marchant, J. M. and Marconi, M. and Marinoni, S. and Marrese, P. M. and Marschalk{\'o}, G. and Marshall, D. J. and {Mart{\'i}n-Fleitas}, J. M. and Martino, M. and Mary, N. and Matijevi{\v c}, G. and Mazeh, T. and McMillan, P. J. and Messina, S. and Michalik, D. and Millar, N. R. and Miranda, B. M. H. and Molina, D. and Molinaro, R. and Molinaro, M. and Moln{\'a}r, L. and Moniez, M. and Montegriffo, P. and Mor, R. and Mora, A. and Morbidelli, R. and Morel, T. and Morgenthaler, S. and Morris, D. and Mulone, A. F. and Muraveva, T. and Musella, I. and Narbonne, J. and Nelemans, G. and Nicastro, L. and Noval, L. and Ord{\'e}novic, C. and {Ordieres-Mer{\'e}}, J. and Osborne, P. and Pagani, C. and Pagano, I. and Pailler, F. and Palacin, H. and Palaversa, L. and Parsons, P. and Pecoraro, M. and Pedrosa, R. and Pentik{\"a}inen, H. and Pichon, B. and Piersimoni, A. M. and Pineau, F. -X. and Plachy, E. and Plum, G. and Poujoulet, E. and Pr{\v s}a, A. and Pulone, L. and Ragaini, S. and Rago, S. and Rambaux, N. and {Ramos-Lerate}, M. and Ranalli, P. and Rauw, G. and Read, A. and Regibo, S. and Reyl{\'e}, C. and Ribeiro, R. A. and Rimoldini, L. and Ripepi, V. and Riva, A. and Rixon, G. and Roelens, M. and {Romero-G{\'o}mez}, M. and Rowell, N. and Royer, F. and {Ruiz-Dern}, L. and Sadowski, G. and Sagrist{\`a} Sell{\'e}s, T. and Sahlmann, J. and Salgado, J. and Salguero, E. and Sarasso, M. and Savietto, H. and Schultheis, M. and Sciacca, E. and Segol, M. and Segovia, J. C. and Segransan, D. and Shih, I. -C. and Smareglia, R. and Smart, R. L. and Solano, E. and Solitro, F. and Sordo, R. and Soria Nieto, S. and Souchay, J. and Spagna, A. and Spoto, F. and Stampa, U. and Steele, I. A. and Steidelm{\"u}ller, H. and Stephenson, C. A. and Stoev, H. and Suess, F. F. and S{\"u}veges, M. and Surdej, J. and Szabados, L. and {Szegedi-Elek}, E. and Tapiador, D. and Taris, F. and Tauran, G. and Taylor, M. B. and Teixeira, R. and Terrett, D. and Tingley, B. and Trager, S. C. and Turon, C. and Ulla, A. and Utrilla, E. and Valentini, G. and {van Elteren}, A. and Van Hemelryck, E. and {van Leeuwen}, M. and Varadi, M. and Vecchiato, A. and Veljanoski, J. and Via, T. and Vicente, D. and Vogt, S. and Voss, H. and Votruba, V. and Voutsinas, S. and Walmsley, G. and Weiler, M. and Weingrill, K. and Wevers, T. and Wyrzykowski, {\L}. and Yoldas, A. and {\v Z}erjal, M. and Zucker, S. and Zurbach, C. and Zwitter, T. and Alecu, A. and Allen, M. and Allende Prieto, C. and Amorim, A. and {Anglada-Escud{\'e}}, G. and Arsenijevic, V. and Azaz, S. and Balm, P. and Beck, M. and Bernstein, H. -H. and Bigot, L. and Bijaoui, A. and Blasco, C. and Bonfigli, M. and Bono, G. and Boudreault, S. and Bressan, A. and Brown, S. and Brunet, P. -M. and Bunclark, P. and Buonanno, R. and Butkevich, A. G. and Carret, C. and Carrion, C. and Chemin, L. and Ch{\'e}reau, F. and Corcione, L. and Darmigny, E. and {de Boer}, K. S. and {de Teodoro}, P. and {de Zeeuw}, P. T. and Delle Luche, C. and Domingues, C. D. and Dubath, P. and Fodor, F. and Fr{\'e}zouls, B. and Fries, A. and Fustes, D. and Fyfe, D. and Gallardo, E. and Gallegos, J. and Gardiol, D. and Gebran, M. and Gomboc, A. and G{\'o}mez, A. and Grux, E. and Gueguen, A. and Heyrovsky, A. and Hoar, J. and Iannicola, G. and Isasi Parache, Y. and Janotto, A. -M. and Joliet, E. and Jonckheere, A. and Keil, R. and Kim, D. -W. and Klagyivik, P. and Klar, J. and Knude, J. and Kochukhov, O. and Kolka, I. and Kos, J. and Kutka, A. and Lainey, V. and LeBouquin, D. and Liu, C. and Loreggia, D. and Makarov, V. V. and Marseille, M. G. and Martayan, C. and {Martinez-Rubi}, O. and Massart, B. and Meynadier, F. and Mignot, S. and Munari, U. and Nguyen, A. -T. and Nordlander, T. and Ocvirk, P. and O'Flaherty, K. S. and Olias Sanz, A. and Ortiz, P. and Osorio, J. and Oszkiewicz, D. and Ouzounis, A. and Palmer, M. and Park, P. and Pasquato, E. and Peltzer, C. and Peralta, J. and P{\'e}turaud, F. and Pieniluoma, T. and Pigozzi, E. and Poels, J. and Prat, G. and Prod'homme, T. and Raison, F. and Rebordao, J. M. and Risquez, D. and {Rocca-Volmerange}, B. and Rosen, S. and {Ruiz-Fuertes}, M. I. and Russo, F. and Sembay, S. and Serraller Vizcaino, I. and Short, A. and Siebert, A. and Silva, H. and Sinachopoulos, D. and Slezak, E. and Soffel, M. and Sosnowska, D. and Strai{\v z}ys, V. and {ter Linden}, M. and Terrell, D. and Theil, S. and Tiede, C. and Troisi, L. and Tsalmantza, P. and Tur, D. and Vaccari, M. and Vachier, F. and Valles, P. and Van Hamme, W. and Veltz, L. and Virtanen, J. and Wallut, J. -M. and Wichmann, R. and Wilkinson, M. I. and Ziaeepour, H. and Zschocke, S.},
  year = 2016,
  month = nov,
  journal = {A\&A},
  volume = {595},
  pages = {A2},
  issn = {0004-6361},
  doi = {10.1051/0004-6361/201629512},
  urldate = {2023-08-21}
}

@article{garduno_galaxy_2021,
  title = {Galaxy {{And Mass Assembly}} ({{GAMA}}): The Interplay between Galaxy Mass, {{SFR}}, and Heavy Element Abundance in Paired Galaxy Sets},
  shorttitle = {Galaxy {{And Mass Assembly}} ({{GAMA}})},
  author = {Gardu{\~n}o, L. E. and {Lara-L{\'o}pez}, M. A. and {L{\'o}pez-Cruz}, O. and Hopkins, A. M. and Owers, M. S. and Pimbblet, K. A. and Holwerda, B. W.},
  year = 2021,
  month = feb,
  journal = {MNRAS},
  volume = {501},
  pages = {2969--2982},
  issn = {0035-8711},
  doi = {10.1093/mnras/staa3799},
  urldate = {2023-05-04}
}

@article{geller_mapping_1989,
  title = {Mapping the Universe},
  author = {Geller, Margaret J. and Huchra, John P.},
  year = 1989,
  month = nov,
  journal = {Science},
  volume = {246},
  pages = {897--903},
  issn = {0036-8075},
  doi = {10.1126/science.246.4932.897},
  urldate = {2021-06-10}
}

@ARTICLE{gomez_first_2003,
       author = {{G{\'o}mez}, Percy L. and {Nichol}, Robert C. and {Miller}, Christopher J. and {Balogh}, Michael L. and {Goto}, Tomotsugu and {Zabludoff}, Ann I. and {Romer}, A. Kathy and {Bernardi}, Mariangela and {Sheth}, Ravi and {Hopkins}, Andrew M. and {Castander}, Francisco J. and {Connolly}, Andrew J. and {Schneider}, Donald P. and {Brinkmann}, Jon and {Lamb}, Don Q. and {SubbaRao}, Mark and {York}, Donald G.},
        title = "{Galaxy Star Formation as a Function of Environment in the Early Data Release of the Sloan Digital Sky Survey}",
      journal = {\apj},
         year = 2003,
        month = feb,
       volume = {584},
       number = {1},
        pages = {210-227},
          doi = {10.1086/345593},
archivePrefix = {arXiv},
       eprint = {astro-ph/0210193},
 primaryClass = {astro-ph},
       adsurl = {https://ui.adsabs.harvard.edu/abs/2003ApJ...584..210G}
}

@article{goto_morphology-density_2003,
  title = {The Morphology-Density Relation in the {{Sloan Digital Sky Survey}}},
  author = {Goto, Tomotsugu and Yamauchi, Chisato and Fujita, Yutaka and Okamura, Sadanori and Sekiguchi, Maki and Smail, Ian and Bernardi, Mariangela and Gomez, Percy L.},
  year = 2003,
  month = dec,
  journal = {MNRAS},
  volume = {346},
  pages = {601--614},
  issn = {0035-8711},
  doi = {10.1046/j.1365-2966.2003.07114.x},
  urldate = {2022-11-21}
}

@article{grutzbauch_relationship_2011,
  title = {The Relationship between Star Formation Rates, Local Density and Stellar Mass up to z {$\sim$} 3 in the {{GOODS NICMOS Survey}}},
  author = {Gr{\"u}tzbauch, R. and Conselice, C. J. and Bauer, A. E. and Bluck, A. F. L. and Chuter, R. W. and Buitrago, F. and Mortlock, A. and Weinzirl, T. and Jogee, S.},
  year = 2011,
  month = dec,
  journal = {MNRAS},
  volume = {418},
  pages = {938--948},
  issn = {0035-8711},
  doi = {10.1111/j.1365-2966.2011.19559.x},
  urldate = {2023-05-03}
}

@article{gunn_infall_1972,
  title = {On the {{Infall}} of {{Matter Into Clusters}} of {{Galaxies}} and {{Some Effects}} on {{Their Evolution}}},
  author = {Gunn, James E. and Gott, III, J. Richard},
  year = 1972,
  month = aug,
  journal = {ApJ},
  volume = {176},
  pages = {1},
  issn = {0004-637X},
  doi = {10.1086/151605},
  urldate = {2021-06-10}
}

@article{gururajan_gas_2025,
  title = {Gas Properties as a Function of Environment in the Proto-Supercluster {{Hyperion}} at z {$\sim$} 2.45},
  author = {Gururajan, G. and Cucciati, O. and Lemaux, B. C. and Talia, M. and Zamorani, G. and Pozzi, F. and Decarli, R. and Forrest, B. and Shen, L. and De Lucia, G. and Fontanot, F. and Bardelli, S. and Baxter, D. C. and Cassar{\`a}, L. P. and {Golden-Marx}, E. and Sikorski, D. and Shah, E. A. and Gal, R. R. and Giavalisco, M. and Giddings, F. and Hathi, N. P. and Hung, D. and Koekemoer, A. M. and Le Brun, V. and Lubin, L. M. and Tasca, L. A. M. and Tresse, L. and Vergani, D. and Zucca, E.},
  year = 2025,
  month = jun,
  journal = {A\&A},
  volume = {698},
  pages = {A312},
  issn = {0004-6361},
  doi = {10.1051/0004-6361/202553926},
  urldate = {2025-07-03}
}

@article{hansen_galaxy_2009,
  title = {The {{Galaxy Content}} of {{SDSS Clusters}} and {{Groups}}},
  author = {Hansen, Sarah M. and Sheldon, Erin S. and Wechsler, Risa H. and Koester, Benjamin P.},
  year = 2009,
  month = jul,
  journal = {ApJ},
  volume = {699},
  pages = {1333--1353},
  issn = {0004-637X},
  doi = {10.1088/0004-637X/699/2/1333},
  urldate = {2022-11-21}
}

@article{harikane_comprehensive_2023,
  title = {A {{Comprehensive Study}} of {{Galaxies}} at z 9-16 {{Found}} in the {{Early JWST Data}}: {{Ultraviolet Luminosity Functions}} and {{Cosmic Star Formation History}} at the {{Pre-reionization Epoch}}},
  shorttitle = {A {{Comprehensive Study}} of {{Galaxies}} at z 9-16 {{Found}} in the {{Early JWST Data}}},
  author = {Harikane, Yuichi and Ouchi, Masami and Oguri, Masamune and Ono, Yoshiaki and Nakajima, Kimihiko and Isobe, Yuki and Umeda, Hiroya and Mawatari, Ken and Zhang, Yechi},
  year = 2023,
  month = mar,
  journal = {ApJS},
  volume = {265},
  pages = {5},
  issn = {0067-0049},
  doi = {10.3847/1538-4365/acaaa9},
  urldate = {2025-03-04}
}

@article{harikane_jwst_2025,
  title = {{{JWST}}, {{ALMA}}, and {{Keck Spectroscopic Constraints}} on the {{UV Luminosity Functions}} at z {$\sim$} 7--14: {{Clumpiness}} and {{Compactness}} of the {{Brightest Galaxies}} in the {{Early Universe}}},
  shorttitle = {{{JWST}}, {{ALMA}}, and {{Keck Spectroscopic Constraints}} on the {{UV Luminosity Functions}} at z {$\sim$} 7--14},
  author = {Harikane, Yuichi and Inoue, Akio K. and Ellis, Richard S. and Ouchi, Masami and Nakazato, Yurina and Yoshida, Naoki and Ono, Yoshiaki and Sun, Fengwu and Sato, Riku A. and Ferrami, Giovanni and Fujimoto, Seiji and Kashikawa, Nobunari and McLeod, Derek J. and {P{\'e}rez-Gonz{\'a}lez}, Pablo G. and Sawicki, Marcin and Sugahara, Yuma and Xu, Yi and Yamanaka, Satoshi and Carnall, Adam C. and Cullen, Fergus and Dunlop, James S. and Egami, Eiichi and Grogin, Norman and Isobe, Yuki and Koekemoer, Anton M. and Laporte, Nicolas and Lee, Chien-Hsiu and Magee, Dan and Matsuo, Hiroshi and Matsuoka, Yoshiki and Mawatari, Ken and Nakajima, Kimihiko and Nakane, Minami and Tamura, Yoichi and Umeda, Hiroya and Yanagisawa, Hiroto},
  year = 2025,
  month = feb,
  journal = {ApJ},
  volume = {980},
  pages = {138},
  issn = {0004-637X},
  doi = {10.3847/1538-4357/ad9b2c},
  urldate = {2025-03-04}
}

@article{harikane_pure_2024,
  title = {Pure {{Spectroscopic Constraints}} on {{UV Luminosity Functions}} and {{Cosmic Star Formation History}} from 25 {{Galaxies}} at z Spec = 8.61-13.20 {{Confirmed}} with {{JWST}}/{{NIRSpec}}},
  author = {Harikane, Yuichi and Nakajima, Kimihiko and Ouchi, Masami and Umeda, Hiroya and Isobe, Yuki and Ono, Yoshiaki and Xu, Yi and Zhang, Yechi},
  year = 2024,
  month = jan,
  journal = {ApJ},
  volume = {960},
  pages = {56},
  issn = {0004-637X},
  doi = {10.3847/1538-4357/ad0b7e},
  urldate = {2025-03-04}
}

@article{harris_array_2020,
  title = {Array Programming with {{NumPy}}},
  author = {Harris, Charles R. and Millman, K. Jarrod and {van der Walt}, St{\'e}fan J. and Gommers, Ralf and Virtanen, Pauli and Cournapeau, David and Wieser, Eric and Taylor, Julian and Berg, Sebastian and Smith, Nathaniel J. and Kern, Robert and Picus, Matti and Hoyer, Stephan and {van Kerkwijk}, Marten H. and Brett, Matthew and Haldane, Allan and {del R{\'i}o}, Jaime Fern{\'a}ndez and Wiebe, Mark and Peterson, Pearu and {G{\'e}rard-Marchant}, Pierre and Sheppard, Kevin and Reddy, Tyler and Weckesser, Warren and Abbasi, Hameer and Gohlke, Christoph and Oliphant, Travis E.},
  year = 2020,
  month = sep,
  journal = {Nature},
  volume = {585},
  number = {7825},
  pages = {357--362},
  issn = {1476-4687},
  doi = {10.1038/s41586-020-2649-2},
  urldate = {2025-02-11},
  copyright = {2020 The Author(s)},
  langid = {english}
}

@article{hashimoto_influence_1998,
  title = {The {{Influence}} of {{Environment}} on the {{Star Formation Rates}} of {{Galaxies}}},
  author = {Hashimoto, Yasuhiro and Oemler, Jr., Augustus and Lin, Huan and Tucker, Douglas L.},
  year = 1998,
  month = may,
  journal = {ApJ},
  volume = {499},
  pages = {589--599},
  issn = {0004-637X},
  doi = {10.1086/305657},
  urldate = {2024-12-19}
}

@article{hasinger_deimos_2018,
  title = {The {{DEIMOS 10K Spectroscopic Survey Catalog}} of the {{COSMOS Field}}},
  author = {Hasinger, G. and Capak, P. and Salvato, M. and Barger, A. J. and Cowie, L. L. and Faisst, A. and Hemmati, S. and Kakazu, Y. and Kartaltepe, J. and Masters, D. and Mobasher, B. and Nayyeri, H. and Sanders, D. and Scoville, N. Z. and Suh, H. and Steinhardt, C. and Yang, Fengwei},
  year = 2018,
  month = may,
  journal = {ApJ},
  volume = {858},
  pages = {77},
  issn = {0004-637X},
  doi = {10.3847/1538-4357/aabacf},
  urldate = {2023-07-19}
}

@article{hernandez-toledo_structural_2005,
  title = {The {{Structural Properties}} of {{Isolated Galaxies}}, {{Spiral-Spiral Pairs}}, and {{Mergers}}: {{The Robustness}} of {{Galaxy Morphology}} during {{Secular Evolution}}},
  shorttitle = {The {{Structural Properties}} of {{Isolated Galaxies}}, {{Spiral-Spiral Pairs}}, and {{Mergers}}},
  author = {{Hern{\'a}ndez-Toledo}, H. M. and {Avila-Reese}, V. and Conselice, C. J. and Puerari, I.},
  year = 2005,
  month = feb,
  journal = {AJ},
  volume = {129},
  pages = {682--697},
  issn = {0004-6256},
  doi = {10.1086/427134},
  urldate = {2023-05-04}
}

@article{hester_ram_2006,
  title = {Ram {{Pressure Stripping}} in {{Clusters}} and {{Groups}}},
  author = {Hester, J. A.},
  year = 2006,
  month = aug,
  journal = {ApJ},
  volume = {647},
  pages = {910--921},
  issn = {0004-637X},
  doi = {10.1086/505614},
  urldate = {2021-06-10}
}

@article{hildebrandt_cfhtlens_2012,
  title = {{{CFHTLenS}}: Improving the Quality of Photometric Redshifts with Precision Photometry},
  shorttitle = {{{CFHTLenS}}},
  author = {Hildebrandt, H. and Erben, T. and Kuijken, K. and {van Waerbeke}, L. and Heymans, C. and Coupon, J. and Benjamin, J. and Bonnett, C. and Fu, L. and Hoekstra, H. and Kitching, T. D. and Mellier, Y. and Miller, L. and Velander, M. and Hudson, M. J. and Rowe, B. T. P. and Schrabback, T. and Semboloni, E. and Ben{\'i}tez, N.},
  year = 2012,
  month = apr,
  journal = {MNRAS},
  volume = {421},
  pages = {2355--2367},
  issn = {0035-8711},
  doi = {10.1111/j.1365-2966.2012.20468.x},
  urldate = {2025-03-03}
}

@article{hine_enhanced_2016,
  title = {An Enhanced Merger Fraction within the Galaxy Population of the {{SSA22}} Protocluster at z = 3.1},
  author = {Hine, N. K. and Geach, J. E. and Alexander, D. M. and Lehmer, B. D. and Chapman, S. C. and Matsuda, Y.},
  year = 2016,
  month = jan,
  journal = {MNRAS},
  volume = {455},
  pages = {2363--2370},
  issn = {0035-8711},
  doi = {10.1093/mnras/stv2448},
  urldate = {2024-09-21}
}

@article{hirschmann_galaxy_2016,
  title = {Galaxy Assembly, Stellar Feedback and Metal Enrichment: The View from the {{GAEA}} Model},
  shorttitle = {Galaxy Assembly, Stellar Feedback and Metal Enrichment},
  author = {Hirschmann, Michaela and De Lucia, Gabriella and Fontanot, Fabio},
  year = 2016,
  month = sep,
  journal = {MNRAS},
  volume = {461},
  pages = {1760--1785},
  issn = {0035-8711},
  doi = {10.1093/mnras/stw1318},
  urldate = {2025-03-01}
}

@article{hogg_dependence_2004,
  title = {The {{Dependence}} on {{Environment}} of the {{Color-Magnitude Relation}} of {{Galaxies}}},
  author = {Hogg, David W. and Blanton, Michael R. and Brinchmann, Jarle and Eisenstein, Daniel J. and Schlegel, David J. and Gunn, James E. and McKay, Timothy A. and Rix, Hans-Walter and Bahcall, Neta A. and Brinkmann, J. and Meiksin, Avery},
  year = 2004,
  month = jan,
  journal = {ApJL},
  volume = {601},
  pages = {L29--L32},
  issn = {0004-637X},
  doi = {10.1086/381749},
  urldate = {2021-06-10}
}

@article{horowitz_second_2022,
  title = {Second {{Data Release}} of the {{COSMOS Ly$\alpha$ Mapping}} and {{Tomography Observations}}: {{The First 3D Maps}} of the {{Detailed Cosmic Web}} at 2.05 {$<$} z {$<$} 2.55},
  shorttitle = {Second {{Data Release}} of the {{COSMOS Ly$\alpha$ Mapping}} and {{Tomography Observations}}},
  author = {Horowitz, Benjamin and Lee, Khee-Gan and Ata, Metin and M{\"u}ller, Thomas and Krolewski, Alex and Prochaska, J. Xavier and Hennawi, Joseph F. and White, Martin and Schlegel, David and Rich, R. Michael and Nugent, Peter E. and Suzuki, Nao and Kashino, Daichi and Koekemoer, Anton M. and Lemaux, Brian C.},
  year = 2022,
  month = dec,
  journal = {ApJS},
  volume = {263},
  pages = {27},
  issn = {0067-0049},
  doi = {10.3847/1538-4365/ac982d},
  urldate = {2025-08-24}
}

@article{hsieh_taiwan_2012,
  title = {{{THE TAIWAN ECDFS NEAR-INFRARED SURVEY}}: {{ULTRA-DEEP J AND KS IMAGING IN THE EXTENDED CHANDRA DEEP FIELD-SOUTH}}},
  shorttitle = {{{THE TAIWAN ECDFS NEAR-INFRARED SURVEY}}},
  author = {Hsieh, Bau-Ching and Wang, Wei-Hao and Hsieh, Chih-Chiang and Lin, Lihwai and Yan, Haojing and Lim, Jeremy and Ho, Paul T. P.},
  year = 2012,
  month = nov,
  journal = {ApJS},
  volume = {203},
  number = {2},
  pages = {23},
  issn = {0067-0049},
  doi = {10.1088/0067-0049/203/2/23},
  urldate = {2023-08-21},
  langid = {english}
}

@article{hung_discovering_2025,
  title = {Discovering {{Large-scale Structure}} at 2 {$<$} z {$<$} 5 in the {{C3VO Survey}}},
  author = {Hung, Denise and Lemaux, Brian C. and Cucciati, Olga and Forrest, Ben and Shah, Ekta A. and Gal, Roy R. and Giddings, Finn and Sikorski, Derek and {Golden-Marx}, Emmet and Lubin, Lori M. and Hathi, Nimish and Zamorani, Giovanni and Shen, Lu and Bardelli, Sandro and Cassar{\`a}, Letizia P. and De Lucia, Gabriella and Fontanot, Fabio and Garilli, Bianca and Guaita, Lucia and Hirschmann, Michaela Monika and Lee, Kyoung-Soo and Newman, Andrew B. and Ramakrishnan, Vandana and Vergani, Daniela and Xie, Lizhi and Zucca, Elena},
  year = 2025,
  month = feb,
  journal = {ApJ},
  volume = {980},
  pages = {155},
  issn = {0004-637X},
  doi = {10.3847/1538-4357/ada616},
  urldate = {2025-02-23}
}

@article{hung_establishing_2020,
  title = {Establishing a New Technique for Discovering Large-Scale Structure Using the {{ORELSE}} Survey},
  author = {Hung, D. and Lemaux, B. C. and Gal, R. R. and Tomczak, A. R. and Lubin, L. M. and Cucciati, O. and Pelliccia, D. and Shen, L. and Le F{\`e}vre, O. and Wu, P.-F. and Kocevski, D. D. and Mei, S. and Squires, G. K.},
  year = 2020,
  month = feb,
  journal = {MNRAS},
  volume = {491},
  pages = {5524--5554},
  issn = {0035-8711},
  doi = {10.1093/mnras/stz3164},
  urldate = {2021-06-06}
}

@article{hung_optical_2021,
  title = {An Optical Observational Cluster Mass Function at z {$\sim$} 1 with the {{ORELSE}} Survey},
  author = {Hung, D. and Lemaux, B. C. and Gal, R. R. and Tomczak, A. R. and Lubin, L. M. and Cucciati, O. and Pelliccia, D. and Shen, L. and Le F{\`e}vre, O. and Zamorani, G. and Wu, P.-F. and Kocevski, D. D. and Fassnacht, C. D. and Squires, G. K.},
  year = 2021,
  month = apr,
  journal = {MNRAS},
  volume = {502},
  pages = {3942--3954},
  issn = {0035-8711},
  doi = {10.1093/mnras/stab300},
  urldate = {2021-06-06}
}

@article{hunter_matplotlib_2007,
  title = {Matplotlib: {{A 2D Graphics Environment}}},
  shorttitle = {Matplotlib},
  author = {Hunter, John D.},
  year = 2007,
  month = may,
  journal = {Computing in Science \& Engineering},
  volume = {9},
  pages = {90--95},
  doi = {10.1109/MCSE.2007.55},
  urldate = {2025-02-11}
}

@article{ilbert_accurate_2006,
  title = {Accurate Photometric Redshifts for the {{CFHT}} Legacy Survey Calibrated Using the {{VIMOS VLT}} Deep Survey},
  author = {Ilbert, O. and Arnouts, S. and McCracken, H. J. and Bolzonella, M. and Bertin, E. and Le F{\`e}vre, O. and Mellier, Y. and Zamorani, G. and Pell{\`o}, R. and Iovino, A. and Tresse, L. and Le Brun, V. and Bottini, D. and Garilli, B. and Maccagni, D. and Picat, J. P. and Scaramella, R. and Scodeggio, M. and Vettolani, G. and Zanichelli, A. and Adami, C. and Bardelli, S. and Cappi, A. and Charlot, S. and Ciliegi, P. and Contini, T. and Cucciati, O. and Foucaud, S. and Franzetti, P. and Gavignaud, I. and Guzzo, L. and Marano, B. and Marinoni, C. and Mazure, A. and Meneux, B. and Merighi, R. and Paltani, S. and Pollo, A. and Pozzetti, L. and Radovich, M. and Zucca, E. and Bondi, M. and Bongiorno, A. and Busarello, G. and {de La Torre}, S. and Gregorini, L. and Lamareille, F. and Mathez, G. and Merluzzi, P. and Ripepi, V. and Rizzo, D. and Vergani, D.},
  year = 2006,
  month = oct,
  journal = {A\&A},
  volume = {457},
  pages = {841--856},
  issn = {0004-6361},
  doi = {10.1051/0004-6361:20065138},
  urldate = {2021-06-06}
}

@article{ilbert_cosmos_2009,
  title = {Cosmos {{Photometric Redshifts}} with 30-{{Bands}} for 2-Deg2},
  author = {Ilbert, O. and Capak, P. and Salvato, M. and Aussel, H. and McCracken, H. J. and Sanders, D. B. and Scoville, N. and Kartaltepe, J. and Arnouts, S. and Le Floc'h, E. and Mobasher, B. and Taniguchi, Y. and Lamareille, F. and Leauthaud, A. and Sasaki, S. and Thompson, D. and Zamojski, M. and Zamorani, G. and Bardelli, S. and Bolzonella, M. and Bongiorno, A. and Brusa, M. and Caputi, K. I. and Carollo, C. M. and Contini, T. and Cook, R. and Coppa, G. and Cucciati, O. and {de la Torre}, S. and {de Ravel}, L. and Franzetti, P. and Garilli, B. and Hasinger, G. and Iovino, A. and Kampczyk, P. and Kneib, J. -P. and Knobel, C. and Kovac, K. and Le Borgne, J. F. and Le Brun, V. and Le F{\`e}vre, O. and Lilly, S. and Looper, D. and Maier, C. and Mainieri, V. and Mellier, Y. and Mignoli, M. and Murayama, T. and Pell{\`o}, R. and Peng, Y. and {P{\'e}rez-Montero}, E. and Renzini, A. and Ricciardelli, E. and Schiminovich, D. and Scodeggio, M. and Shioya, Y. and Silverman, J. and Surace, J. and Tanaka, M. and Tasca, L. and Tresse, L. and Vergani, D. and Zucca, E.},
  year = 2009,
  month = jan,
  journal = {ApJ},
  volume = {690},
  pages = {1236--1249},
  issn = {0004-637X},
  doi = {10.1088/0004-637X/690/2/1236},
  urldate = {2023-08-19}
}

@article{ilbert_mass_2013,
  title = {Mass Assembly in Quiescent and Star-Forming Galaxies since z {$\simeq$} 4 from {{UltraVISTA}}},
  author = {Ilbert, O. and McCracken, H. J. and Le F{\`e}vre, O. and Capak, P. and Dunlop, J. and Karim, A. and Renzini, M. A. and Caputi, K. and Boissier, S. and Arnouts, S. and Aussel, H. and Comparat, J. and Guo, Q. and Hudelot, P. and Kartaltepe, J. and Kneib, J. P. and Krogager, J. K. and Le Floc'h, E. and Lilly, S. and Mellier, Y. and {Milvang-Jensen}, B. and Moutard, T. and Onodera, M. and Richard, J. and Salvato, M. and Sanders, D. B. and Scoville, N. and Silverman, J. D. and Taniguchi, Y. and Tasca, L. and Thomas, R. and Toft, S. and Tresse, L. and Vergani, D. and Wolk, M. and Zirm, A.},
  year = 2013,
  month = aug,
  journal = {A\&A},
  volume = {556},
  pages = {A55},
  issn = {0004-6361},
  doi = {10.1051/0004-6361/201321100},
  urldate = {2023-08-15}
}

@article{kampczyk_environmental_2013,
  title = {Environmental {{Effects}} in the {{Interaction}} and {{Merging}} of {{Galaxies}} in {{zCOSMOS}}},
  author = {Kampczyk, P. and Lilly, S. J. and {de Ravel}, L. and Le F{\`e}vre, O. and Bolzonella, M. and Carollo, C. M. and Diener, C. and Knobel, C. and Kova{\v c}, K. and Maier, C. and Renzini, A. and Sargent, M. T. and Vergani, D. and Abbas, U. and Bardelli, S. and Bongiorno, A. and Bordoloi, R. and Caputi, K. and Contini, T. and Coppa, G. and Cucciati, O. and {de la Torre}, S. and Franzetti, P. and Garilli, B. and Iovino, A. and Kneib, J.-P. and Koekemoer, A. M. and Lamareille, F. and Le Borgne, J.-F. and Le Brun, V. and Leauthaud, A. and Mainieri, V. and Mignoli, M. and Pello, R. and Peng, Y. and Perez Montero, E. and Ricciardelli, E. and Scodeggio, M. and Silverman, J. D. and Tanaka, M. and Tasca, L. and Tresse, L. and Zamorani, G. and Zucca, E. and Bottini, D. and Cappi, A. and Cassata, P. and Cimatti, A. and Fumana, M. and Guzzo, L. and Kartaltepe, J. and Marinoni, C. and McCracken, H. J. and Memeo, P. and Meneux, B. and Oesch, P. and Porciani, C. and Pozzetti, L. and Scaramella, R.},
  year = 2013,
  month = jan,
  journal = {ApJ},
  volume = {762},
  number = {1},
  pages = {43},
  issn = {0004-637X},
  doi = {10.1088/0004-637X/762/1/43},
  urldate = {2025-01-02},
  langid = {english}
}

@article{kauffmann_environmental_2004,
  title = {The Environmental Dependence of the Relations between Stellar Mass, Structure, Star Formation and Nuclear Activity in Galaxies},
  author = {Kauffmann, Guinevere and White, Simon D. M. and Heckman, Timothy M. and M{\'e}nard, Brice and Brinchmann, Jarle and Charlot, St{\'e}phane and Tremonti, Christy and Brinkmann, Jon},
  year = 2004,
  month = sep,
  journal = {MNRAS},
  volume = {353},
  pages = {713--731},
  issn = {0035-8711},
  doi = {10.1111/j.1365-2966.2004.08117.x},
  urldate = {2022-11-22}
}

@ARTICLE{khostovan_cosmos_2025,
       author = {{Khostovan}, Ali Ahmad and {Kartaltepe}, Jeyhan S. and {Salvato}, Mara and {Ilbert}, Olivier and {Casey}, Caitlin M. and {Algera}, Hiddo and {Antwi-Danso}, Jacqueline and {Battisti}, Andrew and {Brinch}, Malte and {Brusa}, Marcella and {Calabro}, Antonello and {Capak}, Peter L. and {Chartab}, Nima and {Cooper}, Olivia R. and {Cox}, Isa G. and {Darvish}, Behnam and {Drakos}, Nicole E. and {Faisst}, Andreas L. and {George}, Matthew R. and {Gozaliasl}, Ghassem and {Harish}, Santosh and {Hasinger}, Gunther and {Hatamnia}, Hossein and {Iovino}, Angela and {Jin}, Shuowen and {Kashino}, Daichi and {Koekemoer}, Anton M. and {Laishram}, Ronaldo and {Lee}, Khee-Gan and {Lertprasertpong}, Jitrapon and {Lilly}, Simon J. and {Liu}, Daizhong and {Masters}, Daniel C. and {Mobasher}, Bahram and {Nagao}, Tohru and {Onodera}, Masato and {Peng}, Yingjie and {Sanders}, David B. and {Sanders}, Ryan L. and {Sattari}, Zahra and {Scoville}, Nick and {Shah}, Ekta A. and {Silverman}, John D. and {Suzuki}, Nao and {Taamoli}, Sina and {Tanaka}, Masayuki and {Tasca}, Lidia A.~M. and {Toft}, Sune and {Toni}, Greta and {Trakhtenbrot}, Benny and {Trump}, Jonathan R. and {Vaccari}, Mattia and {Valentino}, Francesco and {Vanderhoof}, Brittany N. and {Weaver}, John R. and {Yun}, Min S. and {Zavala}, Jorge A.},
        title = "{COSMOS Spectroscopic Redshift Compilation (First Data Release): 488k Redshifts Encompassing Two Decades of Spectroscopy}",
      journal = {arXiv e-prints},
         year = 2025,
        month = feb,
          eid = {arXiv:2503.00120},
        pages = {[arXiv:2503.00120]},
          doi = {10.48550/arXiv.2503.00120},
archivePrefix = {arXiv},
       eprint = {2503.00120},
 primaryClass = {astro-ph.GA},
       adsurl = {https://ui.adsabs.harvard.edu/abs/2025arXiv250300120K}
}

@article{kitzbichler_calibration_2008,
  title = {A Calibration of the Relation between the Abundance of Close Galaxy Pairs and the Rate of Galaxy Mergers},
  author = {Kitzbichler, M. G. and White, S. D. M.},
  year = 2008,
  month = dec,
  journal = {MNRAS},
  volume = {391},
  pages = {1489--1498},
  issn = {0035-8711},
  doi = {10.1111/j.1365-2966.2008.13873.x},
  urldate = {2024-09-21}
}

@article{koekemoer_cosmos_2007,
  title = {The {{COSMOS Survey}}: {{Hubble Space Telescope Advanced Camera}} for {{Surveys Observations}} and {{Data Processing}}},
  shorttitle = {The {{COSMOS Survey}}},
  author = {Koekemoer, A. M. and Aussel, H. and Calzetti, D. and Capak, P. and Giavalisco, M. and Kneib, J. -P. and Leauthaud, A. and Le F{\`e}vre, O. and McCracken, H. J. and Massey, R. and Mobasher, B. and Rhodes, J. and Scoville, N. and Shopbell, P. L.},
  year = 2007,
  month = sep,
  journal = {ApJS},
  volume = {172},
  pages = {196--202},
  issn = {0067-0049},
  doi = {10.1086/520086},
  urldate = {2021-07-05}
}

@article{koyama_evolution_2013,
  title = {On the Evolution and Environmental Dependence of the Star Formation Rate versus Stellar Mass Relation since z {$\sim$} 2},
  author = {Koyama, Yusei and Smail, Ian and Kurk, Jaron and Geach, James E. and Sobral, David and Kodama, Tadayuki and Nakata, Fumiaki and Swinbank, A. M. and Best, Philip N. and Hayashi, Masao and Tadaki, Ken-ichi},
  year = 2013,
  month = sep,
  journal = {MNRAS},
  volume = {434},
  pages = {423--436},
  issn = {0035-8711},
  doi = {10.1093/mnras/stt1035},
  urldate = {2025-02-07}
}

@article{koyama_planck-selected_2021,
  title = {A {{Planck-selected}} Dusty Proto-Cluster at z = 2.16 Associated with a Strong Overdensity of Massive {{H$\alpha$-emitting}} Galaxies},
  author = {Koyama, Yusei and Polletta, Maria del Carmen and Tanaka, Ichi and Kodama, Tadayuki and Dole, Herv{\'e} and Soucail, Genevi{\`e}ve and Frye, Brenda and Lehnert, Matthew and Scodeggio, Marco},
  year = 2021,
  month = may,
  journal = {MNRAS},
  volume = {503},
  pages = {L1--L5},
  issn = {0035-8711},
  doi = {10.1093/mnrasl/slab013},
  urldate = {2025-02-08}
}

@article{kraljic_galaxies_2019,
  title = {Galaxies Flowing in the Oriented Saddle Frame of the Cosmic Web},
  author = {Kraljic, K. and Pichon, C. and Dubois, Y. and Codis, S. and Cadiou, C. and Devriendt, J. and Musso, M. and Welker, C. and Arnouts, S. and Hwang, H. S. and Laigle, C. and Peirani, S. and Slyz, A. and Treyer, M. and Vibert, D.},
  year = 2019,
  month = mar,
  journal = {MNRAS},
  volume = {483},
  pages = {3227--3254},
  issn = {0035-8711},
  doi = {10.1093/mnras/sty3216},
  urldate = {2024-12-31}
}

@article{kriek_mosfire_2015,
  title = {The {{MOSFIRE Deep Evolution Field}} ({{MOSDEF}}) {{Survey}}: {{Rest-frame Optical Spectroscopy}} for \textasciitilde 1500 {{H-selected Galaxies}} at 1.37 {$<$} z {$<$} 3.8},
  shorttitle = {The {{MOSFIRE Deep Evolution Field}} ({{MOSDEF}}) {{Survey}}},
  author = {Kriek, Mariska and Shapley, Alice E. and Reddy, Naveen A. and Siana, Brian and Coil, Alison L. and Mobasher, Bahram and Freeman, William R. and {de Groot}, Laura and Price, Sedona H. and Sanders, Ryan and Shivaei, Irene and Brammer, Gabriel B. and Momcheva, Ivelina G. and Skelton, Rosalind E. and {van Dokkum}, Pieter G. and Whitaker, Katherine E. and Aird, James and Azadi, Mojegan and Kassis, Marc and Bullock, James S. and Conroy, Charlie and Dav{\'e}, Romeel and Kere{\v s}, Du{\v s}an and Krumholz, Mark},
  year = 2015,
  month = jun,
  journal = {ApJS},
  volume = {218},
  pages = {15},
  issn = {0067-0049},
  doi = {10.1088/0067-0049/218/2/15},
  urldate = {2025-08-24}
}

@article{kulkarni_evolution_2019,
  title = {Evolution of the {{AGN UV}} Luminosity Function from Redshift 7.5},
  author = {Kulkarni, Girish and Worseck, G{\'a}bor and Hennawi, Joseph F.},
  year = 2019,
  month = sep,
  journal = {MNRAS},
  volume = {488},
  pages = {1035--1065},
  issn = {0035-8711},
  doi = {10.1093/mnras/stz1493},
  urldate = {2024-11-05}
}

@article{laigle_cosmos2015_2016,
  title = {The {{COSMOS2015 Catalog}}: {{Exploring}} the 1 {$<$} z {$<$} 6 {{Universe}} with {{Half}} a {{Million Galaxies}}},
  shorttitle = {The {{COSMOS2015 Catalog}}},
  author = {Laigle, C. and McCracken, H. J. and Ilbert, O. and Hsieh, B. C. and Davidzon, I. and Capak, P. and Hasinger, G. and Silverman, J. D. and Pichon, C. and Coupon, J. and Aussel, H. and Le Borgne, D. and Caputi, K. and Cassata, P. and Chang, Y. -Y. and Civano, F. and Dunlop, J. and Fynbo, J. and Kartaltepe, J. S. and Koekemoer, A. and Le F{\`e}vre, O. and Le Floc'h, E. and Leauthaud, A. and Lilly, S. and Lin, L. and Marchesi, S. and {Milvang-Jensen}, B. and Salvato, M. and Sanders, D. B. and Scoville, N. and Smolcic, V. and Stockmann, M. and Taniguchi, Y. and Tasca, L. and Toft, S. and Vaccari, Mattia and Zabl, J.},
  year = 2016,
  month = jun,
  journal = {ApJS},
  volume = {224},
  pages = {24},
  issn = {0067-0049},
  doi = {10.3847/0067-0049/224/2/24},
  urldate = {2023-08-18}
}

@article{lambas_galaxy_2003,
  title = {Galaxy Pairs in the {{2dF}} Survey - {{I}}. {{Effects}} of Interactions on Star Formation in the Field},
  author = {Lambas, Diego G. and Tissera, Patricia B. and Alonso, M. Sol and Coldwell, Georgina},
  year = 2003,
  month = dec,
  journal = {MNRAS},
  volume = {346},
  pages = {1189--1196},
  issn = {0035-8711},
  doi = {10.1111/j.1365-2966.2003.07179.x},
  urldate = {2023-05-05}
}

@article{lambas_galaxy_2012,
  title = {Galaxy Interactions. {{I}}. {{Major}} and Minor Mergers},
  author = {Lambas, D. G. and Alonso, S. and Mesa, V. and O'Mill, A. L.},
  year = 2012,
  month = mar,
  journal = {A\&A},
  volume = {539},
  pages = {A45},
  issn = {0004-6361},
  doi = {10.1051/0004-6361/201117900},
  urldate = {2023-05-05}
}

@article{le_fevre_vimos_2013,
  title = {The {{VIMOS VLT Deep Survey}} Final Data Release: A Spectroscopic Sample of 35 016 Galaxies and {{AGN}} out to z \textasciitilde{} 6.7 Selected with 17.5 {$\leq$} {{iAB}} {$\leq$} 24.75},
  shorttitle = {The {{VIMOS VLT Deep Survey}} Final Data Release},
  author = {Le F{\`e}vre, O. and Cassata, P. and Cucciati, O. and Garilli, B. and Ilbert, O. and Le Brun, V. and Maccagni, D. and Moreau, C. and Scodeggio, M. and Tresse, L. and Zamorani, G. and Adami, C. and Arnouts, S. and Bardelli, S. and Bolzonella, M. and Bondi, M. and Bongiorno, A. and Bottini, D. and Cappi, A. and Charlot, S. and Ciliegi, P. and Contini, T. and {de la Torre}, S. and Foucaud, S. and Franzetti, P. and Gavignaud, I. and Guzzo, L. and Iovino, A. and Lemaux, B. and {L{\'o}pez-Sanjuan}, C. and McCracken, H. J. and Marano, B. and Marinoni, C. and Mazure, A. and Mellier, Y. and Merighi, R. and Merluzzi, P. and Paltani, S. and Pell{\`o}, R. and Pollo, A. and Pozzetti, L. and Scaramella, R. and Tasca, L. and Vergani, D. and Vettolani, G. and Zanichelli, A. and Zucca, E.},
  year = 2013,
  month = nov,
  journal = {A\&A},
  volume = {559},
  pages = {A14},
  issn = {0004-6361},
  doi = {10.1051/0004-6361/201322179},
  urldate = {2023-08-13}
}

@article{le_fevre_vimos_2015,
  title = {The {{VIMOS Ultra-Deep Survey}}: \textasciitilde 10 000 Galaxies with Spectroscopic Redshifts to Study Galaxy Assembly at Early Epochs 2 {$<$} z {$\simeq$} 6},
  shorttitle = {The {{VIMOS Ultra-Deep Survey}}},
  author = {Le F{\`e}vre, O. and Tasca, L. A. M. and Cassata, P. and Garilli, B. and Le Brun, V. and Maccagni, D. and Pentericci, L. and Thomas, R. and Vanzella, E. and Zamorani, G. and Zucca, E. and Amorin, R. and Bardelli, S. and Capak, P. and Cassar{\`a}, L. and Castellano, M. and Cimatti, A. and Cuby, J. G. and Cucciati, O. and {de la Torre}, S. and Durkalec, A. and Fontana, A. and Giavalisco, M. and Grazian, A. and Hathi, N. P. and Ilbert, O. and Lemaux, B. C. and Moreau, C. and Paltani, S. and Ribeiro, B. and Salvato, M. and Schaerer, D. and Scodeggio, M. and Sommariva, V. and Talia, M. and Taniguchi, Y. and Tresse, L. and Vergani, D. and Wang, P. W. and Charlot, S. and Contini, T. and Fotopoulou, S. and {L{\'o}pez-Sanjuan}, C. and Mellier, Y. and Scoville, N.},
  year = 2015,
  month = apr,
  journal = {A\&A},
  volume = {576},
  pages = {A79},
  issn = {0004-6361},
  doi = {10.1051/0004-6361/201423829},
  urldate = {2023-08-21}
}

@article{lee_first_2018,
  title = {First {{Data Release}} of the {{COSMOS Ly$\alpha$ Mapping}} and {{Tomography Observations}}: {{3D Ly$\alpha$ Forest Tomography}} at 2.05 {$<$} z {$<$} 2.55},
  shorttitle = {First {{Data Release}} of the {{COSMOS Ly$\alpha$ Mapping}} and {{Tomography Observations}}},
  author = {Lee, Khee-Gan and Krolewski, Alex and White, Martin and Schlegel, David and Nugent, Peter E. and Hennawi, Joseph F. and M{\"u}ller, Thomas and Pan, Richard and Prochaska, J. Xavier and {Font-Ribera}, Andreu and Suzuki, Nao and Glazebrook, Karl and Kacprzak, Glenn G. and Kartaltepe, Jeyhan S. and Koekemoer, Anton M. and Le F{\`e}vre, Olivier and Lemaux, Brian C. and Maier, Christian and Nanayakkara, Themiya and Rich, R. Michael and Sanders, D. B. and Salvato, Mara and Tasca, Lidia and Tran, Kim-Vy H.},
  year = 2018,
  month = aug,
  journal = {ApJS},
  volume = {237},
  pages = {31},
  issn = {0067-0049},
  doi = {10.3847/1538-4365/aace58},
  urldate = {2025-08-24}
}

@article{lee_shadow_2016,
  title = {Shadow of a {{Colossus}}: {{A}} z = 2.44 {{Galaxy Protocluster Detected}} in {{3D Ly$\alpha$ Forest Tomographic Mapping}} of the {{COSMOS Field}}},
  shorttitle = {Shadow of a {{Colossus}}},
  author = {Lee, Khee-Gan and Hennawi, Joseph F. and White, Martin and Prochaska, J. Xavier and {Font-Ribera}, Andreu and Schlegel, David J. and Rich, R. Michael and Suzuki, Nao and Stark, Casey W. and Le F{\`e}vre, Olivier and Nugent, Peter E. and Salvato, Mara and Zamorani, Gianni},
  year = 2016,
  month = feb,
  journal = {ApJ},
  volume = {817},
  pages = {160},
  issn = {0004-637X},
  doi = {10.3847/0004-637X/817/2/160},
  urldate = {2025-02-08}
}

@article{lefevre_commissioning_2003,
       author = {{Le F{\`e}vre}, Oliver and {Saisse}, Michel and {Mancini}, Dario and {Brau-Nogue}, Sylvie and {Caputi}, Oreste and {Castinel}, Louis and {D'Odorico}, Sandro and {Garilli}, Bianca and {Kissler-Patig}, Markus and {Lucuix}, Christian and {Mancini}, Guido and {Pauget}, Alain and {Sciarretta}, Giovanni and {Scodeggio}, Marco and {Tresse}, Laurence and {Vettolani}, Gianpaolo},
        title = "{Commissioning and performances of the VLT-VIMOS instrument}",
    journal = {in Instrument Design and Performance for Optical/Infrared Ground-based Telescopes, eds. M. Iye, \& A. F. M. Moorwood, Proc. SPIE},
         year = 2003,
       editor = {{Iye}, Masanori and {Moorwood}, Alan F.~M.},
       series = {Society of Photo-Optical Instrumentation Engineers (SPIE) Conference Series},
       volume = {4841},
        month = mar,
        pages = {1670-1681},
          doi = {10.1117/12.460959},
       adsurl = {https://ui.adsabs.harvard.edu/abs/2003SPIE.4841.1670L}
}

@article{lehmer_extended_2005,
  title = {The {{Extended Chandra Deep Field-South Survey}}: {{Chandra Point-Source Catalogs}}},
  shorttitle = {The {{Extended Chandra Deep Field-South Survey}}},
  author = {Lehmer, B. D. and Brandt, W. N. and Alexander, D. M. and Bauer, F. E. and Schneider, D. P. and Tozzi, P. and Bergeron, J. and Garmire, G. P. and Giacconi, R. and Gilli, R. and Hasinger, G. and Hornschemeier, A. E. and Koekemoer, A. M. and Mainieri, V. and Miyaji, T. and Nonino, M. and Rosati, P. and Silverman, J. D. and Szokoly, G. and Vignali, C.},
  year = 2005,
  month = nov,
  journal = {ApJS},
  volume = {161},
  pages = {21--40},
  issn = {0067-0049},
  doi = {10.1086/444590},
  urldate = {2024-10-31}
}

@article{lemaux_assembly_2012,
  title = {The {{Assembly}} of the {{Red Sequence}} at z \textasciitilde{} 1: {{The Color}} and {{Spectral Properties}} of {{Galaxies}} in the {{Cl1604 Supercluster}}},
  shorttitle = {The {{Assembly}} of the {{Red Sequence}} at z \textasciitilde{} 1},
  author = {Lemaux, B. C. and Gal, R. R. and Lubin, L. M. and Kocevski, D. D. and Fassnacht, C. D. and McGrath, E. J. and Squires, G. K. and Surace, J. A. and Lacy, M.},
  year = 2012,
  month = feb,
  journal = {ApJ},
  volume = {745},
  pages = {106},
  issn = {0004-637X},
  doi = {10.1088/0004-637X/745/2/106},
  urldate = {2022-11-21}
}

@article{lemaux_chronos_2017,
  title = {Chronos and {{KAIROS}}: {{MOSFIRE}} Observations of Post-Starburst Galaxies in z {$\sim$} 1 Clusters and Groups},
  shorttitle = {Chronos and {{KAIROS}}},
  author = {Lemaux, B. C. and Tomczak, A. R. and Lubin, L. M. and Wu, P.-F. and Gal, R. R. and Rumbaugh, N. and Kocevski, D. D. and Squires, G. K.},
  year = 2017,
  month = nov,
  journal = {MNRAS},
  volume = {472},
  pages = {419--438},
  issn = {0035-8711},
  doi = {10.1093/mnras/stx1579},
  urldate = {2021-06-06}
}

@article{lemaux_persistence_2019,
  title = {Persistence of the Colour-Density Relation and Efficient Environmental Quenching to z {$\sim$} 1.4},
  author = {Lemaux, B. C. and Tomczak, A. R. and Lubin, L. M. and Gal, R. R. and Shen, L. and Pelliccia, D. and Wu, P.-F. and Hung, D. and Mei, S. and Le F{\`e}vre, O. and Rumbaugh, N. and Kocevski, D. D. and Squires, G. K.},
  year = 2019,
  month = nov,
  journal = {MNRAS},
  volume = {490},
  pages = {1231--1254},
  issn = {0035-8711},
  doi = {10.1093/mnras/stz2661},
  urldate = {2021-06-06}
}

@article{lemaux_vimos_2014,
  title = {{{VIMOS Ultra-Deep Survey}} ({{VUDS}}): {{Witnessing}} the Assembly of a Massive Cluster at z \textasciitilde{} 3.3},
  shorttitle = {{{VIMOS Ultra-Deep Survey}} ({{VUDS}})},
  author = {Lemaux, B. C. and Cucciati, O. and Tasca, L. a. M. and Le F{\`e}vre, O. and Zamorani, G. and Cassata, P. and Garilli, B. and Le Brun, V. and Maccagni, D. and Pentericci, L. and Thomas, R. and Vanzella, E. and Zucca, E. and Amor{\'i}n, R. and Bardelli, S. and Capak, P. and Cassar{\`a}, L. P. and Castellano, M. and Cimatti, A. and Cuby, J. G. and {de la Torre}, S. and Durkalec, A. and Fontana, A. and Giavalisco, M. and Grazian, A. and Hathi, N. P. and Ilbert, O. and Moreau, C. and Paltani, S. and Ribeiro, B. and Salvato, M. and Schaerer, D. and Scodeggio, M. and Sommariva, V. and Talia, M. and Taniguchi, Y. and Tresse, L. and Vergani, D. and Wang, P. W. and Charlot, S. and Contini, T. and Fotopoulou, S. and Gal, R. R. and Kocevski, D. D. and {L{\'o}pez-Sanjuan}, C. and Lubin, L. M. and Mellier, Y. and Sadibekova, T. and Scoville, N.},
  year = 2014,
  month = dec,
  journal = {A\&A},
  volume = {572},
  pages = {A41},
  issn = {0004-6361},
  doi = {10.1051/0004-6361/201423828},
  urldate = {2022-12-12},
  langid = {english}
}

@article{lemaux_vimos_2018,
  title = {The {{VIMOS Ultra-Deep Survey}}: {{Emerging}} from the Dark, a Massive Proto-Cluster at z 4.57},
  shorttitle = {The {{VIMOS Ultra-Deep Survey}}},
  author = {Lemaux, B. C. and Le F{\`e}vre, O. and Cucciati, O. and Ribeiro, B. and Tasca, L. A. M. and Zamorani, G. and Ilbert, O. and Thomas, R. and Bardelli, S. and Cassata, P. and Hathi, N. P. and Pforr, J. and Smol{\v c}i{\'c}, V. and Delvecchio, I. and Novak, M. and Berta, S. and McCracken, H. J. and Koekemoer, A. and Amor{\'i}n, R. and Garilli, B. and Maccagni, D. and Schaerer, D. and Zucca, E.},
  year = 2018,
  month = jul,
  journal = {A\&A},
  volume = {615},
  pages = {A77},
  issn = {0004-6361},
  doi = {10.1051/0004-6361/201730870},
  urldate = {2023-08-21}
}

@article{lemaux_vimos_2022,
  title = {The {{VIMOS Ultra Deep Survey}}: {{The}} Reversal of the Star-Formation Rate - Density Relation at 2 {$<$} z {$<$} 5},
  shorttitle = {The {{VIMOS Ultra Deep Survey}}},
  author = {Lemaux, B. C. and Cucciati, O. and Le F{\`e}vre, O. and Zamorani, G. and Lubin, L. M. and Hathi, N. and Ilbert, O. and Pelliccia, D. and Amor{\'i}n, R. and Bardelli, S. and Cassata, P. and Gal, R. R. and Garilli, B. and Guaita, L. and Giavalisco, M. and Hung, D. and Koekemoer, A. and Maccagni, D. and Pentericci, L. and Ribeiro, B. and Schaerer, D. and Shah, E. and Shen, L. and Staab, P. and Talia, M. and Thomas, R. and Tomczak, A. R. and Tresse, L. and Vanzella, E. and Vergani, D. and Zucca, E.},
  year = 2022,
  month = jun,
  journal = {A\&A},
  volume = {662},
  pages = {A33},
  issn = {0004-6361},
  doi = {10.1051/0004-6361/202039346},
  urldate = {2023-05-06}
}

@article{lewis_2df_2002,
  title = {The {{2dF Galaxy Redshift Survey}}: The Environmental Dependence of Galaxy Star Formation Rates near Clusters},
  shorttitle = {The {{2dF Galaxy Redshift Survey}}},
  author = {Lewis, Ian and Balogh, Michael and De Propris, Roberto and Couch, Warrick and Bower, Richard and Offer, Alison and {Bland-Hawthorn}, Joss and Baldry, Ivan K. and Baugh, Carlton and Bridges, Terry and Cannon, Russell and Cole, Shaun and Colless, Matthew and Collins, Chris and Cross, Nicholas and Dalton, Gavin and Driver, Simon P. and Efstathiou, George and Ellis, Richard S. and Frenk, Carlos S. and Glazebrook, Karl and Hawkins, Edward and Jackson, Carole and Lahav, Ofer and Lumsden, Stuart and Maddox, Steve and Madgwick, Darren and Norberg, Peder and Peacock, John A. and Percival, Will and Peterson, Bruce A. and Sutherland, Will and Taylor, Keith},
  year = 2002,
  month = aug,
  journal = {MNRAS},
  volume = {334},
  pages = {673--683},
  issn = {0035-8711},
  doi = {10.1046/j.1365-8711.2002.05558.x},
  urldate = {2021-06-09}
}

@article{lidman_evidence_2012,
  title = {Evidence for Significant Growth in the Stellar Mass of Brightest Cluster Galaxies over the Past 10 Billion Years},
  author = {Lidman, C. and Suherli, J. and Muzzin, A. and Wilson, G. and Demarco, R. and Brough, S. and Rettura, A. and Cox, J. and DeGroot, A. and Yee, H. K. C. and Gilbank, D. and Hoekstra, H. and Balogh, M. and Ellingson, E. and Hicks, A. and Nantais, J. and Noble, A. and Lacy, M. and Surace, J. and Webb, T.},
  year = 2012,
  month = nov,
  journal = {MNRAS},
  volume = {427},
  pages = {550--568},
  issn = {0035-8711},
  doi = {10.1111/j.1365-2966.2012.21984.x},
  urldate = {2025-01-03}
}

@article{lilly_zcosmos_2007,
  title = {{{zCOSMOS}}: {{A Large VLT}}/{{VIMOS Redshift Survey Covering}} 0 {$<$} z {$<$} 3 in the {{COSMOS Field}}},
  shorttitle = {{{zCOSMOS}}},
  author = {Lilly, S. J. and Le F{\`e}vre, O. and Renzini, A. and Zamorani, G. and Scodeggio, M. and Contini, T. and Carollo, C. M. and Hasinger, G. and Kneib, J. -P. and Iovino, A. and Le Brun, V. and Maier, C. and Mainieri, V. and Mignoli, M. and Silverman, J. and Tasca, L. A. M. and Bolzonella, M. and Bongiorno, A. and Bottini, D. and Capak, P. and Caputi, K. and Cimatti, A. and Cucciati, O. and Daddi, E. and Feldmann, R. and Franzetti, P. and Garilli, B. and Guzzo, L. and Ilbert, O. and Kampczyk, P. and Kovac, K. and Lamareille, F. and Leauthaud, A. and Le Borgne, J. -F. and McCracken, H. J. and Marinoni, C. and Pello, R. and Ricciardelli, E. and Scarlata, C. and Vergani, D. and Sanders, D. B. and Schinnerer, E. and Scoville, N. and Taniguchi, Y. and Arnouts, S. and Aussel, H. and Bardelli, S. and Brusa, M. and Cappi, A. and Ciliegi, P. and Finoguenov, A. and Foucaud, S. and Franceschini, A. and Halliday, C. and Impey, C. and Knobel, C. and Koekemoer, A. and Kurk, J. and Maccagni, D. and Maddox, S. and Marano, B. and Marconi, G. and Meneux, B. and Mobasher, B. and Moreau, C. and Peacock, J. A. and Porciani, C. and Pozzetti, L. and Scaramella, R. and Schiminovich, D. and Shopbell, P. and Smail, I. and Thompson, D. and Tresse, L. and Vettolani, G. and Zanichelli, A. and Zucca, E.},
  year = 2007,
  month = sep,
  journal = {ApJS},
  volume = {172},
  pages = {70--85},
  issn = {0067-0049},
  doi = {10.1086/516589},
  urldate = {2022-11-22}
}

@article{lim_flamingo_2024,
  title = {The {{FLAMINGO}} Simulation View of Cluster Progenitors Observed in the Epoch of Reionization with {{JWST}}},
  author = {Lim, Seunghwan and Tacchella, Sandro and Schaye, Joop and Schaller, Matthieu and Helton, Jakob M. and Kugel, Roi and Maiolino, Roberto},
  year = 2024,
  month = aug,
  journal = {MNRAS},
  volume = {532},
  pages = {4551--4569},
  issn = {0035-8711},
  doi = {10.1093/mnras/stae1790},
  urldate = {2025-02-07}
}

@article{lin_pan-starrs1_2014,
  title = {The {{Pan-STARRS1 Medium-Deep Survey}}: {{The Role}} of {{Galaxy Group Environment}} in the {{Star Formation Rate}} versus {{Stellar Mass Relation}} and {{Quiescent Fraction}} out to z \textasciitilde{} 0.8},
  shorttitle = {The {{Pan-STARRS1 Medium-Deep Survey}}},
  author = {Lin, Lihwai and Jian, Hung-Yu and Foucaud, Sebastien and Norberg, Peder and Bower, R. G. and Cole, Shaun and {Arnalte-Mur}, Pablo and Chen, Chin-Wei and Coupon, Jean and Hsieh, Bau-Ching and Heinis, Sebastien and Phleps, Stefanie and Chen, Wen-Ping and Lee, Chien-Hsiu and Burgett, William and Chambers, K. C. and Denneau, L. and Draper, P. and Flewelling, H. and Hodapp, K. W. and Huber, M. E. and Kaiser, N. and Kudritzki, R. -P. and Magnier, E. A. and Metcalfe, N. and Price, Paul A. and Tonry, J. L. and Wainscoat, R. J. and Waters, C.},
  year = 2014,
  month = feb,
  journal = {ApJ},
  volume = {782},
  pages = {33},
  issn = {0004-637X},
  doi = {10.1088/0004-637X/782/1/33},
  urldate = {2025-02-07}
}

@article{lin_where_2010,
  title = {Where Do {{Wet}}, {{Dry}}, and {{Mixed Galaxy Mergers Occur}}? {{A Study}} of the {{Environments}} of {{Close Galaxy Pairs}} in the {{DEEP2 Galaxy Redshift Survey}}},
  shorttitle = {Where Do {{Wet}}, {{Dry}}, and {{Mixed Galaxy Mergers Occur}}?},
  author = {Lin, Lihwai and Cooper, Michael C. and Jian, Hung-Yu and Koo, David C. and Patton, David R. and Yan, Renbin and Willmer, Christopher N. A. and Coil, Alison L. and Chiueh, Tzihong and Croton, Darren J. and Gerke, Brian F. and Lotz, Jennifer and Guhathakurta, Puragra and Newman, Jeffrey A.},
  year = 2010,
  month = aug,
  journal = {ApJ},
  volume = {718},
  pages = {1158--1170},
  issn = {0004-637X},
  doi = {10.1088/0004-637X/718/2/1158},
  urldate = {2025-01-02}
}

@article{liu_what_2023,
  title = {What Boost Galaxy Mergers in Two Massive Galaxy Protoclusters at z = 2.24?},
  author = {Liu, Shuang and Zheng, Xian Zhong and Shi, Dong Dong and Cai, Zheng and Fan, Xiaohui and Wang, Xin and Yuan, Qirong and Xu, Haiguang and Pan, Zhizheng and Liu, Wenhao and Qin, Jianbo and Zhang, Yuheng and Wen, Run},
  year = 2023,
  month = aug,
  journal = {MNRAS},
  volume = {523},
  pages = {2422--2439},
  issn = {0035-8711},
  doi = {10.1093/mnras/stad1543},
  urldate = {2023-08-16}
}

@article{long_emergence_2020,
  title = {Emergence of an {{Ultrared}}, {{Ultramassive Galaxy Cluster Core}} at z = 4},
  author = {Long, Arianna S. and Cooray, Asantha and Ma, Jingzhe and Casey, Caitlin M. and Wardlow, Julie L. and Nayyeri, Hooshang and Ivison, R. J. and Farrah, Duncan and Dannerbauer, Helmut},
  year = 2020,
  month = aug,
  journal = {ApJ},
  volume = {898},
  pages = {133},
  issn = {0004-637X},
  doi = {10.3847/1538-4357/ab9d1f},
  urldate = {2023-05-03}
}

@article{lopez-sanjuan_alhambra_2015,
  title = {The {{ALHAMBRA}} Survey: Accurate Merger Fractions Derived by {{PDF}} Analysis of Photometrically Close Pairs},
  shorttitle = {The {{ALHAMBRA}} Survey},
  author = {{L{\'o}pez-Sanjuan}, C. and Cenarro, A. J. and Varela, J. and Viironen, K. and Molino, A. and Ben{\'i}tez, N. and {Arnalte-Mur}, P. and Ascaso, B. and {D{\'i}az-Garc{\'i}a}, L. A. and {Fern{\'a}ndez-Soto}, A. and {Jim{\'e}nez-Teja}, Y. and M{\'a}rquez, I. and Masegosa, J. and Moles, M. and Povi{\'c}, M. and Aguerri, J. A. L. and Alfaro, E. and {Aparicio-Villegas}, T. and Broadhurst, T. and {Cabrera-Ca{\~n}o}, J. and Castander, F. J. and Cepa, J. and Cervi{\~n}o, M. and {Crist{\'o}bal-Hornillos}, D. and Del Olmo, A. and Gonz{\'a}lez Delgado, R. M. and Husillos, C. and Infante, L. and Mart{\'i}nez, V. J. and Perea, J. and Prada, F. and Quintana, J. M.},
  year = 2015,
  month = apr,
  journal = {A\&A},
  volume = {576},
  pages = {A53},
  issn = {0004-6361},
  doi = {10.1051/0004-6361/201424913},
  urldate = {2023-07-22}
}

@article{lopez-sanjuan_massiv_2013,
  title = {{{MASSIV}}: {{Mass Assembly Survey}} with {{SINFONI}} in {{VVDS}}. {{V}}. {{The}} Major Merger Rate of Star-Forming Galaxies at 0.9 \&lt; z \&lt; 1.8 from {{IFS-based}} Close Pairs},
  shorttitle = {{{MASSIV}}},
  author = {{L{\'o}pez-Sanjuan}, C. and Le F{\`e}vre, O. and Tasca, L. a. M. and Epinat, B. and Amram, P. and Contini, T. and Garilli, B. and {Kissler-Patig}, M. and Moultaka, J. and Paioro, L. and Perret, V. and Queyrel, J. and Tresse, L. and Vergani, D. and Divoy, C.},
  year = 2013,
  month = may,
  journal = {A\&A},
  volume = {553},
  pages = {A78},
  issn = {0004-6361},
  doi = {10.1051/0004-6361/201220286},
  urldate = {2022-12-02},
  langid = {english}
}

@article{lotz_caught_2013,
  title = {Caught in the {{Act}}: {{The Assembly}} of {{Massive Cluster Galaxies}} at z = 1.62},
  shorttitle = {Caught in the {{Act}}},
  author = {Lotz, Jennifer M. and Papovich, Casey and Faber, S. M. and Ferguson, Henry C. and Grogin, Norman and Guo, Yicheng and Kocevski, Dale and Koekemoer, Anton M. and Lee, Kyoung-Soo and McIntosh, Daniel and Momcheva, Ivelina and Rudnick, Gregory and Saintonge, Amelie and Tran, Kim-Vy and {van der Wel}, Arjen and Willmer, Christopher},
  year = 2013,
  month = aug,
  journal = {ApJ},
  volume = {773},
  pages = {154},
  issn = {0004-637X},
  doi = {10.1088/0004-637X/773/2/154},
  urldate = {2025-01-02}
}

@article{lubin_observations_2009,
  title = {The {{Observations}} of {{Redshift Evolution}} in {{Large-Scale Environments}} ({{ORELSE}}) {{Survey}}. {{I}}. {{The Survey Design}} and {{First Results}} on {{CL}} 0023+0423 at z = 0.84 and {{RX J1821}}.6+6827 at z = 0.82},
  author = {Lubin, L. M. and Gal, R. R. and Lemaux, B. C. and Kocevski, D. D. and Squires, G. K.},
  year = 2009,
  month = jun,
  journal = {AJ},
  volume = {137},
  pages = {4867--4883},
  issn = {0004-6256},
  doi = {10.1088/0004-6256/137/6/4867},
  urldate = {2024-10-31}
}

@article{madau_cosmic_2014,
  title = {Cosmic {{Star-Formation History}}},
  author = {Madau, Piero and Dickinson, Mark},
  year = 2014,
  month = aug,
  journal = {ARA\&A},
  volume = {52},
  pages = {415--486},
  issn = {0066-4146},
  doi = {10.1146/annurev-astro-081811-125615},
  urldate = {2023-08-16}
}

@article{mantha_major_2018,
  title = {Major Merging History in {{CANDELS}}. {{I}}. {{Evolution}} of the Incidence of Massive Galaxy-Galaxy Pairs from z = 3 to z {$\sim$} 0},
  author = {Mantha, Kameswara Bharadwaj and McIntosh, Daniel H. and Brennan, Ryan and Ferguson, Henry C. and Kodra, Dritan and Newman, Jeffrey A. and Rafelski, Marc and Somerville, Rachel S. and Conselice, Christopher J. and Cook, Joshua S. and Hathi, Nimish P. and Koo, David C. and Lotz, Jennifer M. and Simmons, Brooke D. and Straughn, Amber N. and Snyder, Gregory F. and Wuyts, Stijn and Bell, Eric F. and Dekel, Avishai and Kartaltepe, Jeyhan and Kocevski, Dale D. and Koekemoer, Anton M. and Lee, Seong-Kook and Lucas, Ray A. and Pacifici, Camilla and Peth, Michael A. and Barro, Guillermo and Dahlen, Tomas and Finkelstein, Steven L. and Fontana, Adriano and Galametz, Audrey and Grogin, Norman A. and Guo, Yicheng and Mobasher, Bahram and Nayyeri, Hooshang and {P{\'e}rez-Gonz{\'a}lez}, Pablo G. and Pforr, Janine and Santini, Paola and Stefanon, Mauro and Wiklind, Tommy},
  year = 2018,
  month = apr,
  journal = {MNRAS},
  volume = {475},
  pages = {1549--1573},
  issn = {0035-8711},
  doi = {10.1093/mnras/stx3260},
  urldate = {2022-11-03}
}

@article{mcconachie_spectroscopic_2022,
  title = {Spectroscopic {{Confirmation}} of a {{Protocluster}} at z = 3.37 with a {{High Fraction}} of {{Quiescent Galaxies}}},
  author = {McConachie, Ian and Wilson, Gillian and Forrest, Ben and Marsan, Z. Cemile and Muzzin, Adam and Cooper, M. C. and Annunziatella, Marianna and Marchesini, Danilo and Chan, Jeffrey C. C. and Gomez, Percy and Abdullah, Mohamed H. and Saracco, Paolo and Nantais, Julie},
  year = 2022,
  month = feb,
  journal = {ApJ},
  volume = {926},
  pages = {37},
  issn = {0004-637X},
  doi = {10.3847/1538-4357/ac2b9f},
  urldate = {2023-05-12}
}

@article{mccracken_ultravista_2012,
  title = {{{UltraVISTA}}: A New Ultra-Deep near-Infrared Survey in {{COSMOS}}},
  shorttitle = {{{UltraVISTA}}},
  author = {McCracken, H. J. and {Milvang-Jensen}, B. and Dunlop, J. and Franx, M. and Fynbo, J. P. U. and Le F{\`e}vre, O. and Holt, J. and Caputi, K. I. and Goranova, Y. and Buitrago, F. and Emerson, J. P. and Freudling, W. and Hudelot, P. and {L{\'o}pez-Sanjuan}, C. and Magnard, F. and Mellier, Y. and M{\o}ller, P. and Nilsson, K. K. and Sutherland, W. and Tasca, L. and Zabl, J.},
  year = 2012,
  month = aug,
  journal = {A\&A},
  volume = {544},
  pages = {A156},
  issn = {0004-6361},
  doi = {10.1051/0004-6361/201219507},
  urldate = {2023-05-02}
}

@article{mcgee_accretion_2009,
  title = {The Accretion of Galaxies into Groups and Clusters},
  author = {McGee, Sean L. and Balogh, Michael L. and Bower, Richard G. and Font, Andreea S. and McCarthy, Ian G.},
  year = 2009,
  month = dec,
  journal = {MNRAS},
  volume = {400},
  number = {2},
  pages = {937--950},
  issn = {0035-8711},
  doi = {10.1111/j.1365-2966.2009.15507.x},
  urldate = {2025-01-02},
  langid = {english}
}

@article{mclean_design_2010,
  title = {Design and Development of {{MOSFIRE}}: The Multi-Object Spectrometer for Infrared Exploration at the {{Keck Observatory}}},
  shorttitle = {Design and Development of {{MOSFIRE}}},
  author = {McLean, Ian S. and Steidel, Charles C. and Epps, Harland and Matthews, Keith and Adkins, Sean and Konidaris, Nicholas and Weber, Bob and Aliado, Ted and Brims, George and Canfield, John and Cromer, John and Fucik, Jason and Kulas, Kristin and Mace, Greg and Magnone, Ken and Rodriguez, Hector and Wang, Eric and Weiss, Jason},
  year = 2010,
  month = jul,
  journal = {Proc. SPIE},
  volume = {7735},
  pages = {77351E},
  doi = {10.1117/12.856715},
  urldate = {2023-08-18}
}

@article{mclean_mosfire_2012,
  title = {{{MOSFIRE}}, the Multi-Object Spectrometer for Infra-Red Exploration at the {{Keck Observatory}}},
  author = {McLean, Ian S. and Steidel, Charles C. and Epps, Harland W. and Konidaris, Nicholas and Matthews, Keith Y. and Adkins, Sean and Aliado, Theodore and Brims, George and Canfield, John M. and Cromer, John L. and Fucik, Jason and Kulas, Kristin and Mace, Greg and Magnone, Ken and Rodriguez, Hector and Rudie, Gwen and Trainor, Ryan and Wang, Eric and Weber, Bob and Weiss, Jason},
  year = 2012,
  month = sep,
  journal = {Ground-based and Airborne Instrumentation for Astronomy IV, Proc. SPIE},
  volume = {8446},
  pages = {84460J},
  doi = {10.1117/12.924794},
  urldate = {2023-08-18}
}

@article{mcnab_gogreen_2021,
  title = {The {{GOGREEN}} Survey: Transition Galaxies and the Evolution of Environmental Quenching},
  shorttitle = {The {{GOGREEN}} Survey},
  author = {McNab, Karen and Balogh, Michael L. and {van der Burg}, Remco F. J. and Forestell, Anya and Webb, Kristi and Vulcani, Benedetta and Rudnick, Gregory and Muzzin, Adam and Cooper, M. C. and McGee, Sean and Biviano, Andrea and Cerulo, Pierluigi and Chan, Jeffrey C. C. and De Lucia, Gabriella and Demarco, Ricardo and Finoguenov, Alexis and Forrest, Ben and Golledge, Caelan and Jablonka, Pascale and Lidman, Chris and Nantais, Julie and Old, Lyndsay and {Pintos-Castro}, Irene and Poggianti, Bianca and Reeves, Andrew M. M. and Wilson, Gillian and Yee, Howard K. C. and Zaritsky, Dennis},
  year = 2021,
  month = nov,
  journal = {MNRAS},
  volume = {508},
  pages = {157--174},
  issn = {0035-8711},
  doi = {10.1093/mnras/stab2558},
  urldate = {2022-01-18}
}

@article{mesa_interacting_2014,
  title = {Interacting Galaxies: Corotating and Counter-Rotating Systems with Tidal Tails},
  shorttitle = {Interacting Galaxies},
  author = {Mesa, Valeria and Duplancic, Fernanda and Alonso, Sol and Coldwell, Georgina and Lambas, Diego G.},
  year = 2014,
  month = feb,
  journal = {MNRAS},
  volume = {438},
  pages = {1784--1793},
  issn = {0035-8711},
  doi = {10.1093/mnras/stt2317},
  urldate = {2023-05-04}
}

@article{miyazaki_hyper_2018,
  title = {Hyper {{Suprime-Cam}}: {{System}} Design and Verification of Image Quality},
  shorttitle = {Hyper {{Suprime-Cam}}},
  author = {Miyazaki, Satoshi and Komiyama, Yutaka and Kawanomoto, Satoshi and Doi, Yoshiyuki and Furusawa, Hisanori and Hamana, Takashi and Hayashi, Yusuke and Ikeda, Hiroyuki and Kamata, Yukiko and Karoji, Hiroshi and Koike, Michitaro and Kurakami, Tomio and Miyama, Shoken and Morokuma, Tomoki and Nakata, Fumiaki and Namikawa, Kazuhito and Nakaya, Hidehiko and Nariai, Kyoji and Obuchi, Yoshiyuki and Oishi, Yukie and Okada, Norio and Okura, Yuki and Tait, Philip and Takata, Tadafumi and Tanaka, Yoko and Tanaka, Masayuki and Terai, Tsuyoshi and Tomono, Daigo and Uraguchi, Fumihiro and Usuda, Tomonori and Utsumi, Yousuke and Yamada, Yoshihiko and Yamanoi, Hitomi and Aihara, Hiroaki and Fujimori, Hiroki and Mineo, Sogo and Miyatake, Hironao and Oguri, Masamune and Uchida, Tomohisa and Tanaka, Manobu M. and Yasuda, Naoki and Takada, Masahiro and Murayama, Hitoshi and Nishizawa, Atsushi J. and Sugiyama, Naoshi and Chiba, Masashi and Futamase, Toshifumi and Wang, Shiang-Yu and Chen, Hsin-Yo and Ho, Paul T. P. and Liaw, Eric J. Y. and Chiu, Chi-Fang and Ho, Cheng-Lin and Lai, Tsang-Chih and Lee, Yao-Cheng and Jeng, Dun-Zen and Iwamura, Satoru and Armstrong, Robert and Bickerton, Steve and Bosch, James and Gunn, James E. and Lupton, Robert H. and Loomis, Craig and Price, Paul and Smith, Steward and Strauss, Michael A. and Turner, Edwin L. and Suzuki, Hisanori and Miyazaki, Yasuhito and Muramatsu, Masaharu and Yamamoto, Koei and Endo, Makoto and Ezaki, Yutaka and Ito, Noboru and Kawaguchi, Noboru and Sofuku, Satoshi and Taniike, Tomoaki and Akutsu, Kotaro and Dojo, Naoto and Kasumi, Kazuyuki and Matsuda, Toru and Imoto, Kohei and Miwa, Yoshinori and Suzuki, Masayuki and Takeshi, Kunio and Yokota, Hideo},
  year = 2018,
  month = jan,
  journal = {PASJ},
  volume = {70},
  pages = {S1},
  issn = {0004-6264},
  doi = {10.1093/pasj/psx063},
  urldate = {2023-08-19}
}

@article{momcheva_3d-hst_2016,
  title = {The {{3D-HST Survey}}: {{Hubble Space Telescope WFC3}}/{{G141 Grism Spectra}}, {{Redshifts}}, and {{Emission Line Measurements}} for \textasciitilde{} 100,000 {{Galaxies}}},
  shorttitle = {The {{3D-HST Survey}}},
  author = {Momcheva, Ivelina G. and Brammer, Gabriel B. and {van Dokkum}, Pieter G. and Skelton, Rosalind E. and Whitaker, Katherine E. and Nelson, Erica J. and Fumagalli, Mattia and Maseda, Michael V. and Leja, Joel and Franx, Marijn and Rix, Hans-Walter and Bezanson, Rachel and Da Cunha, Elisabete and Dickey, Claire and F{\"o}rster Schreiber, Natascha M. and Illingworth, Garth and Kriek, Mariska and Labb{\'e}, Ivo and Ulf Lange, Johannes and Lundgren, Britt F. and Magee, Daniel and Marchesini, Danilo and Oesch, Pascal and Pacifici, Camilla and Patel, Shannon G. and Price, Sedona and Tal, Tomer and Wake, David A. and {van der Wel}, Arjen and Wuyts, Stijn},
  year = 2016,
  month = aug,
  journal = {ApJS},
  volume = {225},
  pages = {27},
  issn = {0067-0049},
  doi = {10.3847/0067-0049/225/2/27},
  urldate = {2023-05-02}
}

@article{moneti_vizier_2023,
  title = {{{VizieR Online Data Catalog}}: {{The}} Fourth {{UltraVISTA}} Data Release ({{DR4}}) ({{Moneti}}+, 2019)},
  shorttitle = {{{VizieR Online Data Catalog}}},
  author = {Moneti, A. and McCracken, H. J. and Hudelot, W. and Rouberol, O. and Herent, O. and Mellier, Y. and Dunlop, J. and Le Fevre, O. and Franx, M. and Fynbo, J. and Bowler, R. and Caputi, K. and Kauffmann, O. and {Milvang-Jensen}, B. and {Gonzalez-Fernandez}, C. and {Gonzalez-Solares}, E. and Irwin, M. and Lewis, J. and Blake, R. and Cross, N. and Read, M. and Sutorius, E.},
  year = 2023,
  month = jan,
  journal = {VizieR Online Data Catalog},
  pages = {II/373},
  urldate = {2023-08-19}
}

@article{moore_galaxy_1996,
  title = {Galaxy Harassment and the Evolution of Clusters of Galaxies},
  author = {Moore, Ben and Katz, Neal and Lake, George and Dressler, Alan and Oemler, Augustus},
  year = 1996,
  month = feb,
  journal = {Nature},
  volume = {379},
  pages = {613--616},
  issn = {0028-0836},
  doi = {10.1038/379613a0},
  urldate = {2021-06-10}
}

\begin{appendix}

\section{Redshift PDF sampling}
\label{app:zpdfs}

\begin{figure}[t]
    \centering
    \includegraphics[width=1\linewidth]{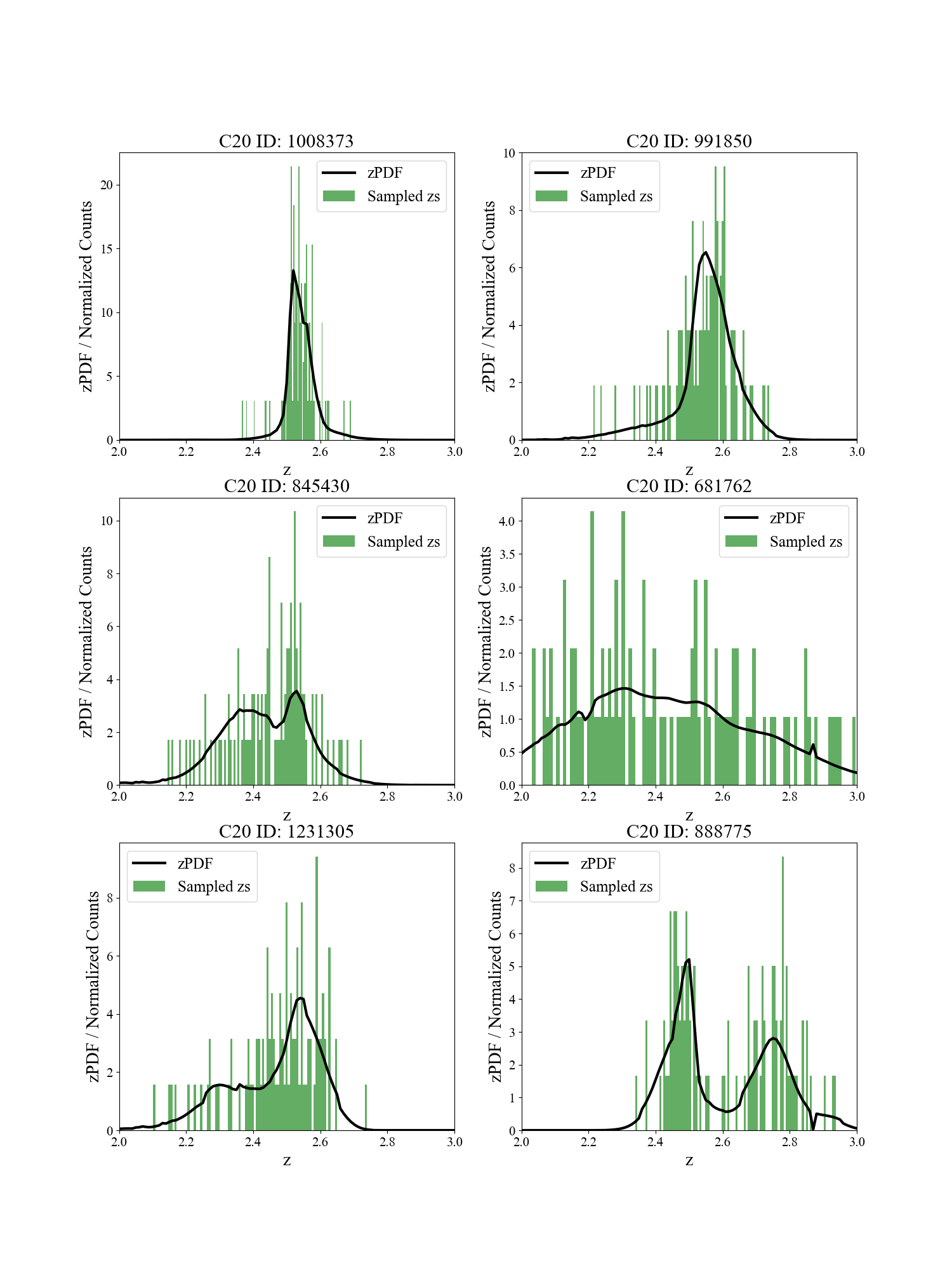}
    \caption{Redshift PDFs and sampled redshifts from MC process for 6 sources from $2<z<3$ in our final sample (COSMOS2020 IDs: 1008373, 991850, 845430, 681762, 1231305, and 888775). The corresponding zPDFs from COSMOS2020 (Classic Catalog \texttt{LePhare}) are plotted in black while the selected redshifts from our MC process are shown in the green histograms (with the number of bins $=$ the number of redshifts selected from $2<z<3$ to better show each chosen redshift). Our 100 MC iterations accurately sample a variety of PDF shapes and capture secondary peaks. For C20 object \#681762 (shown in the middle right panel), 6 chosen redshifts fall outside of $2<z<3$ (which is expected as that PDF has portions outside $2<z<3$).}
    \label{fig:ex_zPDFs}
\end{figure}

As a majority of sources in our final sample have only a photometric redshift available, we account for the relatively wide errors on their photo-zs by drawing from the corresponding COSMOS2020 \texttt{LePhare} redshift PDF for each iteration of our MC process (we also occasionally draw from the COSMOS2020 \texttt{LePhare} zPDF for sources with spectroscopic and/or grism redshifts, see Section \ref{sec:zMC}). To demonstrate that our 100 MC iterations capture the uncertainty of wider zPDFs and secondary redshift peaks, we include here Figure \ref{fig:ex_zPDFs}. In this figure, we plot the zPDF of 6 semi-randomly selected photometric (only) sources from our final sample (COSMOS2020 IDs: 1008373, 991850, 845430, 681762, 1231305, and 888775). These sources were selected to show varying levels of complexity in zPDF shape and all have a redshift peak(s) in our relevant redshift range of $2<z<3$. Also included are histograms of the 100 redshifts sampled from the zPDF during our MC process described in Section \ref{sec:zMC} (with a number of bins $=$ the number of redshifts selected from $2<z<3$) to show how our process samples each PDF. We find that our 100 MCs sample secondary peaks and capture the overall shape of the zPDF. 

While increasing the number of MC iterations could improve our sampling the underlying zPDFs, this will not improve the recovery of true companion galaxies (i.e., it will not improve the overall recovered fraction nor improve the purity or completeness of identified companion galaxies). Satisfying the stringent line-of-sight criteria of $\Delta v < 1000$ km/s (or $\Delta z \lesssim 0.01$) for companion galaxies is difficult, especially when both sources are photometric and need to simultaneously scatter to the $\sim$same redshift. When we examine the purity and completeness of identified companion galaxies within our 100 MC trials using the mock lightcone (see Section \ref{sec:cone}), we find that the average standard deviation in the purity is $\sim0.9$\% and $\sim0.3$\% for the completeness (these values are obtained by taking the average of the median $\sigma$ of all 5 SzFs used for testing) and thus with 100 MC iterations we are well sampling the distributions of purity and completeness. With such little scatter in the recovered purity and completeness, increasing the number of MC iterations will not meaningfully change our results.

\end{appendix}

\end{document}